\documentclass[%
 preprint,
 amsmath,amssymb,
 aps,
]{revtex4-1}

\usepackage{graphicx}% Include figure files
\usepackage{dcolumn}% Align table columns on decimal point
\usepackage{bm}% bold math
\usepackage{hyperref}% add hypertext capabilities
\usepackage{dblfloatfix}
\usepackage{float}
\usepackage[T1]{fontenc}
\usepackage{xcolor}
\usepackage{cases}
\usepackage{bbm}
\usepackage{cleveref}
\begin{document}

\title{ Cusp of  non - Gaussian density of  particles for a  diffusing diffusivity model
}% Force line breaks with \\
\author{M. Hidalgo-Soria}
\email{mariohidalgosoria@gmail.com}
\author{E. Barkai}%
% \email{Eli.Barkai@biu.ac.il}
\affiliation{%
 Department of Physics, Institute of Nanotechnology and Advanced Materials, Bar-Ilan University, Ramat-Gan 5290002, Israel\\
}%
\author{S. Burov}
\affiliation{%
Department of Physics, Bar-Ilan University, Ramat-Gan 5290002, Israel
}%

\date{\today}
\begin{abstract}
We study a two state ``jumping diffusivity'' model for a Brownian process alternating between two different diffusion constants,  $D_{+}>D_{-}$,  with random waiting times in both states  whose distribution is rather general. In the limit of long measurement times  Gaussian behavior with an effective diffusion coefficient is recovered. We show that for equilibrium initial conditions and when the limit of the diffusion coefficient $D_-\to0$ is taken, the short time  behavior leads to a cusp, namely a non - analytical behavior, in the distribution of the displacements $P(x,t)$ for $x\longrightarrow 0$. Visually this cusp, or tent-like shape, resembles  similar behavior  found in many experiments of diffusing particles in disordered environments, such as glassy systems and intracellular media. This general result depends only on the existence of finite mean values of the waiting times at the different states of the model. Gaussian statistics in the long time limit is achieved due to ergodicity and convergence of the distribution of the temporal occupation fraction in state $D_{+}$ to a $\delta$-function. The short time behavior  of the same quantity converges  to a uniform distribution, which leads to the non - analyticity in $P(x,t)$. We demonstrate how super - statistical framework is a zeroth order short time expansion of $P(x,t)$,  in the number of transitions,  that does not yield the cusp like shape. The latter, considered as the key feature  of experiments in the field, is found with the first  correction in perturbation theory. 
\end{abstract}

\pacs{Valid PACS appear here}
                             
\maketitle
\section{\label{sec:1} \textbf{INTRODUCTION}} 
The emergence of  non-Gaussian features for the positional probability density function (PDF) of particle spreading, denoted  $P(x,t)$,  in a disordered environment is a common attribute that arises in many different physical and biological systems.
Specifically a tent like shape of the PDF, in the semi-log scale,  together with a linear time dependence of the mean square displacement (MSD) appear for    diffusion in glassy systems ~\cite{kob2007}, biological cells ~\cite{hapca2009,wang2009,leptos2009,granick2012,spako2017,Sabri2020} and colloidal suspensions ~\cite{Weeks2000,Kegel2000,Yael2020,Lava2020}. This tent shape, sometimes fitted with a Laplace distribution $ P(x,t) \sim \exp(- C \vert x \vert)$  with $C$ a constant, suggests that the decay of the PDF is exponential. This feature is becoming a more frequent observation for the spreading of molecules. Phenomenological approaches are diffusing diffusivity models, in which non-Gaussianity  is obtained by coupled stochastic  differential equations with random diffusion coefficients ~\cite{chuby2014,aki2016,chechk2017,vittoria2018,LanG2018,Sposini_2018,jakub_2019,GSMO2020,Wang2020,Sabri2020}, and  path integrals formalism for Brownian motion in the presence of a sink ~\cite{Seb2016}. More recently, theoretical frameworks describing this behavior emerged from continuous time random walk (CTRW) approaches employing large deviations theory ~\cite{bb2019,WBB2020,pozo2020} and microscopical models like molecular dynamics of tracer particles in polymer networks ~\cite{Raja2016,KRaja2019} and interacting particles with fluctuating sizes ~\cite{Flavio2019,HSB2020,yin2021}, the so called Hitchhiker model ~\cite{HSB2020}. 

While in some of the systems the non-Gaussian behavior disappears when the measurement time is made long enough, the short time tent-like decay of the PDF seems to be a universal phenomenon ~\cite{bb2019}.   It is then natural to ask if there is some sort of universality which can be deduced for the temporal limit of short times. Within  the diffusing diffusivity models for the large $x$ limit, exponentially decaying propagators have been observed by employing a dichotomous process for the diffusivity ~\cite{aki2016}. The latter model consists  of  a ``fast'' and a ``slow'' phases, each one with a diffusion coefficient $D_{+}$ and $D_{-}$ respectively \cite{aki2016,Sebas2020}. Furthermore, the appearance of a cusp at small displacements also has been reported in different diffusive approaches like the Sinai model ~\cite{BOU1990},  employing the quenched trap model  ~\cite{Month2003,BB2011PRL,BB2012PRE,Luo2018,Luo2019,Post2020} or spatial dependence in the diffusivity ~\cite{Regev2016}, within the L\'{e}vy - Lorentz gas model  ~\cite{Radice2019}  and using the fractional Fokker - Planck equation ~\cite{Barkai2001}. Important to notice is that the cusp found in ~\cite{BOU1990,Barkai2001,Month2003,BB2011PRL,BB2012PRE,Regev2016,Radice2019,Luo2019} is within the context of anomalous diffusion in the MSD sense, and those presented in \cite{Luo2018,Luo2019,Post2020} are for normal diffusive systems.

It is worth mentioning that several systems in nature exhibit, or can be reduced to a dichotomous process. Examples of two state systems include  nuclear magnetic imaging to measure the diffusion of  heterogeneous molecules ~\cite{krager1985}, diffusion in glassy materials ~\cite{kob2007}, blinking quantum dots ~\cite{MarB2004,Ahar2019}, diffusion in single molecules  tracking experiments ~\cite{leptos2009,Sabri2020} and protein conformational dynamics ~\cite{Yama2020}. Other approaches for analyzing two state systems were also devised over the years, see heterogeneous molecular transport ~\cite{krager1985} telegraphic noise ~\cite{Ahar2019}, L{\`e}vy Flights ~\cite{Kana2020} and CTRW models ~\cite{kob2007,bb2019}. 

In this work we deal with a two state jumping diffusivity model with equilibrium initial conditions, \textit{i.e.}  we assume that the process started long  before the measurement began.  
The long measurement time behavior of the positional PDF for this model is Gaussian and is independent of the specifics of the waiting times at the different diffusive states. A rather unexpected result is achieved for the opposite temporal regime. We obtain that the behavior in the limit of the short measurement times,  the shape of the positional PDF of the molecule spreading in the two state jumping diffusivity model attains a cusp or a general tent-like shape.
Our result is based on  the statistics of the temporal occupation fraction of the diffusivity states, the latter is defined as the time spent in state $D_{+}$ over the total measurement time. 
The Gaussian behavior in the long measurement time is dictated by the $\delta$-function shape of the distribution of this temporal occupation fraction, a feature that is solely based on the ergodic properties of the system. 
We show that in the limit of short measurement time the distribution of the temporal occupation fraction attains a uniform distribution, that leads to the mentioned cusp behavior of $P(x,t)$.  The uniformity of the occupation fraction is a general result in the sense that it does not depend on the statistics of the waiting times in the two states, the latter can be arbitrary. The non - Gaussian behavior of $P(x,t)$ for short measurement times is similarly general  as the Gaussian behavior of the propagator  for long times.
We then show that our  approach reproduces the  results of a specific representative system with  exponentially distributed waiting times.

Our manuscript is organized as follows, in Section~\ref{sec:2} we introduce  the jumping diffusivity model and the  initial conditions utilized in this work. In Section~\ref{sec:3} we develop our theory for  the statistics of the occupation time in the short measurement time limit. For which the PDFs of the waiting times in states $D_{\pm}$ are rather general.  The obtained  behavior of the occupation fraction is used in order to describe the non-Gaussian features of $P(x,t)$, \textit{i.e.} its cusp shape,  that is observed in this model. In Section~\ref{sec:4} we corroborate our previous results for a system with exponentially distributed waiting times.  In section~\ref{sec:45} we discuss briefly how these theoretical results differ from those found within the super - statistical approach \cite{chuby2014,Seb2016,chechk2017} and further how our approach may be applicable in experiments. Finally in Section~\ref{sec:5} we present a summary of our results,  and we discuss briefly recent work of Postnikov \textit{et al.} ~\cite{Post2020} who considered a model with quenched disorder, emphasizing the  importance of equilibrium initial conditions. The main derivations are given in the corresponding Appendixes.  

%%%%%%%%%%%%%%%%%%%%%%%%%%%%%%%%%%%%%%%%%%%%%%%%%%%%%%%%%%%%%%%%%%%%%%%%%%%%%%%%%%%%%%%%%%%%%%%%%%%%%
\subsection{\label{sec:2} The model}  
\begin{figure}[H]
\centering
\includegraphics[width=10.5 cm]{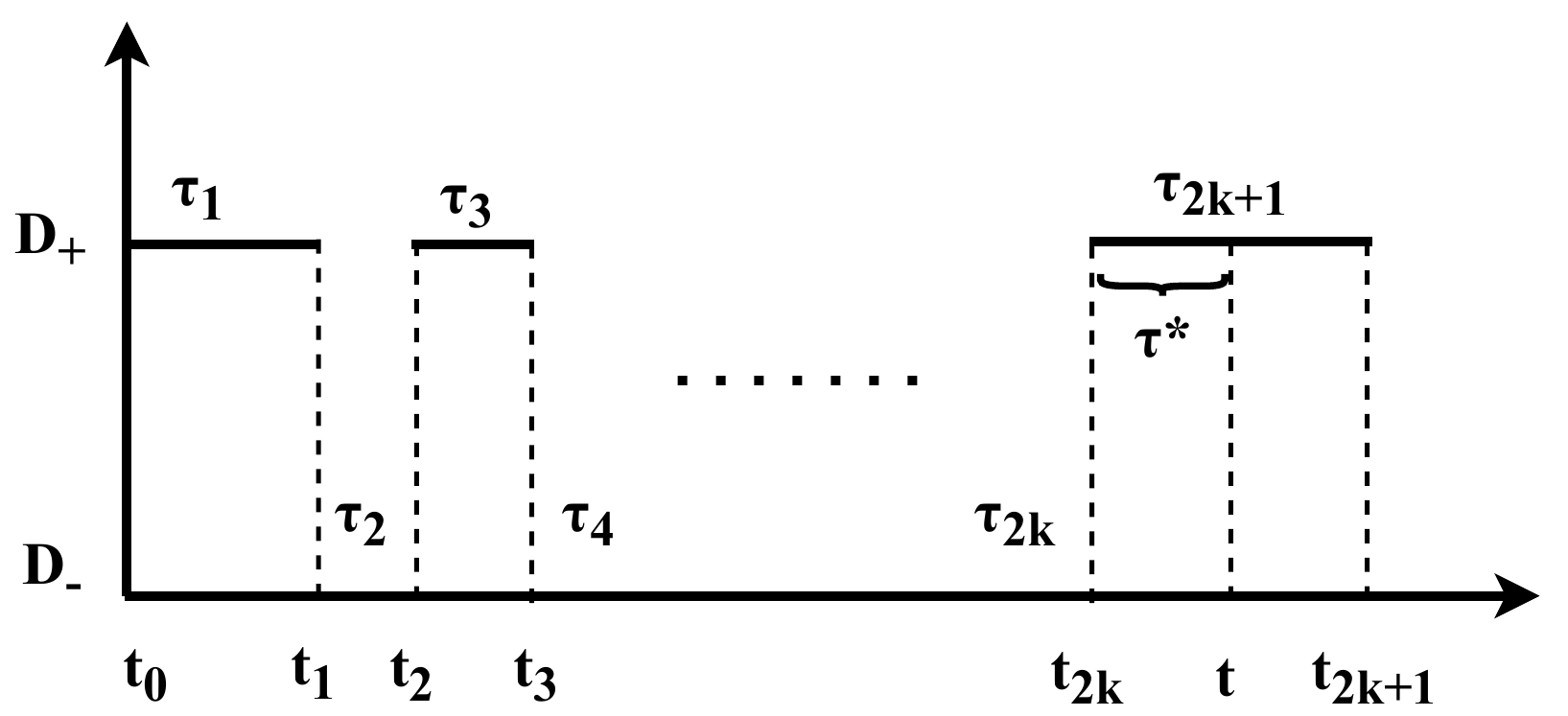}
\caption{Alternating process for the diffusivity, starting from the state `$+$' and $N=2k+1$. For the case of equilibrium initial conditions exposed in section ~\ref{sec:3}, for $N=1$ $\tau_{1}$ works as the forward recurrence time with PDF Eq.~\eqref{eq:pdfforward}.\label{fig:2D}}
\end{figure}

We consider a two state renewal model, with a stochastic diffusion field $D(t)$ for a particle in a random medium. The position of the particle is following a diffusion process given by $dx(t)/dt=\sqrt{2D(t)}\xi$. With $D(t)\in\lbrace D_{+},D_{-} \rbrace$ a dichotomous model, considering the case when $D_{-}<D_{+}$ and $\xi$ a standard white noise, \textit{i.e.} with mean zero, variance one  and delta correlated. As an example of the dynamics of the model, at a given time the particle  follows a pure diffusion process  with a diffusion coefficient $D_{+}>0$ during a period $\tau$. After this time period has elapsed,  the diffusion coefficient jumps and during the next time  interval the particle diffuses with diffusion coefficient $D_{-}$. The waiting times at each state $D_{\pm}$  are distributed according to a general PDF $\psi_{\pm}(\tau)$, with mean waiting times $\langle\tau\rangle_{\pm}$. 
The subscript $\pm$ denotes whether the waiting times are defined for the $D_{+}$ or $D_{-}$ states. In the following we present the two-state model with $D_{-}=0$, while the case with $D_{+}>D_{-}>0$ is analyzed in the Appendix~\ref{sec:A0}.  In Figure~\ref{fig:2DX}  we show representative trajectories for the  position at time $t$, $x(t)$. While in Figure~\ref{fig:2D} we present the same  for $D(t)$ and we show the notation we use. 

\begin{figure}[H]
\centering
\includegraphics[width=8.5 cm]{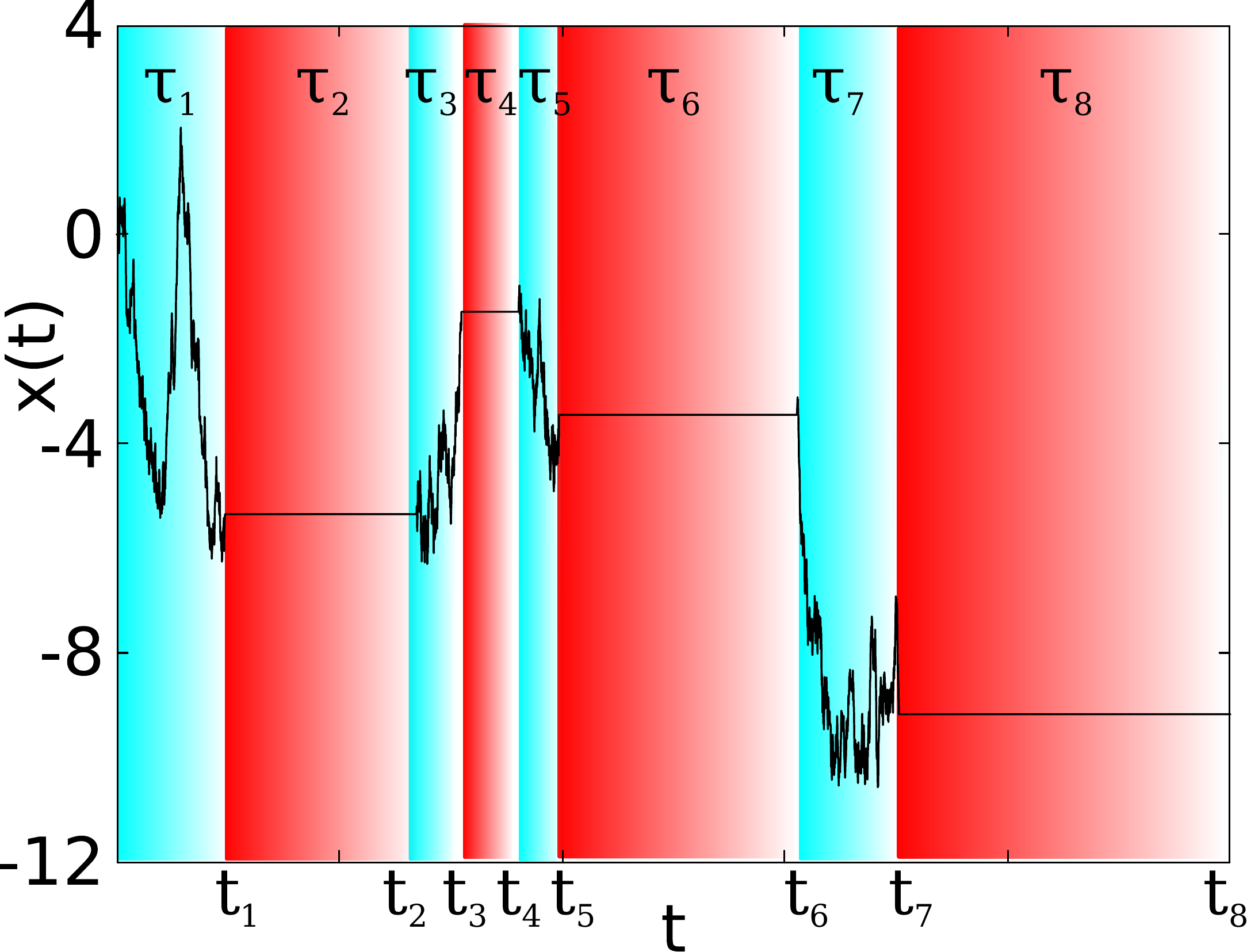}
\caption{Typical trajectory of $x(t)$ given by Eq.~\eqref{eq:PDTp} with $D_{+}=10$ (blue regions), and $D_{-}=0$ (red regions). For this trajectory exponential waiting times with $\langle \tau \rangle_{+} =1$ and $\langle \tau \rangle_{-} =5$ were used.\label{fig:2DX}}
\end{figure}   

We define $T_\pm$ as the occupation time in state ``$\pm$'', namely the total amount of time that the process diffuses with $D_{+}$ or $D_{-}$ during $t$. Jumps between states $D_{+}$ and $D_{-}$ occur at random times $t_{1}$, $t_{2}$, etc., until a final measurement time $t$ and clearly $t = T_{+} + T_{-}$. The intervals of time between each jump are defined by $\tau_{1}=t_{1}$, $\tau_{2}=t_{2}-t_{1}$, $\tau_{3}=t_{3}-t_{2}$, etc., see Figure~\ref{fig:2D}. Then the occupation times in each state, when started from $D_{+}$, are explicitly provided by 
\begin{eqnarray}\label{eq:Tpo}
T_{+}&=&\tau_{1}+\tau_{3}+\ldots+\tau_{N} \nonumber\\
T_{-}&=&\tau_{2}+\tau_{4}+\ldots+\tau_{N-1}+\tau^{\ast} \nonumber\,\,\,\ if \,\,\,\ N=2k+1,\label{eq:Tmo}\\
T_{+}&=&\tau_{1}+\tau_{3}+\ldots+\tau_{N-1}+\tau^{\ast} \nonumber,\label{eq:Tpe}\\
T_{-}&=&\tau_{2}+\tau_{4}+\ldots+\tau_{N} \,\,\,\,\,\,\,\,\,\,\,\,\,\,\,\,\,\,\,\,\,\,\,\ if \,\,\,\ N=2k,
\end{eqnarray}
where $N$ is the random number of transitions  that were performed between the two states during the measurement time $t$ and $k$ is an integer. The measurement time $t$ and $N$ satisfy  $t\geq t_{N}$, with $t_{N}=\tau_{1}+\tau_{2}+\ldots + \tau_{N}$, \textit{i.e.} the exact time when the $N$th jump was performed. The backward recurrence time $\tau^{\ast}$ is defined by $\tau^{\ast}=t-t_{N}$ ~\cite{GL2001}. Each waiting time $\tau_i$ follows $\tau_{i}=t_{i}-t_{i-1}$ with $i\in(1,N)$. For this particular initial condition, odd values of $i$ in $\tau_{i}$ refer to waiting times at $D_{+}$ and even values of $i$ to waiting times during which the diffusion coefficient is  $D_{-}$ (see Figure~\ref{fig:2D}). Expressions similar to Eq.~\eqref{eq:Tpo} are also obtained when the process starts from $D_{-}$, see Eq.~\eqref{eq:Tpodm} . 

Since the particle is diffusing with a constant diffusion constant $D_+$ for time $\tau_1$, when starting from $D_+$, the position $x(\tau_1)$ is simply $x(\tau_1)=\sqrt{2D_+\tau_1}\xi_1$, where $\xi_1$ is a zero mean Gaussian random variable with $\langle \xi_1^2\rangle=1$. When at the state with diffusion constant $D_-=0$, the particle is not moving, therefore $x(t_2)-x(t_1)=0$ and $x(t_3)-x(t_2)=\sqrt{2D_+\tau_3}\xi_3$, where $\xi_3$ is a zero mean Gaussian random variable with $\langle \xi_3^2\rangle=1$ independent of $\xi_1$. Generally,  $x(t_i)-x(t_{i-1})=\sqrt{2D_\pm\tau_i}\xi_i$, where all $\xi_i$ are independent zero mean Gaussian random variables that satisfy $\langle \xi_i^2\rangle=1$. By using Eq.~\eqref{eq:Tpo} and exploiting the properties of summation of independent Gaussian variables we obtain that the position at general time $t$ is provided by
%%%%%%%%%%%%%%%%%%%%%%%%%%%%%%%%%%%%%
\begin{eqnarray}\label{eq:PDTp}
x(t)=\sqrt{2D_{+}T_{+}}\xi,
\end{eqnarray}
%%%%%%%%%%%%%%%%%%%%%%%%%%%%%%%%%%%%%
when $D_{+}>0$ and $D_{-}=0$. Eq.~\eqref{eq:PDTp} holds irrespective of the state at $t=0$. We see that the particles' position is a product of two independent random variables, the square root of the time staying at the state $D_{+}$ times a standard Gaussian  random variable. 

In the following we consider a situation in which the process has started long before the measurement began, \textit{i.e.} at $t=0$ the process was already running for a very long time.  In this way the measurement begins from an initial condition in which the system is  in equilibrium, meaning that the probability to start from $D_{+}$ is   $\langle\tau\rangle_{+} /[\langle\tau\rangle_{+} + \langle\tau\rangle_{-}]$ and accordingly the probability to start from $D_{-}$ is  $\langle\tau\rangle_{-} /[\langle\tau\rangle_{+} + \langle\tau\rangle_{-}]$ (see ~\cite{GL2001,Aki2019}).  For this set-up the PDF of the occupation time $T_{+}$, $f_t(T_+)$, is determined by the contribution to start from $D_{+}$ and the contribution to start from $D_{-}$,   yielding 
%%%%%%%%%%%%%%%%%%%%%%%%%%%%%%%%%%%%%%%%
\begin{eqnarray}\label{eq:pdfTp}
f_{t}(T_{+})=\frac{\langle\tau\rangle_{+} }{ \langle\tau \rangle_{+} + \langle\tau \rangle_{-}} f^{+}_{t}(T_{+})+ \frac{\langle\tau\rangle_{-} }{ \langle\tau \rangle_{+} + \langle\tau \rangle_{-}}f^{-}_{t}(T_{+}),
\end{eqnarray}
%%%%%%%%%%%%%%%%%%%%%%%%%%%%%%%%%%%%%%%%%%
here $f^{\pm}_{t}(T_{+})$ is the PDF of $T_{+}$ for measurement time $t$, given that the process has started from $\pm$. 
%%%%%%%%%%%%%%%%%%%%%%%%%%%%%%%%%%%%%%%%%%%%%%%%%%
Since $D_{-}=0$, Eq.~\eqref{eq:PDTp} dictates that the positional PDF, provided that the system has occupied the state with $D_{+}$ for a time $T_{+}$, is given by
\begin{eqnarray}\label{eq:pxTpc1}
P(x\vert T_{+})=\frac{e^{-\frac{x^{2}}{4 D_{+} T_{+}} } }{\sqrt{4\pi D_{+}T_{+}}}.
\end{eqnarray}
The propagator of the system is obtained via integrating over all possible values of the occupation time $T_{+}$, whose PDF is  $f_{t}(T_{+})$ Eq.~\eqref{eq:pdfTp}, yielding    
\begin{eqnarray}\label{eq:pxT}
P(x,t)&=&\displaystyle \int \limits_{0}^{t} P(x\vert T_{+})f_{t}(T_{+})dT_{+}.
\end{eqnarray} 
%%%%%%%%%%%%%%%%%%%%%%%%%%
Likewise we can work with the temporal  occupation fraction, which  is defined by $p_{+}=T_{+}/t$ with $ 0 \leq p_{+}\leq 1$. In this case the positional PDF for a specific value of $p_{+}$ follows
%%%%%%%%%%%%%%%%%%%%%%%%%%%%%%%%%%%%%%%%
\begin{eqnarray}\label{eq:pxp}
P(x\vert p_{+})=\frac{e^{-\frac{x^{2}}{4 D_{+} tp_{+}} } }{\sqrt{4\pi D_{+}tp_{+}}}.
\end{eqnarray}
%%%%%%%%%%%%%%%%%%%%%%%%%%%%%%%%%%%%%%%%
and the propagator is obtained similarly to Eq.~\eqref{eq:pxT}, but using the PDF of $p_{+}$, which we denote by $g_{t}(p_{+})$, 
%%%%%%%%%%%%%%%%%%%%%%%%%%%%%%%%%%%%
\begin{eqnarray}\label{eq:pxpp}
P(x,t)&=&\displaystyle \int \limits_{0}^{1} P(x\vert p_{+})g_{t}(p_{+})dp_{+}.
\end{eqnarray} 
%%%%%%%%%%%%%%%%%%%%%%%%%%
Since the properties of $P(x|T_+)$ or $P(x|p_+)$ are known, the task of computing the propagator completely depends on our ability to calculate the PDF of $T_+$ or $p_+$. In the following section we address this problem.
\subsection{\label{sec:3} The general case: Arbitrary distribution of waiting times}
Two regimes of the process are of special interest. The long and the short limits of the measurement time $t$. The two different limits involve different considerations when computing the PDFs of the occupation time ($T_+$) and fraction ($p_+)$. We first handle the regime of small $t$ and then we treat the $t\to\infty$ limit.  
\subsubsection{Short time regime}
The PDF of the occupation time $T_+$ is defined by Eq.~\eqref{eq:pdfTp}. We condition on the number of transitions $N$ and each term $f_t^\pm (T_+)$ is provided by
\begin{eqnarray}
f_{t}^{\pm}(T_{+})=\displaystyle \sum \limits_{N=0}^{\infty} f_{t}^{\pm}(T_{+}\vert N)Q_{t}^{\pm}(N) , 
\label{eqn:subordS}
\end{eqnarray}
where $Q^{\pm}_{t}(N)$ is the probability to perform exactly $N$ transitions during $t$ when the process started at $\pm$. $f_{t}^{\pm}(T_{+}\vert N)$ is the PDF of $T_{+}$  when exactly $N$ transitions were performed (during $t$), and the process has started from $\pm$. This conditional probability is obtained by counting the number of  trajectories of temporal span $t$ that started from the $\pm$ state and performed exactly $N$ transitions, out of the total number of trajectories that started from the $\pm$ state and for which  the diffusion spent a total time $T_{+}$ at this state. Utilizing Eq.~\eqref{eqn:subordS}, we rewrite Eq.~\eqref{eq:pdfTp} as

\begin{eqnarray}\label{eq:fTpQN}
f_{t}(T_{+})=\frac{\langle \tau\rangle_{+} }{ \langle \tau \rangle_{+} + \langle \tau \rangle_{-} } \displaystyle \sum \limits _{N=0} ^{\infty} f_{t}^{+}(T_{+}\vert N) Q_{t}^{+}(N) + \frac{\langle \tau\rangle_{-} }{ \langle \tau \rangle_{+} + \langle \tau \rangle_{-} } \displaystyle \sum \limits _{N=0} ^{\infty} f_{t}^{-}(T_{+}\vert N) Q_{t}^{-}(N).
\end{eqnarray}
%%%%%%%%%%%%%%%%%%%%%%%%%%%%%%%%%%%%%%%%%%%%%%%%%%%%%%%
Since we consider a renewal process, the expression for $Q^{\pm}_{t}(N) $ is known in the Laplace space ~\cite{MarB2004}, as $ \hat{Q}^{\pm}_{s}(N)=\mathcal{L}\lbrace Q^{\pm}_{t}(N) \rbrace =\int_0^\infty Q_t^\pm(N)\exp(-ts)\,dt$, for any general $\hat{\psi}_{\pm}(s)=\mathcal{L}^{}\lbrace \psi_{\pm}(\tau) \rbrace$. Concretely, $Q_t^\pm(N)$ is obtained by taking into account all the  possibilities to perform $N$ jumps up to time $t_N<t$, and no additional jumps during the backward recurrence time $\tau^*$. This sums up to a convolution of $N+1$ random variables. 
It is important to notice that since we assume equilibrium initial conditions, $\tau_1$,  that is measured  from $t=0$, is only a part of a full renewal event and is termed  the forward  recurrence time . The PDF of $\tau_1$ for the $\pm$ state, $f_{eq}^\pm(\tau_1)$, is provided by (see ~\cite{GL2001}) 
%%%%%%%%%%%%%%%%%%%%%%%%%%%%%%%%%%%%%
\begin{equation}\label{eq:pdfforward}
f_{eq}^{\pm}(\tau_1)=
\left(
1-\int_0^{\tau_{1}} \psi_\pm(\tau)\,d\tau
\right)/\langle\tau\rangle_\pm
\end{equation}
%%%%%%%%%%%%%%%%%%%%%%%%%%%%%%%%%%%%%
and in the Laplace space $\mathcal{L}\lbrace f_{eq}^\pm(\tau_1) \rbrace = (1-\hat{\psi}_\pm(s))\Big/\langle \tau \rangle_\pm s$. This initial condition stems from the equilibrium of the underlying process, in which we do not have a jump at the initial time ($t_{0}=0$ in Figure~\ref{fig:2D}). In the literature ~\cite{cox1962,GL2001,MarB2004,BBel2005,aki2016} the case where the renewal process starts at $t=0$ is called ordinary or non-equilibrium, and as we will see below by following our approach this does not yield any universal features for $P(x,t)$, hence the assumption of an equilibrium process is important in our methodology, (see discussion about non-equilibrium initial conditions in Appendix~\ref{sec:AB}). 

The probability of not performing any jumps during $\tau^*$ is equivalent to the probability of obtaining a waiting time $\tau_{N+1}>\tau^*$, i.e $1-\int_0^{\tau^*}\psi_\pm(\tau)\,d\tau$. 
Eventually, by implementing the initial equilibrium condition we obtain  
\begin{eqnarray}\label{eq:QsNS}
\hat{Q}^{\pm}_{s}(0)&=& \frac{1- \frac{1-\hat{\psi}_{\pm}(s)}{\langle \tau\rangle_{\pm} s}}{s},\nonumber\\
\hat{Q}^{\pm}_{s}(1)&=& \Bigg(\frac{1-\hat{\psi}_{\pm}(s)}{\langle \tau\rangle_{\pm} s} \Bigg) \Bigg( \frac{1-\hat{\psi}_{\mp}(s)}{s}\Bigg)\nonumber,\\
\hat{Q}^{\pm}_{s}(2)&=& \Bigg(\frac{1-\hat{\psi}_{\pm}(s)}{\langle \tau\rangle_{\pm} s} \Bigg)  \hat{\psi}_{\mp}(s) \Bigg( \frac{1-\hat{\psi}_{\pm}(s)}{s}\Bigg)\nonumber,\\
\hat{Q}^{\pm}_{s}(3)&=& \Bigg(\frac{1-\hat{\psi}_{\pm}(s)}{\langle \tau\rangle_{\pm} s} \Bigg)  \hat{\psi}_{\mp}(s) \psi_{\pm}(s) \Bigg( \frac{1-\hat{\psi}_{\mp}(s)}{s}\Bigg).
\end{eqnarray}  
In all the equations above on the right hand side we have a multiplication of functions in the Laplace space, this implies convolutions as we transform from $s$ to $t$. The first term in the multiplication on the right hand side of Eq.~\eqref{eq:QsNS}  obviously stems from the equilibrium initial condition under study. We assume that the PDF of the waiting times is analytic for $\tau \to 0$, thus we can express ${\psi}_\pm(\tau)$ as \cite{bb2019,WBB2020}
\begin{eqnarray}\label{eq:STpsi} 
\psi_{\pm}(\tau) \sim C^{\pm} _{A_{\pm}}\tau^{A_{\pm}} + C^{\pm} _{A_{\pm}+1}\tau^{A_{\pm}+1} + \ldots,
\end{eqnarray}
with $A_{\pm}\geq 0$ an integer number. As an example consider the case with exponential waiting times, i.e. $\psi_{\pm}(\tau)=\psi(\tau)=\exp(-\tau/\langle \tau \rangle)/\langle \tau \rangle$, namely the waiting times at the $D_{\pm}$ states are identically distributed. Its analytic expansion is $\psi(\tau)\sim 1/\langle \tau \rangle - \tau / \langle \tau \rangle^{2}$, with $A_{\pm}=0$, $C^{\pm}_{A_{\pm}}=1/\langle \tau \rangle$ and $C^{\pm}_{A_{\pm}+1}=1/\langle \tau \rangle^{2}$.   The analyticity of $\psi_{\pm}(\tau)$   Eq.~\eqref{eq:STpsi},  is a very mild demand which covers a wide range of sojourn times distributions. Since we are interested in the small $t$ limit, the corresponding behavior in the Laplace space is found for $s\to \infty$, where the leading terms of $\hat{\psi}_{\pm}(s)$ are ~\cite{bb2019}
\begin{eqnarray}\label{eq:psils}
\hat{\psi}_{\pm}(s) \sim \frac{\Gamma(A_{\pm}+1) C^{\pm}_{A_{\pm}}}{s^{A_{\pm}+1}}+\frac{\Gamma(A_{\pm}+2) C^{\pm}_{A_{\pm} +1}}{s^{A_{\pm}+2}}+ \ldots,
\end{eqnarray}
For the mentioned example with exponential waiting times  $\hat{\psi}(s)\sim 1/[\langle \tau \rangle s]$.
Using Eq.~\eqref{eq:psils} for $\hat{Q}^{\pm}_{s}(N)$ we obtain that in the $s\to\infty$ limit, corresponding to the short time limit which is at the focus of our interest
%%%%%%%%%%%%%%%%%%%%%%%%%%%%%%%%%%%%
\begin{eqnarray}
\hat{Q}^{\pm}_{s}(0) &\sim & \frac{1}{s}-\frac{1}{ \langle \tau \rangle_{\pm} s^{2}} + \frac{\Gamma(A_{\pm}+1) C^{\pm}_{\pm}}{\langle \tau\rangle_{\pm} s ^{A_{\pm} +3}} + \ldots \nonumber, \\
\hat{Q}^{\pm}_{s}(1)&\sim &\frac{1}{\langle \tau\rangle_{\pm} s^{2}}- \frac{2 C^{\pm}_{A_{\pm}} \Gamma(A_{\pm}+1)}{\langle \tau\rangle_{\pm} s ^{A_{\pm} +3}}+\ldots \nonumber\\
\hat{Q}^{\pm}_{s}(2)&\sim & \frac{\Gamma(A_{\mp}+1)C^{\mp}_{A_{\mp}}}{ \langle\tau \rangle_{\pm} s^{A_{\mp}+3}}+ \ldots \nonumber\\
\hat{Q}^{\pm}_{s}(3)&\sim & \frac{\Gamma(A_{\pm}+1) \Gamma(A_{\mp}+1) C^{\pm}_{A_{\pm}} C^{\mp}_{A_{\mp}}}{\langle \tau \rangle_{\pm} s^{A_{\pm}+A_{\mp}+4}} + \ldots .
\end{eqnarray} 
%%%%%%%%%%%%%%%%%%%%%%%%%%%%%%%%%%%
We see that the leading terms for all  $\hat{Q}_s^{\pm}(N)$ with $N>1$ are of the order $1/s^\gamma$ with $\gamma>2$. Thus in the small $t$ limit,  terms with $N>1$ contain contributions that scale like $t^{\gamma-1}$ and are negligible with respect to the $N\in\lbrace 0,1 \rbrace$ cases. Therefore only the first two $Q_t^{\pm}(N)$s are taken into account, \textit{i.e.}
\begin{eqnarray}\label{eq:Q0}
Q^{\pm}_{t}(0)&\sim & 1- \frac{t}{\langle \tau\rangle_{\pm}},\\
Q^{\pm}_{t}(1)&\sim & \frac{t}{\langle \tau\rangle_{\pm}}. \label{eq:Q1}
\end{eqnarray} 
This is an expected result, as for short times only contributions from a single transition and zero transitions are important. By calculating $Q_t^{\pm}(N)$ we advanced towards obtaining the behavior of the PDF of $T_+$, according to Eq.~\eqref{eq:fTpQN} in order to complete this mission one needs to compute the relevant contributions of $f_t^{\pm}(T_+\vert N)$ in the $t\to 0$ limit. First we see that the conditional distribution $f^{\pm}_{t}(T_{+}\vert 0) $ depends only on the starting state. There are only two types of trajectories that has performed $0$ transitions, \textit{i.e.} for all the time they have been either at $D_{+}$ or at $D_{-}$. Consequently 
\begin{eqnarray}\label{eq:fTpN1p}
 f^{+}_{t}(T_{+} \vert 0  )&=&\delta(t-T_{+} ), \\
   f^{-}_{t}(T_{+} \vert 0  )&=&	\delta(T_{+}).\label{eq:fTpN1m}
\end{eqnarray}
The calculation of $f^{\pm}_{t}(T_{+}\vert N)$ is obtained by conditioning over the first event. If starting from the $+$ state, the process will spend a time $\tau_1$ at this state before jumping to the $-$ state. $\tau_1$ can attain any value $0\leq \tau_1 \leq T_+$ and for the remaining time $t-\tau_1$ the process has to perform one transition less. In general without regarding the initial conditions of the problem, an integration over all possible $\tau_1$'s provides the relation
%%%%%%%%%%%%%%%%%%%%%%%%%%%%%%%%%%%%%%%%%%%%%%%
\begin{equation}
    \label{eq:fplusNgen1}
    f_t^+(T_+|N^{\prime}+1) = \int_0^{T_+} \frac{1}{B_+} 
    \psi_+(\tau_1) f_{t-\tau_1}^{-}(T_+-\tau_1|N^{\prime})\,d\tau_1
\end{equation}
%%%%%%%%%%%%%%%%%%%%%%%%%%%%%%%%%%%%%%%%%%%%%%%
with $N^{\prime} +1 =N$,  and  $B_+$ a normalization factor. For instance for $N=1$ we have that $\int_0^t \psi_+(\tau_1)/B_{+}\,d\tau_1=1$, that stems from the fact that we consider only trajectories of time span $t$. The corresponding formula for $f_t^-(T_+|N^{\prime}+1)$ is %%%%%%%%%%%%%%%%%%%%%%%%%%%%%%%%%%%%%%%%%%%%%%%%
\begin{equation}
    \label{eq:fplusNgen2}
    f_t^-(T_+|N^{\prime}+1) = \int_0^{t-T_+} \frac{1}{B_-}     \psi_-(\tau_1) f_{t-\tau_1}^{+}(T_+|N^{\prime})\,d\tau_1.
\end{equation}
%%%%%%%%%%%%%%%%%%%%%%%%%%%%%%%%%%%%%%%%%%%%%%%
Since we are assuming equilibrium initial conditions, the $\psi_\pm$ in the $N^{\prime}+1$ element of the iterative forms (Eq.~\eqref{eq:fplusNgen1} and Eq.~\eqref{eq:fplusNgen2}) must be replaced by $f_{eq}^\pm$ (Eq.~\eqref{eq:pdfforward}). As was already noted above, only the $N=0$ and $N=1$ are of interest in the small $t$ limit, then according to Eq.~\eqref{eq:fTpN1p} - Eq.~\eqref{eq:fplusNgen2}  and Eq.~\eqref{eq:pdfforward} we get for $N=1$
%%%%%%%%%%%%%%%%%%%%%%%%%%%%%%%%%%%%
\begin{eqnarray}\label{eq:fTpNE1}
f^{+}_{t}(T_{+} \vert 1  )&=&	 \frac{f_{eq}^{+}(T_+)} { \int_0^t f_{eq}^+(t')\,dt'},   \\ 
f^{-}_{t}(T_{+} \vert 1  )&=&	   \frac{f_{eq}^{-}(t-T_+)} {\int_0^t f_{eq}^-(t')\,dt'}. \label{eq:fTpNE12}
\end{eqnarray}
%%%%%%%%%%%%%%%%%%%%%%%%%%%%%%%%%%%%%

Using the small time approximation of $\psi_{\pm}(\tau)$ Eq.~\eqref{eq:STpsi}, in Eq.~\eqref{eq:fTpNE1} and Eq.~\eqref{eq:fTpNE12},  we obtain  that, independently of the starting state, 
\begin{eqnarray}\label{eq:fTpN12}
f^{\pm}_{t}(T_{+}\vert 1) \sim \frac{1}{t}.
\end{eqnarray}
The  $1/t$ dependence  comes from the integral factors in Eq.~\eqref{eq:fTpNE1} and Eq.~\eqref{eq:fTpNE12}, all the other terms in the numerator and denominator simply cancel out. See Appendix~\ref{sec:AB} for a complementary derivation of Eq.~\eqref{eq:fTpN12} using the definition of the joint PDF of $T_+$ and $N$. Gathering Eq.~\eqref{eq:Q0}, Eq.~\eqref{eq:Q1}, Eq.~\eqref{eq:fTpN1p}, Eq.~\eqref{eq:fTpN1m} and Eq.~\eqref{eq:fTpN12} in Eq.~\eqref{eq:fTpQN}  we  find that
\begin{eqnarray}\label{eq:fTpQNst}
f_{t}(T_{+}) &\sim &  \frac{\langle \tau \rangle_{+} }{ \langle \tau \rangle_{+} + \langle \tau \rangle_{-} } \Bigg(1- \frac{t}{\langle \tau \rangle_{+}} \Bigg) \delta(t-T_{+}) \nonumber \\
&+ & \frac{\langle \tau \rangle_{-} }{ \langle \tau \rangle_{+} + \langle \tau \rangle_{-} } \Bigg(1- \frac{t}{\langle \tau \rangle_{-}} \Bigg) \delta(T_{+}) 
+  \frac{2}{ \langle \tau_{+} \rangle + \langle \tau_{-}\rangle }.
\end{eqnarray}
The PDF of the occupation fraction is obtained trivially from Eq.~\eqref{eq:fTpQNst} by changing variables to $p_{+}=T_{+}/t$  
\begin{eqnarray}\label{eq:fppQNst}
g_{t}(p_{+}) &\sim &  \frac{\langle \tau \rangle_{+} }{ \langle \tau \rangle_{+} + \langle \tau \rangle_{-} } \Bigg(1- \frac{t}{\langle \tau \rangle_{+}} \Bigg) \delta(1-p_{+}) \nonumber \\
&+ & \frac{\langle \tau \rangle_{-} }{ \langle \tau \rangle_{+} + \langle \tau \rangle_{-} } \Bigg(1- \frac{t}{\langle \tau \rangle_{-}} \Bigg) \delta(p_{+}) 
+  \frac{2t}{ \langle \tau \rangle_{+} + \langle \tau\rangle_{-} }.
\end{eqnarray}
The third term in Eq.~\eqref{eq:fTpQNst} and Eq.~\eqref{eq:fppQNst} is uniform, \textit{i.e.} terms which are independent of $T_{+}$ or $p_{+}$, and this  is the first main result of this paper. All the additional terms and contributions to the PDF of $p_{+}$ only introduce terms that depend on higher orders of $t$ and thus negligible in the small $t$ limit. This means that for equilibrium initial conditions, regardless of the exact form of  $\psi_{\pm}(\tau)$, the PDF of $p_{+}$ (Eq.~\eqref{eq:fppQNst}) is always uniform for $0<p_{+}<1$. 
This  general uniform behavior of the PDF of the occupation fraction is applicable for an extremely large class of waiting times PDFs  $\psi_{\pm}(\tau)$.  As a remark, the connection between the conditional PDF of $T^+$, $f_{t}^{\pm}(T_{+}\vert N)$, and the joint PDF of $T^+$ and $N$, $f_{t}^{\pm}(T_{+}, N)$ is discussed in Sec.~\ref{sec:AB2}.
In the following it is shown that this uniformity leads to universal features of the propagator in the limit of small $t$.
In sections ~\ref{sec:41} and ~\ref{sec:42} we treat  particular examples (with exponential waiting times) that are exactly tractable, without any simplifications or assumptions. The results agree perfectly with the general form in Eq.~\eqref{eq:fppQNst}.  It is important to notice that our approximations affect only the form of $g_t(p_{+})$ and do not affect $P(x\vert p_{+})$. This allows us to obtain the behavior of $P(x,t)$ for any $-\infty < x < \infty$, as is shown below.

\paragraph{\label{sec:31} \textbf{$P(x,t)$ for arbitrary waiting times }\\}

In order to obtain the positional PDF for small $t$ we combine Eq.~\eqref{eq:pxp}, Eq.~\eqref{eq:pxpp} and  Eq.~\eqref{eq:fppQNst}, which after integration gives 

\begin{eqnarray}\label{eq:pxstc1}
P(x,t)&=&\frac{ \langle\tau \rangle _{+}}{\langle\tau\rangle_{+} + \langle\tau\rangle_{-} }  \Bigg(1-\frac{t}{\langle \tau \rangle_{+} } \Bigg) \frac{e^{-\frac{x^{2}}{4D_{+}t}}}{\sqrt{4\pi D_{+} t}}+\frac{ \langle\tau \rangle_{-}}{\langle\tau\rangle_{+} + \langle\tau\rangle_{-} }  \Bigg(  1- \frac{t}{\langle \tau\rangle_{-}}\Bigg)\delta(x) \nonumber \\
&+& \Bigg( \frac{2 t}{\langle \tau \rangle_{+} + \langle \tau \rangle_{-}}\Bigg)\Bigg\lbrace\frac{e^{-\frac{x^{2}}{4D_{+}t} }}{\sqrt{\pi D_{+}t}} - \frac{\vert x \vert}{ 2D_{+}t} \Big( 1- Erf\Big( \frac{\vert x \vert}{\sqrt{4D_{+}t}} \Big)\Big)\Bigg\rbrace . 
\end{eqnarray}
Considering $x\neq 0$,  in the limit of $x\longrightarrow 0$ when  $\exp(-x^{2}/4D_{+}t)\sim 1-x^{2}/4D_{+}t $ and $1-Erf( \vert x \vert / \sqrt{4D_{+}t})\sim 1- 2\vert x \vert / \sqrt{4\pi D_{+} t}$. After substituting in  Eq.~\eqref{eq:pxstc1} it turns into 
\begin{eqnarray}\label{eq:Pxstsxc1}
P(x,t)& \sim &  \frac{(3t + \langle \tau \rangle_{+} ) }{\sqrt{4\pi D_{+}t} [\langle \tau\rangle_{+} + \langle \tau \rangle_{-} ]} - \frac{\vert x \vert}{D_{+} [ \langle \tau\rangle_{+} + \langle \tau \rangle_{-}]} + K_{1} x^{2}, 
\end{eqnarray} 
with $K_{1}=(5t- \langle \tau \rangle_{+} )/ [8( \langle \tau \rangle_{+} + \langle \tau \rangle_{-} ) \sqrt{\pi}(D_{+}t)^{\frac{3}{2}}] $. We can see that in Eq.~\eqref{eq:Pxstsxc1} there is a linear dependence on $\vert x \vert$ in the vicinity of $x=0$. This means that for short enough measurement times the PDF of $x$ will always have a tent like shape, irrespective of the distributions  $\psi_{\pm}$ that were chosen (see Figure ~\ref{fig:pdfxUG} below). Only the mean sojourn times affect the shape. This is a general result for the short time regime, and it is based on the general fact that the PDF of the temporal occupation fraction is uniform for $0<p_+<1$. Concretely, at short times when $\vert x \vert$ is small, the decay of $P(x,t)$ will always resemble  an exponential one. For large $\vert x \vert$ the form of $P(x,t)$ must be Gaussian, due to the fact that this limit is determined by the instances when no transition to $D_{-}$ was ever made and the transport is controlled by diffusion with $D_+$. But, if we only look at the particles that have moved, \textit{i.e.} we get rid of the delta function at $x=0$ in Eq.~\eqref{eq:pxstc1}. We can relate this dynamics with some experiments which condition the measurements on the movement of the particles. This procedure is called population splitting see ~\cite{Schulz2013,Schulz2014}. Technically, if $D_{-}>0$ the cusp is not found,  however as long as $D_{-}/D_{+} << 1$ the tent like shape will be found, for further details see Appendix~\ref{sec:A0}

\subsubsection{\label{sec:32}Long time regime }

In the limit $t \longrightarrow \infty$, the PDF of the temporal occupation fraction $g_{t}(p_{+})$ follows a different but also a general form. As mentioned  we are focusing on the case where both $\psi_{\pm}$ have finite first moments,  $\langle \tau\rangle_{\pm} >0$. In the long time limit, ergodicity is satisfied, namely  the equivalence of \textit{ensemble} and temporal averages is attained. Particularly in this case the \textit{ensemble} average of the  occupation fraction  at $D_{+}$ is equal to the temporal average which is defined by  the fraction of average waiting times at  $D_{+}$ and $D_{-}$, \textit{i.e.} $\langle p_{+} \rangle = \langle \tau \rangle_{+}/[\langle \tau \rangle_{+}+\langle \tau \rangle_{-}]$ (see Appendix ~\ref{sec:A3}). Thus in the long time limit $g_{t}(p_{+})$ converges to a $\delta$ - function,
\begin{eqnarray}\label{eq:fpltC1}
g_{t}(p_{+})\xrightarrow[t\rightarrow \infty ]{ }\delta\Bigg(p_{+} - \frac{\langle\tau \rangle_{+}}{ \langle \tau\rangle_{+} + \langle \tau\rangle_{-}}\Bigg).
\end{eqnarray}

Since ergodicity prevails, by using Eq.~\eqref{eq:fpltC1} in Eq.~\eqref{eq:pxpp} the positional PDF   gets the form 
\begin{eqnarray}\label{eq:PGstc1}
 P(x,t)= \sqrt{  \frac{\langle \tau \rangle_{+} + \langle \tau\rangle_{-} }{4\pi D_{+} t \langle \tau\rangle_{+}}  } e^{- \frac{x^{2} (\langle \tau \rangle_{+} + \langle \tau \rangle_{-}  )}{4D_{+} t \langle \tau\rangle_{+}}}.
 \end{eqnarray}
 %%%%%%%%%%%%%%%%%%%%%%%%%%%%
 In the long time limit the positional PDF given by  Eq.~\eqref{eq:PGstc1} represents a Gaussian propagator with an effective diffusion coefficient $D_{+} \langle \tau \rangle_{+}/[\langle \tau \rangle_{+} + \langle \tau \rangle_{-}]$. Since $\langle \tau \rangle_{+}/[\langle \tau \rangle_{+} + \langle \tau \rangle_{-}]<1$, the effective diffusion coefficient  is always smaller compared with $D_{+}$.   
 Indeed this slow-down is an expected result due to the portion of the time that the particle spends in the state with $D_{-}=0$ and basically not moving during this period.

\begin{figure}[H]
\centering
\includegraphics[width=6.5 cm]{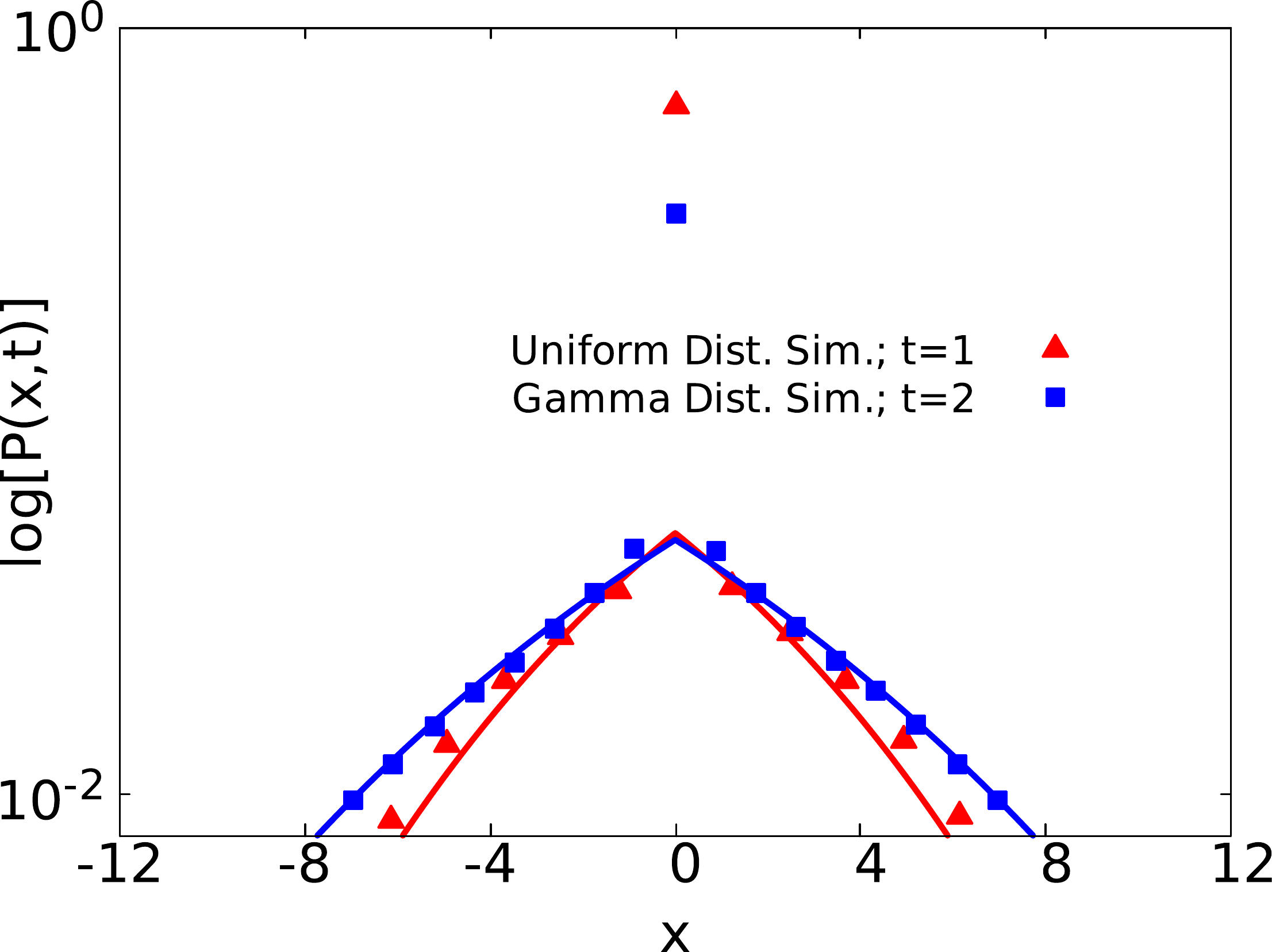}
\includegraphics[width=6.5 cm]{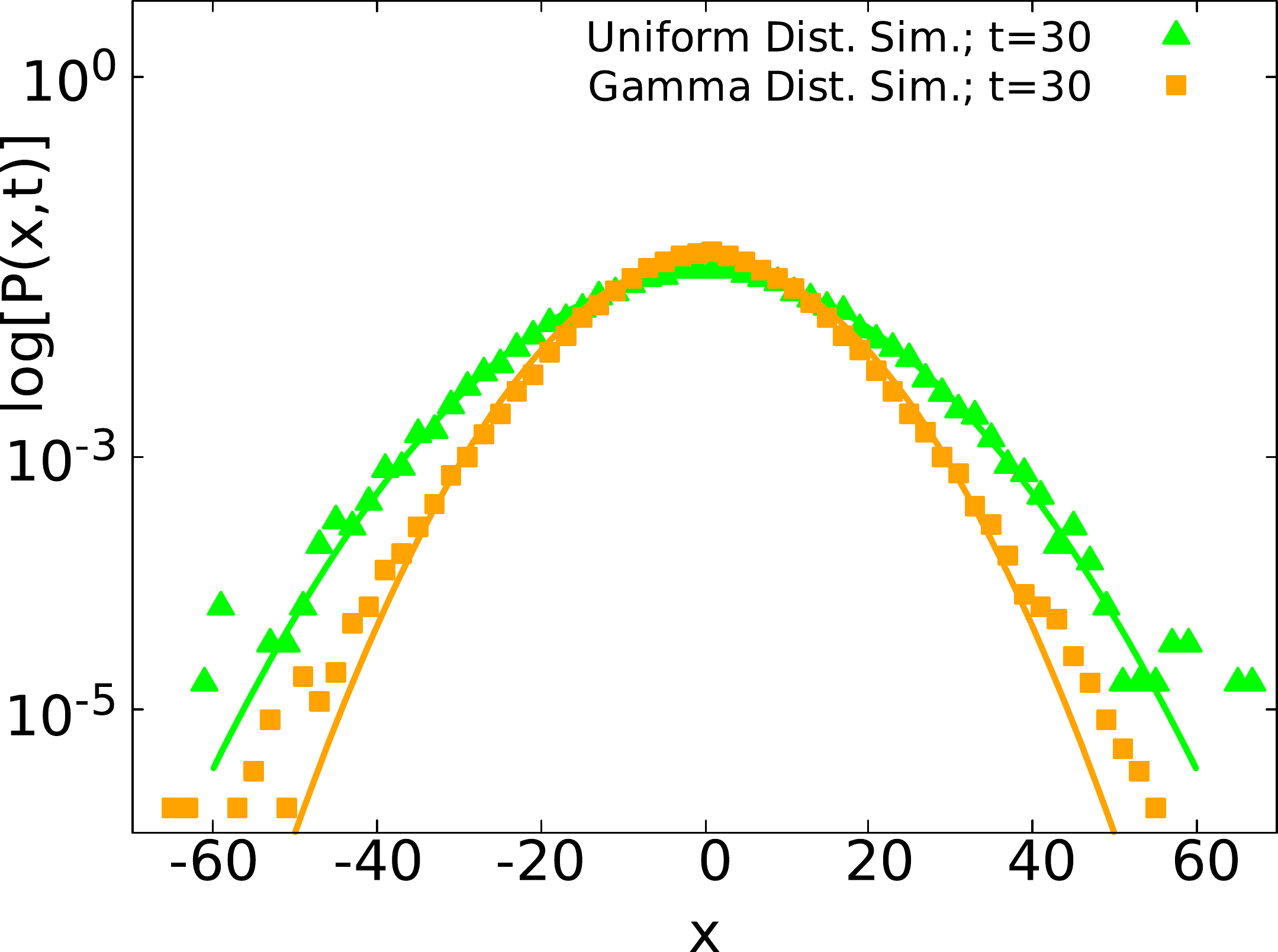}
\caption{Distribution of displacements $P(x,t)$ in semi-log scale, obtained by simulations, for a two state system with  uniform distributed waiting times and  gamma distributed waiting times. 
The left panel presents short time results where a tent like shape is clearly visible and an non-analytical feature obvious, while the right panel exhibits Gaussian statistics for long times. Left: $P(x,t)$ for  $t=1$ for  $\tau \sim U(0,5)$ at $D_{+}$ and $\tau \sim U(0,10)$ at $D_{-}$ (red triangles). With $\langle \tau \rangle_{+}=2.5 < \langle \tau \rangle _{-} =5$. And for $t=2$ with $\tau\sim Gamma(0.5,8)$ at $D_{+}$ and  $\tau\sim Gamma(0.5,12)$ at $D_{-}$ (blue squares), such that $\langle \tau \rangle_{+}=4 < \langle \tau \rangle _{-} =6$. Both cases fit with Eq.~\eqref{eq:pxstc1} (red and blue solid lines) with a tent like shape. In both normalized histograms  at $x=0$ there is a peak representing the Dirac delta function in Eq.~\eqref{eq:pxstc1}. Right: $P(x,t)$ for $t=30$ and  waiting times uniformly distributed  (green triangles) with the same parameters as above and for gamma distributed waiting times  (orange squares) with $\tau \sim Gamma(2,1)$ at $D_{+}$, $\tau \sim Gamma(8,1)$ at $D_{-}$, and $\langle \tau \rangle_{+}=2<\langle \tau \rangle_{-}=8 $. We employed the last set of parameters in the gamma distributed waiting times in order to avoid an overlapping between curves.   $P(x,t)$ converges to the Gaussian statistics Eq.~\eqref{eq:PGstc1} (green and orange solid lines) . In all the presented cases $D_{+}=10$ and $D_{-}=0$ were used.\label{fig:pdfxUG}}
\end{figure}   

\subsubsection{\label{sec:33}Simulations }

The two general limits of $g_{t}(p_{+})$ Eq.~\eqref{eq:fppQNst} and Eq.~\eqref{eq:fpltC1} produce two different prevailing distributions of $P(x,t)$ Eq.~\eqref{eq:pxstc1} and Eq.~\eqref{eq:PGstc1}. In Figure~\ref{fig:pdfxUG} we compare analytical formulas Eq.~\eqref{eq:pxstc1} and Eq.~\eqref{eq:PGstc1} (solid lines) with simulations of different two state models. One with uniform distributed waiting times $\tau \sim U(0,5)$ for $D_{+}$ and  $\tau \sim U(0,10)$ for $D_{-}$ and with $t=1$ (red triangles) and $t=30$ (green triangles), such that $\langle \tau \rangle_{+}=2.5  < \langle \tau \rangle_{-}=5$. Here the notation $\tau\sim U(a,b)$ means that $\tau$ has a uniform distribution with $a$ and $b$ the minimum and maximum values respectively. And the other with gamma distributed waiting times, such that  $\tau \sim Gamma(k,\theta)$. The latter notation denotes that $\tau$ has a gamma distribution with $k$ its shape parameter and $\theta$ the corresponding scale parameter. In this case the  PDF follows 
\begin{eqnarray}\label{eq:taugamma}
\psi_{\pm}(\tau)=\frac{\tau^{k-1}e^{-\frac{\tau}{\theta}} }{\Gamma(k) \theta^{k}},
\end{eqnarray}
particularly the PDF of the gamma distribution Eq.~\eqref{eq:taugamma} implies a cumulative distribution function $F(\tau)=\gamma(k,\tau/\theta)/ \Gamma(k)$, with $\gamma(x,y)$ the incomplete gamma function and $\Gamma(x)$ the standard gamma function. For the latter case we used  $\tau\sim Gamma(0.5,8)$ at $D_{+}$ and  $\tau \sim Gamma(0.5,12)$ at $D_{-}$, for $t=2$ (blue squares) and $t=30$ (orange squares), with $\langle \tau \rangle_{+}=4  < \langle \tau \rangle_{-}=6$. 
As we can see in the short time regime, for uniform and gamma distributed waiting times (red triangles and blue squares) $P(x,t)$ has a tent shape for short displacements and it agrees with Eq.~\eqref{eq:pxstc1}, joined with a peak at $x=0$ due to the Dirac delta function in Eq.~\eqref{eq:pxstc1}. For $t=30$ (green triangles and orange squares)  each case of $P(x,t)$ converges to Gaussian statistics (Eq.~\eqref{eq:PGstc1}).

The cusp we have found for small $\vert x \vert$ implies that we may approximate the distribution on a small scale with a Laplace like distribution, $P(x,t)\sim \exp( - C \vert x\vert)$. However, clearly this does not hold globally for large $x$, see Fig.~\ref{fig:pdfxUGC} in Appendix~\ref{sec:AC}. Still within the interval of short displacements, due to the presence of the delta peak at the origin we expect for this span a considerable contribution on the normalization of $P(x,t)$. Particularly, we find that the  area underneath the curve for the case of uniformly distributed waiting times (red line) in Fig.~\ref{fig:pdfxUG} at the left panel, has a value of $0.88$ for $x\in (-4,4)$. And the corresponding area within the same figure but for gamma distributed waiting times (blue curve) has a value of 0.89 for $x\in(-8,8).$

\subsection{\label{sec:4} Exponentially distributed waiting times}
In this section we obtain $g_{t}(p_{+})$ for a specific distribution of waiting times, but using different methods which let us corroborate the validity of our general approach described above. We analyze the case of exponential waiting times  in  states with $D_{+}$ and $D_{-}$,  each waiting time following a PDF  given by 
\begin{eqnarray}\label{eq:pdfp}
\psi_{\pm}(\tau)&=&\frac{e^{-\frac{\tau}{\langle\tau\rangle_{\pm} }}}{\langle \tau \rangle_{\pm}}. \label{eq:pdfm}
\end{eqnarray}
We show first the case of a two state system with the same mean waiting times  and then investigate the complimentary case. In  Appendix~\ref{sec:A1}  we analyze both cases for non-equilibrium initial conditions, \textit{e.g.} a system starting from $D_{+}$. 

\subsubsection{\label{sec:41} Equal mean waiting times $\langle \tau \rangle_{+} =\langle \tau \rangle_{-}$}

Let us consider a system with  $ \langle \tau \rangle_{+}=\langle\tau \rangle_{-}=\langle\tau\rangle$. We know that the temporal fraction occupation $p_{+}$ and $T_{+}$ can be related to the difference of occupation times defined by $S_{t}=T_{+}-T_{-}$, as $S_{t}=2T_{+}-t=2p_{+}t-t$ ~\cite{GL2001}.
In this section we analyze the double Laplace transform of the PDF of $S_{t}$, called $\phi_t(S_t)$
with Laplace pairs $S_{t}\Leftrightarrow v$ and $t\Leftrightarrow s$.  In ~\cite{GL2001} $\phi_t(S_t)$ is provided by
%%%%%%%%%%%%%%%%%%%%%%%%%%%%%%%%%%%%%%%
\begin{eqnarray}\label{eq:fsvS}
\hat{\phi}_{s}(v)=\frac{s[1-\psi(s+v)\psi(s-v)] +v[\psi(s+v)-\psi(s-v) ]}{(s^{2} - v^{2}) [1- \psi(s+v)\psi(s-v)]}.
\end{eqnarray}
%%%%%%%%%%%%%%%%%%%%%%%%%%%%%%%%%%%%%%%%
The  Laplace transform of $\psi(\tau)$ in Eq.~\eqref{eq:pdfp} is given by $\mathcal{L}\Big\lbrace \psi(\tau) \Big\rbrace= \hat{\psi}(s)=  \frac{1}{1+\langle\tau\rangle s}$. Substituting $\hat{\psi}(s)$ in Eq.~\eqref{eq:fsvS} we obtain 
\begin{eqnarray}\label{eq:fSvs1}
\hat{\phi}_{s}(v) = \frac{s+2\langle\tau\rangle}{s^{2}+2\langle\tau\rangle s -v^{2}}.
\end{eqnarray}

\begin{figure}[H]
\centering
\includegraphics[width=6.5 cm]{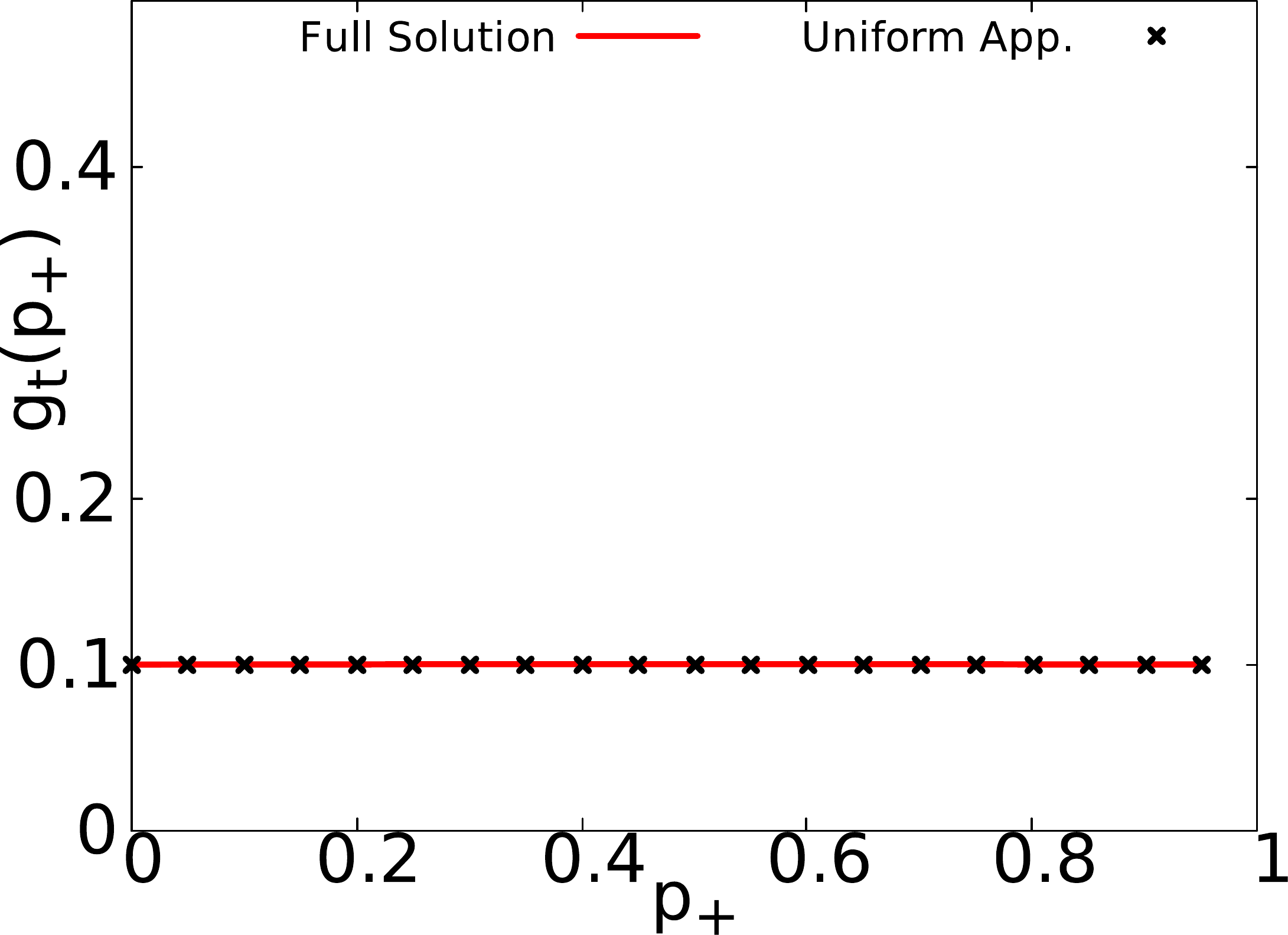}
\includegraphics[width=6.5 cm]{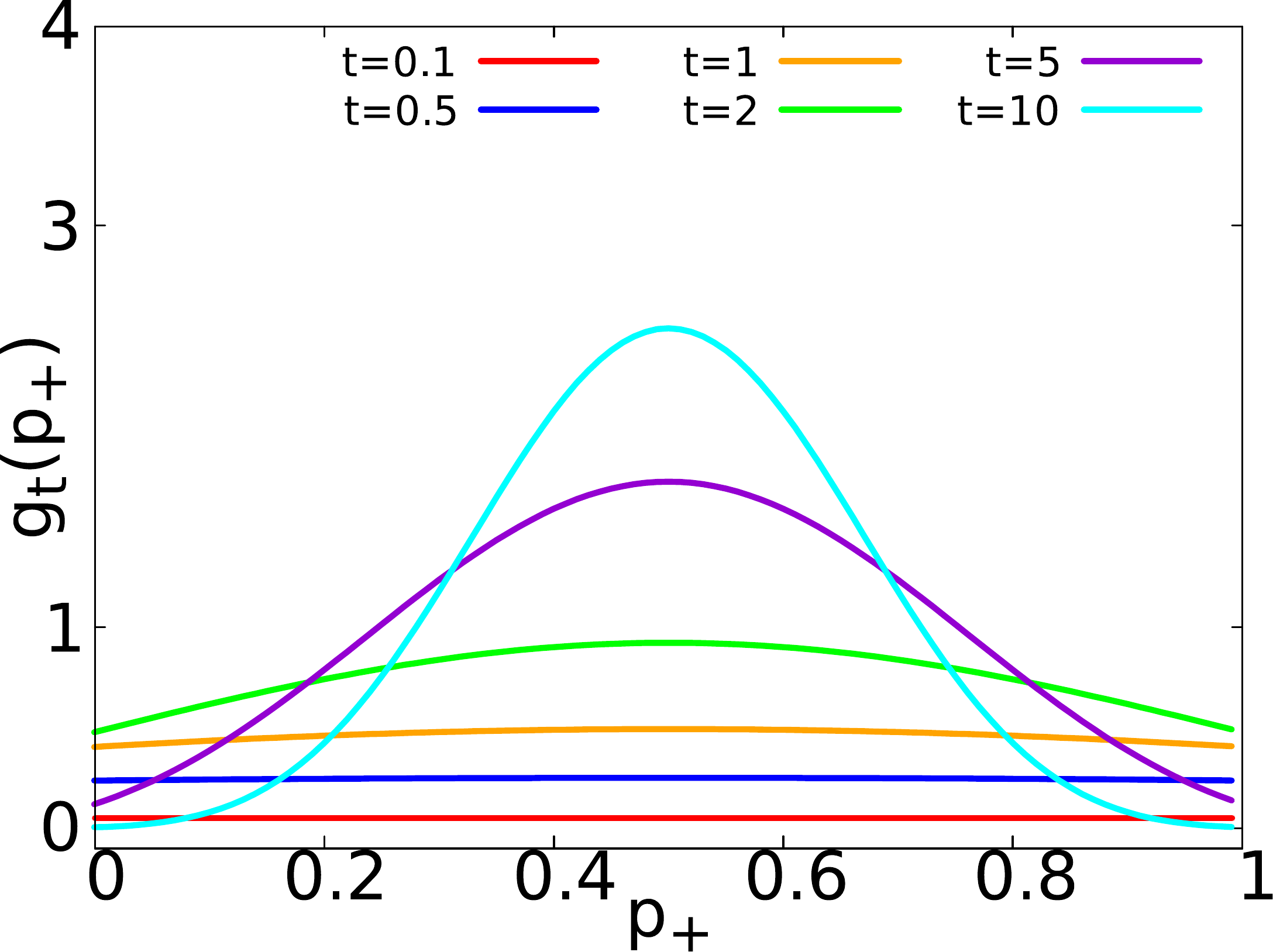}
\caption{Left:  Comparison between $g_{t}(p_{+})$ Eq.~\eqref{eq:MWpdfp}  (red solid line) and the short time uniform approximation Eq.~\eqref{eq:fppu} (black asterisks) for exponentially distributed waiting times Eq.~\eqref{eq:pdfp} with $\langle \tau \rangle_{\pm}=\langle \tau \rangle=1$ and $t=0.1$. Right: $g_{t}(p_{+})$ Eq.~\eqref{eq:MWpdfp}  for $\langle \tau \rangle=1$ and $t\in\lbrace 0.1,0.5,1,2,5,10\rbrace$.\label{fig:pdfp}}
\end{figure}

In Appendix ~\ref{sec:A2} an analytical expression for the PDF of $S_{t}$ is found, \textit{i.e.} inverse Laplace transform of Eq.~\eqref{eq:fSvs1} is performed (see Eq.~\eqref{eq:pdfSsr}). Then by remembering that the temporal occupation fraction in the plus state $p_{+}$ is related to the difference of occupation times as $S_{t}=2p_{+}t-t$. We can employ  Eq.~\eqref{eq:pdfSsr}  for obtaining the PDF of $p_+$ which is given by
%%%%%%%%%%%%%%%%%%%%%%%%%%%%%%%%%%%%%%%%%%%%%%%
\begin{eqnarray}\label{eq:MWpdfp}
g_{t}(p_{+})&=&\frac{1}{2}e^{-\frac{t}{\langle \tau \rangle}}\Bigg\lbrace \delta(1-p_{+})+\delta(p_{+}) \Bigg\rbrace \nonumber\\
&+&\frac{t \Theta(t-\vert 2p_{+}t-t\vert)e^{-\frac{t}{\langle \tau \rangle}}}{\langle \tau \rangle}\Bigg[ I_{0}\Big(\frac{2t\sqrt{p_{+}(1-p_{+})}}{\langle \tau \rangle}\Big)+ \frac{I_{1}\Big( \frac{2t\sqrt{p_{+}(1-p_{+})}}{\langle \tau \rangle}\Big)}{2\sqrt{p_{+} (1-p_{+})}}\Bigg].
\end{eqnarray}
%%%%%%%%%%%%%%%%%%%%%%%%%%%%%%%%%%%%%%%%%%%%%%
A similar expression for a system with non-equilibrium initial conditions (always starting from $D_{+}$) is found in Appendix ~\ref{sec:A1}. 
%\paragraph{\textbf{Short time approximation of $g_{t}(p_{+})$ for $\langle \tau \rangle_{+} = \langle \tau \rangle_{-}$}\\}
% 
Expanding Eq.~\eqref{eq:MWpdfp} in the short time limit $t\longrightarrow 0$, \textit{i.e.} $t<< \langle \tau\rangle$, Eq.~\eqref{eq:MWpdfp}  can be approximated by  a uniform distribution 
%%%%%%%%%%%%%%%%%%%%%%%%%%%%%%%%%%%%%%%%%
\begin{eqnarray}\label{eq:fppu}
g_{t}(p_{+})\sim\frac{e^{-\frac{t}{\langle \tau \rangle}}}{2} \Big\lbrace \delta(1-p_{+})+\delta(p_{+}) \Big\rbrace +  \frac{t}{\langle \tau \rangle}. 
\end{eqnarray}
%%%%%%%%%%%%%%%%%%%%%%%%%%%%%%%%%%%%%%%%%%
For $\langle \tau \rangle_{+} = \langle \tau \rangle_{-} =\langle \tau \rangle$, Eq.~\eqref{eq:fppu} agrees with Eq.~\eqref{eq:fppQNst}  obtained by the general approach of Sec.~\ref{sec:3}. In the left panel of Figure~\ref{fig:pdfp} we show the short time approximation of $g_{t}(p_{+})$ (Eq.~\eqref{eq:fppu}) compared with the general formula in Eq.~\eqref{eq:MWpdfp}, it is evident that both results agree perfectly. In the right panel of Figure~\ref{fig:pdfp} we show Eq.~\eqref{eq:MWpdfp} for short and long measurement times.  $g_{t}(p_{+})$ evolves from a uniform distribution to a peaked distribution centered at its mean value $p_{+}=1/2$ (see Appendix ~\ref{sec:A2} for a deduction of the central moments of $g_{t}(p_{+})$). 

\paragraph{\textbf{Positional distribution function}\\}

An analytical expression for the positional distribution function $P(x,t)$ (given by  Eq.~\eqref{eq:pxpp}), with $g_t(p_+)$ provided by  Eq.~\eqref{eq:MWpdfp}, can be deduced  by  using the series representation of the modified  Bessel functions , $I_{\nu}(y)= \sum \limits _{k=0}^{\infty} (\frac{y}{2})^{2k+\nu} / [k! \Gamma(\nu +k+1)]$. The integration in Eq.~\eqref{eq:pxpp} yields 
%%%%%%%%%%%%%%%%%%%%%%%%%%%%%%%%%%%%%%%%
\begin{eqnarray}
&P(x,t)=\frac{e^{-\frac{t}{\langle \tau \rangle}- \frac{x^{2}}{4Dt}}}{2\sqrt{4\pi D t}} +\frac{\delta(x) e^{-\frac{t}{ \langle\tau \rangle}}}{2}+\frac{te^{-\frac{t}{\langle \tau \rangle}-\frac{x^{2}}{4Dt}}}{2\langle \tau \rangle \sqrt{4\pi Dt}}\Bigg\lbrace \displaystyle \sum \limits _{k=0}^{\infty}\frac{(-1)^{k}\pi}{k!}\Big( \frac{t}{\langle \tau \rangle}\Big)^{2k}  \Bigg[ \frac{_1F_1\big(k+1;\frac{1}{2}-k;\frac{x^{2}}{4DT} \big)}{\Gamma\big(2k+\frac{3}{2} \big) \Gamma\big( \frac{1}{2}-k\big)}  \nonumber \\
&- \frac{\big( \frac{x^{2}}{4Dt} \big)^{k+\frac{1}{2}}\,\ _1F_1\big( 2k+\frac{3}{2};k+\frac{3}{2};\frac{x^{2}}{4Dt}\big)}{\Gamma\big( k+1\big)\Gamma\big( k+\frac{3}{2}\big)}\Bigg] + \frac{1}{2} \displaystyle \sum \limits_{k=0}^{\infty} \frac{(-1)^{k}\pi}{(k+1)!} \Big(\frac{t}{\langle \tau \rangle} \Big)^{2k+1}   \Bigg[ \frac{_1F_1\big( k+1;\frac{1}{2}-k;\frac{x^{2}}{4Dt} \big)}{\Gamma\big(2k+\frac{3}{2} \big) \Gamma\big( \frac{1}{2}-k\big)} \nonumber \\
&- \frac{\big( \frac{x^{2}}{4Dt} \big)^{k+\frac{1}{2}} \,\ _1F_1\big(2k+\frac{3}{2};k+\frac{3}{2};\frac{x^{2}}{4Dt}\big)}{\Gamma\big(k+1\big) \Gamma\big(k+\frac{3}{2} \big)}\Bigg] \Bigg\rbrace, 
\end{eqnarray}
%%%%%%%%%%%%%%%%%%%%%%%%%%%%%%%%%%%%%%%%%%%%%%%
with $_1F_1(a;b;z)$ the confluent hypergeometric function of the first kind. Nonetheless, in the short time limit we can use the uniform approximation of $g_{t}(p_{+})$ (Eq.~\eqref{eq:fppu}),  then   Eq.~\eqref{eq:pxpp} provides 
%%%%%%%%%%%%%%%%%%%%%%%%%%%%%%%%%%%%%%%%%%
\begin{eqnarray}\label{eq:pxsru}
P(x,t)\sim  \frac{e^{-\frac{t}{ \langle\tau \rangle } - \frac{x^{2}}{4D_{+}t}}}{2\sqrt{4\pi D_{+}t}}+\frac{\delta(x) e^{-\frac{t}{ \langle\tau \rangle}}}{2}+ \frac{t}{ \langle\tau\rangle}\Bigg\lbrace \frac{2 e^{-\frac{x^{2}}{4D_{+}t}}}{ \sqrt{4\pi D_{+} t}}- \frac{\vert x \vert}{2 D_{+} t}\Bigg[ 1 - Erf\Bigg( \frac{\vert x \vert}{\sqrt{4D_{+}t}}\Bigg)  \Bigg] \Bigg\rbrace,
\end{eqnarray} 
%%%%%%%%%%%%%%%%%%%%%%%%%%%%%%%%%%%%%%%%%%%%%%%
which agrees with the result obtained above  in Eq.~\eqref{eq:pxstc1}, when  $\langle \tau \rangle_{+}=\langle \tau\rangle_{-}$ and for $t\longrightarrow 0$. Since in that limit $\exp(-t/\langle \tau \rangle) \sim 1- t/\langle \tau \rangle$. Particularly  for $x\neq 0$ and taking $x \longrightarrow 0$, Eq.~\eqref{eq:pxsru}  yields  a tent shaped propagator described  by
%%%%%%%%%%%%%%%%%%%%%%%%%%%%%%%%%%%%%%%%%%%%%%
\begin{eqnarray}\label{eq:pxsrusx}
P(x,t)\sim \frac{3t+ \langle \tau \rangle}{4\langle \tau \rangle \sqrt{\pi D_{+}t}}  - \frac{\vert x \vert}{2D_{+}\langle \tau \rangle }+ K_{2}x^{2}, 
\end{eqnarray} 
with $K_{2}= (5t-\langle \tau \rangle)/[16 \langle \tau \rangle \sqrt{\pi} (D_{+}t)^{\frac{3}{2}}]$ and in concordance with Eq.~\eqref{eq:Pxstsxc1}. On the other hand, within this short time limit, for large displacements $x\longrightarrow \infty$, the two terms between curly braces in Eq.~\eqref{eq:pxsru} cancel each other, and only the first term in Eq.~\eqref{eq:pxsru} is left (when $x\neq 0$). This is due to the expansion of $1-Erf(z) \sim \exp(-z^{2})/ (\sqrt{\pi} z) $ for $z\longrightarrow \infty$, in our case $z=\vert x \vert / \sqrt{4D_{+}t}$. Then Eq.~\eqref{eq:pxsru}  can be approximated by  
\begin{eqnarray}\label{eq:pxsrulx}
 P(x,t) \underset{x \rightarrow \infty}{\sim}   \frac{e^{-\frac{t}{ \langle\tau \rangle } - \frac{x^{2}}{4D_{+}t}}}{2\sqrt{4\pi D_{+}t}}.
\end{eqnarray}
This Gaussian behavior of $P(x,t)$  at the tails is expected. The large $\vert x \vert$ limit is dominated by trajectories for which no transitions to $D_{-}$ were performed and a pure diffusion process with $D_{+}$ occurs.   

\begin{figure}[H]
\centering
\includegraphics[width=6.5 cm]{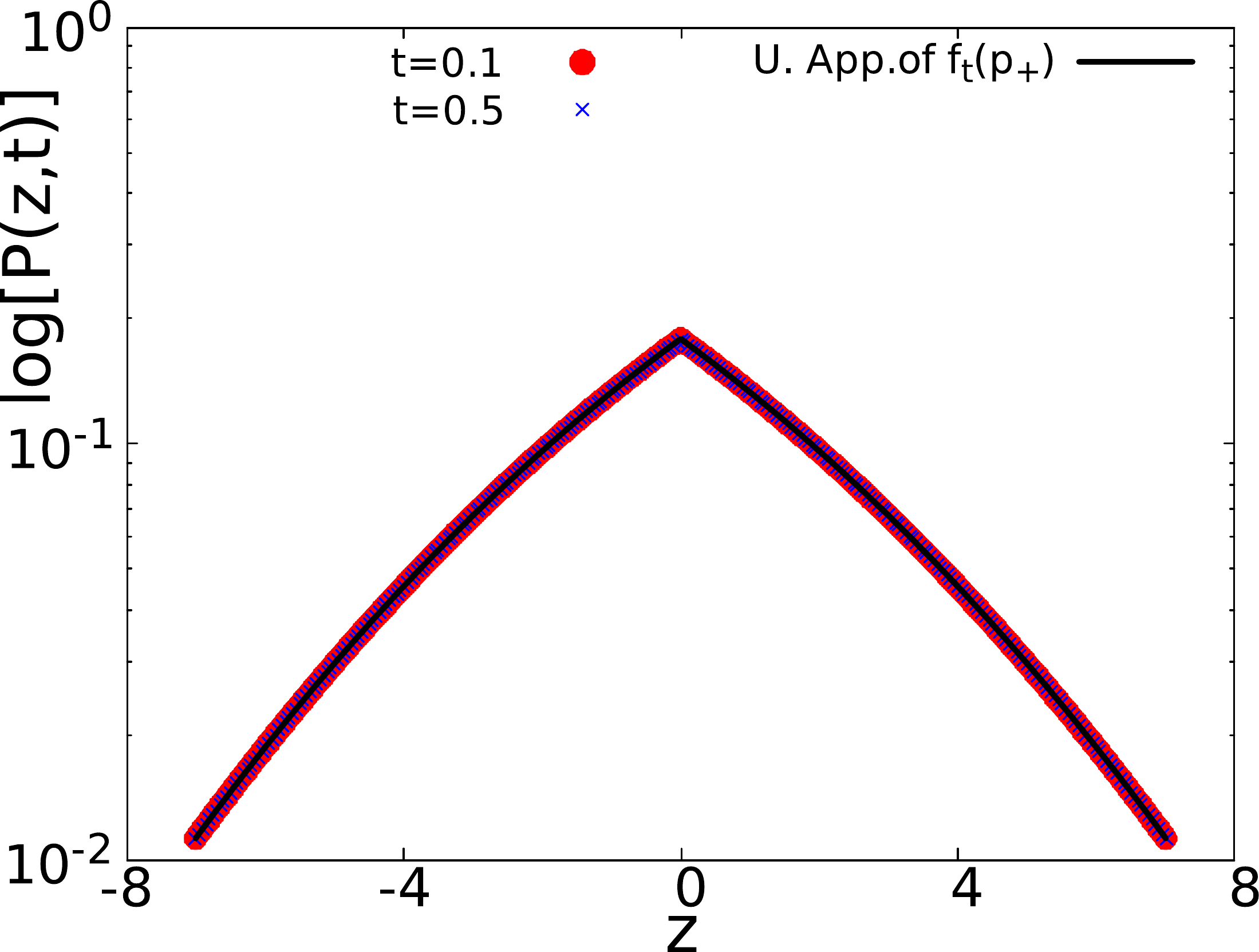}
\includegraphics[width=6.5 cm]{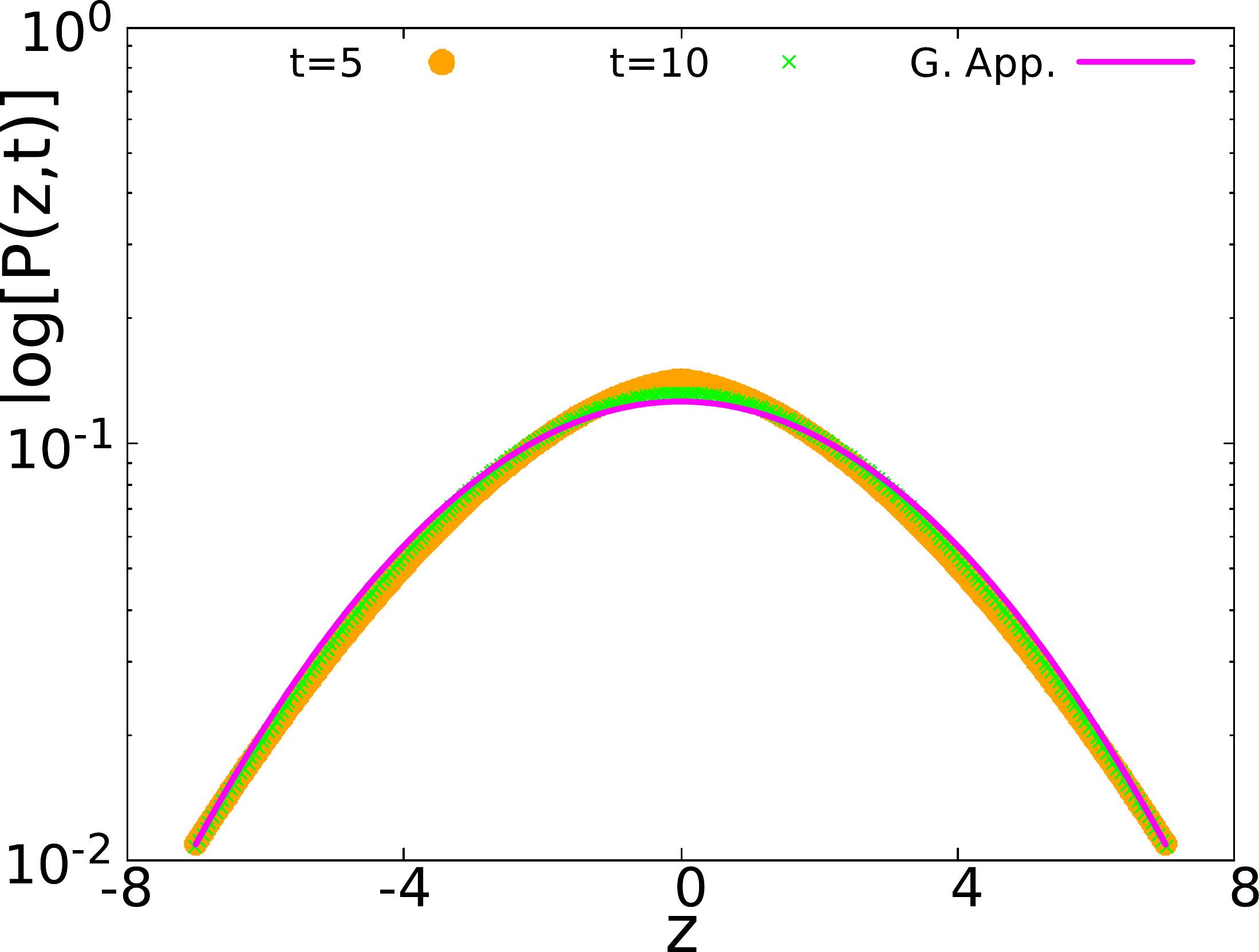}
\caption{$P(z,t)$ in semi-log scale, with $z=x/\sqrt{t}$. Left: For short times $t=0.1$ (red circles) and $t=0.5$ (blue crosses) $P(z,t)$ is represented by Eq.~\eqref{eq:pxsru} ( black solid line) with a tent like shape. Right: The same for large times $t=5$ (orange circles) and $t=10$ (green crosses), $P(z,t)$ converges to the Gaussian distribution Eq.~\eqref{eq:pxtltsr} (magenta solid line). In all the cases $D_{+}=10$, $D_{-}=0$, and $\langle \tau \rangle=1$ were used. \label{fig:pdfxSR}}
\end{figure}  

When $t>>\langle \tau \rangle$, ergodicity  is satisfied  and  therefore the system on average visits the two states the same amount of time. Namely, the \textit{ensemble} average of $p_{+}$ is equal to the corresponding fraction of the average waiting times. In this case when $\langle \tau\rangle_{+}=\langle \tau\rangle_{-}$, the occupation fraction is concentrated at $p_{+}=1/2$. Thus the PDF of $p_{+}$ is represented by the delta function 
\begin{eqnarray}\label{eq:fpltsr}
g_{t}(p_{+})\xrightarrow[t\rightarrow \infty ]{ }\delta\Big(p_{+} - \frac{1}{2} \Big).
\end{eqnarray}
Substituting Eq.~\eqref{eq:fpltsr} in Eq.~\eqref{eq:pxpp} we recover Gaussian statistics for the displacements
\begin{eqnarray}\label{eq:pxtltsr}
P(x,t)\sim \frac{e^{-\frac{x^{2}}{2D_{+}t} }}{\sqrt{2\pi D_{+}t}}.
\end{eqnarray}
In Figure~\ref{fig:pdfxSR} we present the two different limit distributions  for $P(x,t)$ in the short time limit $t=0.1$ (red circles) and $t=0.5$ (blue crosses) Eq.~\eqref{eq:pxsru} and the Gaussian limit  for $t=5$ (orange circles) and $t=10$ (green crosses) Eq.~\eqref{eq:pxtltsr}, for the normalized variable $z=x/\sqrt{t}$. As we can see the displacements for short times follow a tent  shape (black solid line) and a Gaussian one in the long time limit (magenta solid line). 

\subsubsection{\label{sec:42} Different mean waiting times $\langle \tau \rangle_{+} \neq \langle \tau \rangle_{-}$}

Relaxing the assumption of equal mean waiting times for exponentially distributed sojourn times in the model,  we have that $\langle \tau \rangle_{+}\neq \langle \tau \rangle_{-}$, with waiting times following  Eq.~\eqref{eq:pdfm}. 
As mentioned, for equilibrium initial conditions the PDF of $T_{+}$ is given by Eq.~\eqref{eq:pdfTp}. Let $\hat{f}_{s}^{\pm}(u)$ be the double Laplace transform of $f_{t}^{\pm}(T_{+})$, defined as $\hat{f}_{s}^{\pm}(u)= \int_0^\infty\int_0^\infty f_{t}(T_{+})\exp(-uT_+ -st)\,dT_+\,dt$. Then the different terms of the PDF of $T_{+}$ in Eq.~\eqref{eq:pdfTp} are provided in Laplace space,   by ~\cite{MarB2004,BBel2005,Aki2019} 
\begin{eqnarray}\label{eq:fTpdr}
\hat{f}^{+}_{s}(u)&=&\Bigg\lbrace \hat{\psi}_{+}(s+u)\Big[ \frac{1-\hat{\psi}_{-}(s)}{s}\Big] + \frac{1-\hat{\psi}_{+}(s+u)}{s+u}\Bigg\rbrace\frac{1}{1-\hat{\psi}_{+}(s+u)\hat{\psi}_{-}(s)}, \\
\hat{f}^{-}_{s}(u)&=&\Bigg\lbrace \hat{\psi}_{-}(s)\Big[ \frac{1-\hat{\psi}_{+}(s+u)}{s+u}\Big] + \frac{1-\hat{\psi}_{-}(s)}{s}\Bigg\rbrace\frac{1}{1-\hat{\psi}_{+}(s+u)\hat{\psi}_{-}(s)}.\label{eq:fTmdr}
\end{eqnarray} 
Summing up Eq.~\eqref{eq:fTpdr} and Eq.~\eqref{eq:fTmdr} according to Eq.~\eqref{eq:pdfTp}, we obtain, for exponentially distributed waiting times, 
%%%%%%%%%%%%%%%%%%%%%%%%%%%%%%%%%%%%%%%%
\begin{eqnarray}\label{eq:fsudrm}
\hat{f}_{s}(u)=\frac{\langle \tau\rangle_{-}^{2} +\langle \tau \rangle_{+}^{2} (1+\langle \tau \rangle_{-} s) +\langle \tau\rangle_{+} \langle \tau\rangle_{-} [2+ \langle \tau\rangle_{-} (s+u)] }{(\langle \tau \rangle_{+} + \langle \tau \rangle_{-} )[  \langle \tau \rangle_{-} s +  \langle \tau \rangle_{+} (1+ \langle \tau \rangle_{-} s)(s+u) ]}.
\end{eqnarray}
Taking the double inverse Laplace transform  of Eq.~\eqref{eq:fsudrm} with respect to $u\Leftrightarrow T_{+}$ and $s \Leftrightarrow t$ and changing variables to $p_{+}=T_{+}/t$, we obtain the PDF for $p_+$ (see details in Appendix ~\ref{sec:A3})

\begin{eqnarray}\label{eq:pdfpdr}
&g_{t}(p_{+})=\frac{\langle \tau \rangle_{-}  e^{-\frac{t}{\langle \tau \rangle_{-}}}}{ \langle \tau \rangle_{+} + \langle \tau\rangle_{-}}\delta(p_{+})+\frac{ \langle \tau\rangle_{+} e^{-\frac{t}{\langle \tau \rangle_{+}} }}{ \langle \tau\rangle_{+} + \langle \tau\rangle_{-}}\delta(1-p_{+})+\frac{2t}{\langle \tau \rangle_{+} + \langle \tau \rangle_{-}} \Bigg\lbrace I_{0}\Bigg( 2t \sqrt{\frac{p_{+}(1-p_{+})}{\langle \tau\rangle_{+} \langle \tau \rangle_{-}}}\Bigg) \nonumber \\
&+ \Bigg[ \frac{(1-p_{+}) \sqrt{\langle \tau\rangle_{+} \langle \tau \rangle_{-} }}{\langle \tau \rangle_{+}} + \frac{p_{+}\sqrt{\langle \tau \rangle_{+} \langle \tau \rangle_{-}}}{\langle \tau\rangle_{-}} \Bigg]  \frac{I_{1}\Big( 2t \sqrt{\frac{p_{+}(1-p_{+})}{\langle \tau\rangle_{+} \langle \tau \rangle_{-}}}\Big)}{2\sqrt{p_{+}(1-p_{+})}} \Bigg\rbrace e^{- \frac{tp_{+}}{\langle \tau\rangle_{+}} - \frac{t(1-p_{+})}{\langle \tau \rangle_{-}}} .
\end{eqnarray}
For the case when $\langle \tau \rangle_{+} = \langle \tau \rangle_{-}= \langle \tau \rangle$, Eq.~\eqref{eq:pdfpdr} recovers Eq.~\eqref{eq:MWpdfp} obtained by the methods reported in ~\cite{MW1996,ML2017}.   The case of non-equilibrium initial conditions is shown in Appendix ~\ref{sec:A1}.

In the short time regime,  strictly speaking when $t<<\langle \tau\rangle_{\pm}$, by expanding Eq.~\eqref{eq:pdfpdr} for $t\longrightarrow 0$, $g_{t}(p_{+})$ can be approximated by the uniform distribution  
\begin{eqnarray}\label{eq:fppudr}
g_{t}(p_{+})\sim \frac{\langle \tau \rangle_{-} e^{-\frac{t}{\langle \tau \rangle_{-}}}}{ \langle \tau \rangle_{+}  + \langle \tau \rangle_{-}} \delta(p_{+})+ \frac{\langle \tau \rangle_{+} e^{- \frac{t}{\langle \tau \rangle_{+} }}}{ \langle \tau \rangle_{+} + \langle \tau \rangle_{-}}\delta(1-p_{+}) +  \frac{2t}{ \langle \tau \rangle_{+} + \langle \tau \rangle_{-} }. 
\end{eqnarray}
As mentioned above, Eq.~\eqref{eq:fppQNst} that was deduced for general PDFs of waiting times, encloses the particular case of Eq.~\eqref{eq:fppudr}. For the uniform approximation of $g_t(p^+)$  (Eq.~\eqref{eq:fppudr}) the positional PDF  (Eq.~\eqref{eq:pxpp}) is

\begin{eqnarray} \label{eq:pxdru}
P(x,t)&\sim & \frac{\langle \tau\rangle_{+} e^{- \frac{t}{\langle \tau \rangle_{+}}- \frac{x^{2}}{4 D_{+} t}}   }{ (\langle \tau \rangle_{+} +\langle \tau\rangle_{-} ) \sqrt{4 \pi D_{+} t}}    + \frac{\langle \tau \rangle_{-}}{ \langle \tau\rangle_{+} + \langle \tau \rangle_{-}} e^{-\frac{t}{\langle \tau \rangle_{-}} } \delta(x) \nonumber \\
&+& \frac{2te^{-\frac{x^{2}}{4D_{+}t}}}{( \langle \tau \rangle_{+} +\langle \tau \rangle_{-} ) \sqrt{\pi D_{+} t}} - \frac{\vert x \vert}{D_{+} ( \langle \tau \rangle_{+} + \langle \tau \rangle_{-})} \Bigg[ 1- Erf\Big( \frac{\vert x \vert}{\sqrt{4 D_{+}t}}\Big) \Bigg],
\end{eqnarray} 
%%%%%%%%%%%%%%%%%%%%%%%%%%%%%%%%%%%%%%
which agrees with the general case described by Eq.~\eqref{eq:pxstc1}.

\begin{figure}[H]
\centering
\includegraphics[width=6.5 cm]{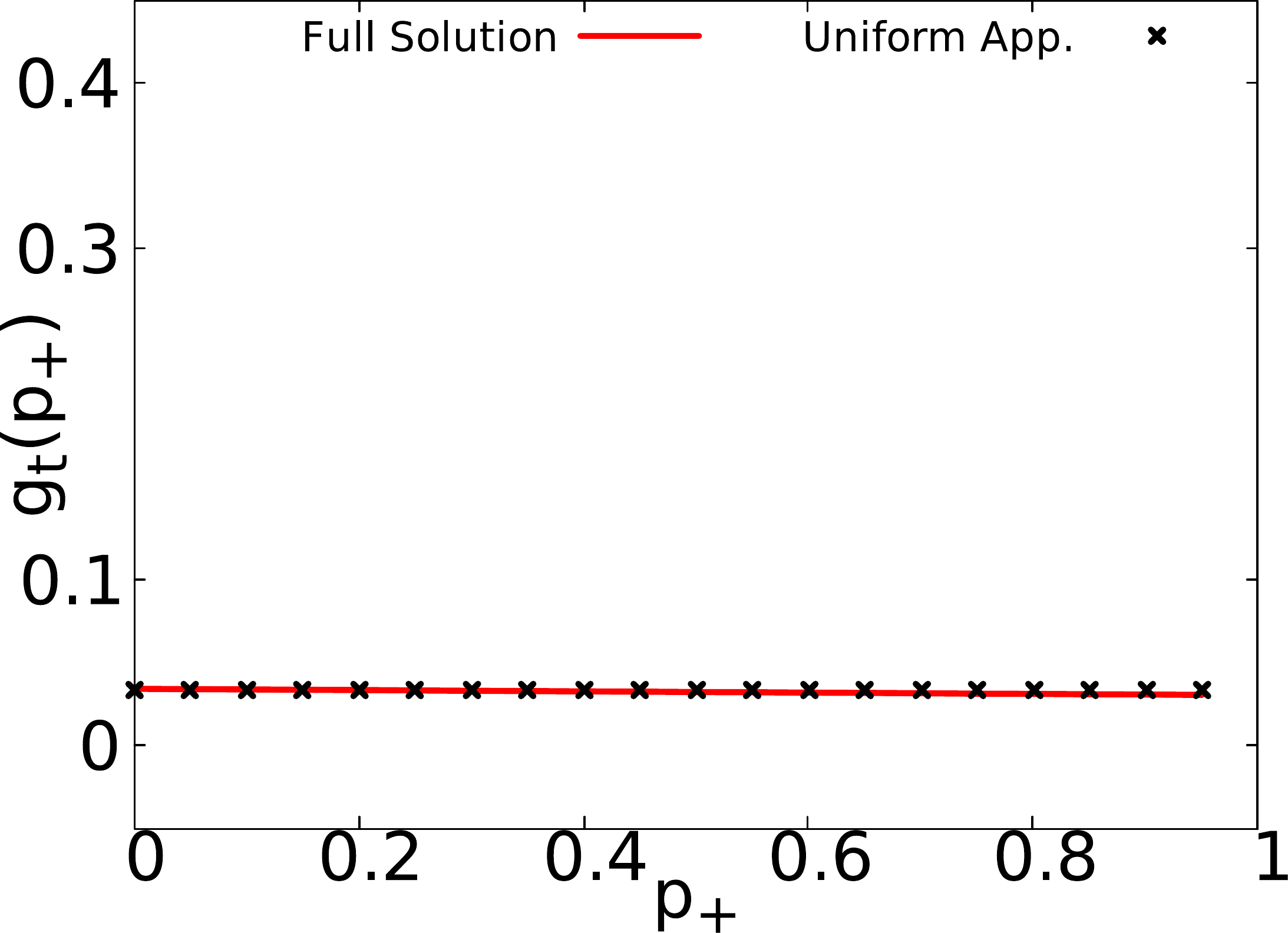}
\includegraphics[width=6.5 cm]{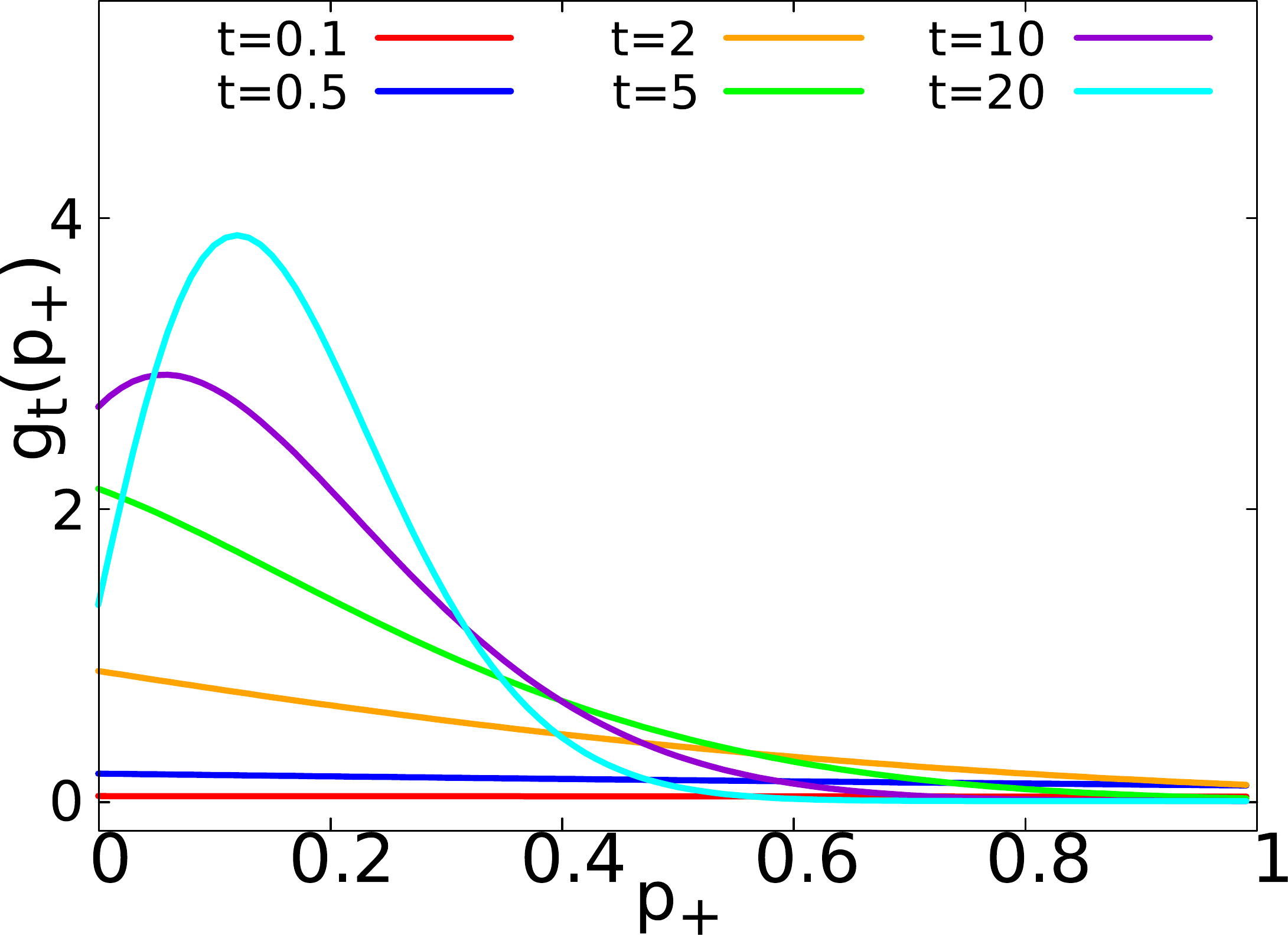}
\caption{Left:  Comparison between $g_{t}(p_{+})$ Eq.~\eqref{eq:pdfpdr}  (red solid line)  and the uniform approximation Eq.~\eqref{eq:fppudr} (black asterisks) for $\langle \tau \rangle_{+}=1$, $\langle \tau \rangle_{-}=5$ and $t=0.1$. Right: $g_{t}(p_{+})$ Eq.~\eqref{eq:pdfpdr}  for $\langle \tau \rangle_{+}=1$, $\langle \tau \rangle_{-}=5$  and $t\in\lbrace 0.1,0.5,2,5,10,20\rbrace$. \label{fig:pdfpdr}}
\end{figure} 

Similar to Sec.~\ref{sec:3}, in the limit $t \longrightarrow \infty$, the PDF of the occupation fraction $g_{t}(p_{+})$ follows Eq~\eqref{eq:fpltC1}. And the PDF of the displacements in the long time regime is given by Eq.~\eqref{eq:pxpp} and Eq.~\eqref{eq:PGstc1}, recovering Gaussianity.

In Figure~\ref{fig:pdfpdr} we show $g_{t}(p_{+})$ for exponential waiting times with $\langle \tau \rangle_{+}=1$ and $\langle \tau \rangle_{-}=5$, in the left panel we compare the uniform approximation of Eq.~\eqref{eq:fppudr}  (black asterisks) with the full solution Eq.~\eqref{eq:pdfpdr} (red solid line), observing an excellent agreement. In the right panel of Figure~\ref{fig:pdfpdr} the behavior of $g_{t}(p_{+})$ (as provided  by Eq.~\eqref{eq:pdfpdr}) is displayed. As we can see it starts with a uniform distribution for short times and then it evolves to a peaked distribution centered at $p_{+}= \langle \tau \rangle_{+} / (\langle \tau \rangle_{+} + \langle \tau \rangle_{-})=1/6$.  As shown in Appendix ~\ref{sec:A1}, for   non-equilibrium initial condition, the PDF of $p_+$  is still uniform  within the short time regime. See also Appendix~\ref{sec:AB} for other similar cases.

\begin{figure}[H]
\centering
\includegraphics[width=6.5 cm]{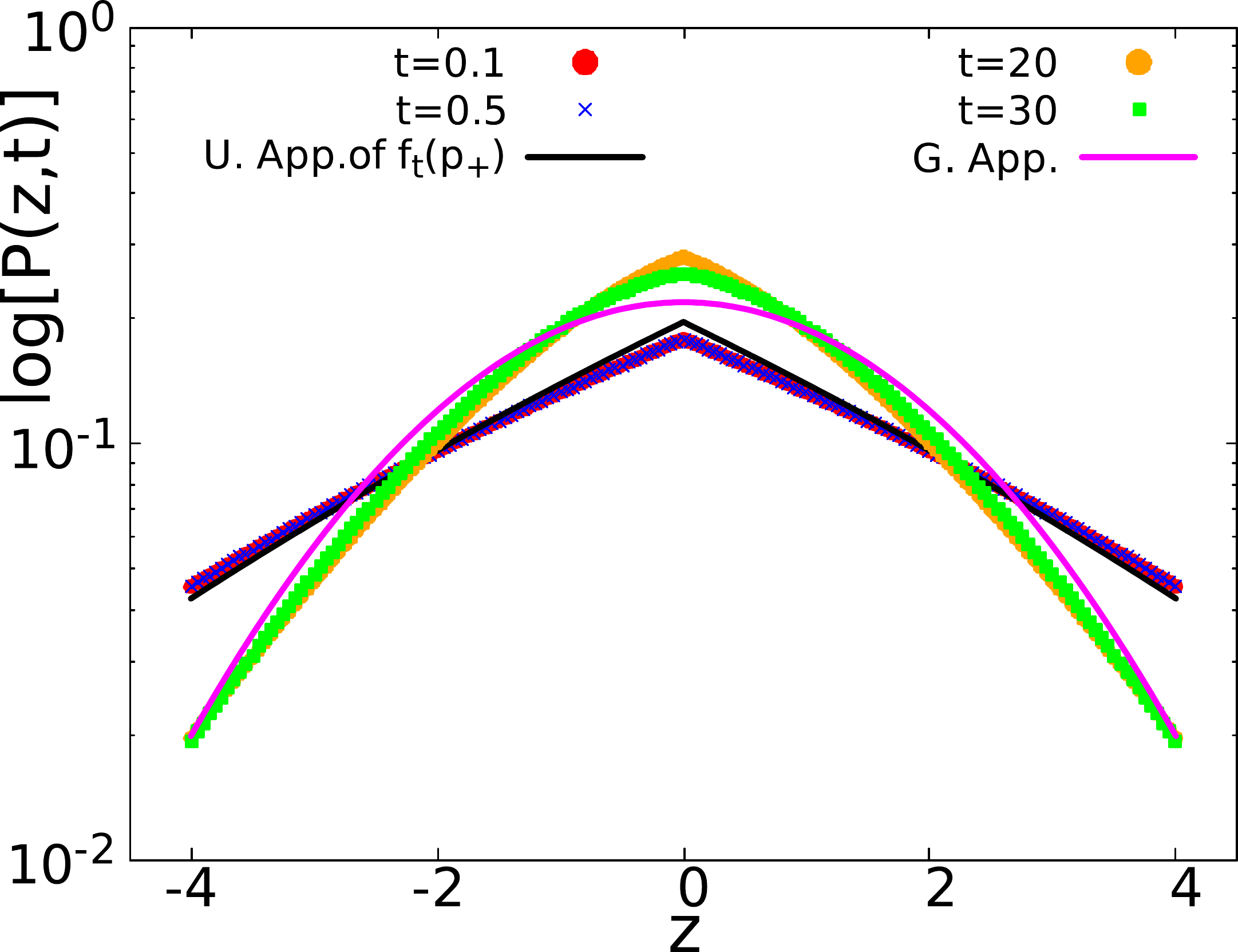}
\caption{For a system with, $\langle \tau\rangle_{+}=1 $ and $\langle \tau\rangle_{-}=5$, $P(z,t)$ in semi-log scale, with $z=x/\sqrt{t}$. For short times $t=0.1$ (red circles) and $t=0.5$ (blue crosses) $P(z,t)$ is represented by Eq.~\eqref{eq:pxdru} ( black solid line) with a tent like shape. For large times $t=20$ (orange circles) and $t=30$ (green diamonds) $P(z,t)$ converges to the Gaussian statistics Eq.~\eqref{eq:PGstc1} (magenta solid line). In all the cases $D_{+}=10$ and $D_{-}=0$  were used. Compared with Figure~\ref{fig:pdfxSR} in this case the Gaussian curve is above the tent curve, contrary to the case with equal mean waiting times. This is because the coefficient of the Gaussian curve Eq.~\eqref{eq:PGstc1} is bigger compared with the weight of the delta peak in Eq.~\eqref{eq:pxdru}. In   Figure~\ref{fig:pdfxSR} we have the opposite, the weight of the corresponding delta function in Eq.\eqref{eq:pxsru} is bigger compared with the Gaussian Eq.~\eqref{eq:pxtltsr}. \label{fig:pdfxDR}}
\end{figure}

Finally in Figure~\ref{fig:pdfxDR} we show the corresponding positional spreading for the normalized variable $z=x/\sqrt{t}$. As we can see in the short time $t=0.1$ (red circles) and $t=0.5$ (blue crosses) $P(z,t)$ (given by Eq.~\eqref{eq:pxdru}) attains a tent-like shape. In the long run $t=20$ (orange circles) and $t=30$ (green squares) $P(z,t)$ has a Gaussian distribution given by Eq.~\eqref{eq:PGstc1}.  
%%%%%%%%%%%%%%%%%%%%%%%%%%%%%%%%%%%%%%%%%%%

\section{Discussion}

\subsection{\label{sec:45} The histogram of the diffusion coefficient as extracted from experimental data}

\subsubsection{Super - Statistics}
We have found that at $x=0$,  $P(x,t)$ exhibits a cusp. A mathematically
similar non - analytical 
behavior is found using an approach called super - statistics ~\cite{BECK2003,chuby2014,Seb2016,chechk2017}, which was
used to explain laboratory observations.  
This framework postulates that the distribution of diffusion constants in the system
is exponential, namely
 $P(D) = \exp(- D/\langle D \rangle)/ \langle D \rangle$ for $D> 0$ and $\langle D \rangle$ the average diffusivity.
Then the diffusion follows a Gaussian process with a random $D$.  This
approach gives 
\begin{eqnarray}\label{eq:DDL}
 P(x,t) = \displaystyle \int_{0} ^{\infty}  \frac{e^{-\frac{x^{2}}{4Dt} }}{\sqrt{4\pi D t}}\frac{e^{-\frac{D}{\langle D \rangle}}}{\langle D \rangle}d D = \frac{e^{- \frac{\vert x \vert }{\langle D \rangle t}}}{4 \langle D \rangle t}.
\end{eqnarray}
Here on the right hand side we have the Laplace PDF, which was
used by Laplace in 1774 ~\cite{Laplace1774} to describe his  linear law of errors ~\cite{wilson1923}. Also as in our case, within the super - statistics method  we see in Eq.~\eqref{eq:DDL} a non - analytical behavior since $P(x,t) \sim C_{1} - C_{2} \vert x \vert $,  for small $x$ and with $C_{1}, C_{2}$ constants. Our work does not support the Laplace law, see Eq.~\eqref{eq:pxstc1} and  Eq.~\eqref{eq:Pxstsxc1}. But maybe more importantly the whole approach presented in this manuscript  differs from the super - statistical approach in the following way. In our model we have two diffusion constants, $D_+$ and $D_{-}=0$ (see Appendix ~\ref{sec:A0} for the case when $D_{-} \neq 0$). Hence the PDF of diffusion constants is $P(D) = a \delta (D) + b \delta( D - D_+)$, with $a,b\geq 0$. It follows that the super - statistical approach predicts that the diffusing packet $P(x,t)$ is a sum of a delta function corresponding to non - moving particles and a Gaussian packet describing the movers.  Thus when the non - moving particles are excluded we have perfect Gaussian behavior. This is actually correct, to leading order, for very short times. Thus the super - statistical approach gives the correct $t\longrightarrow 0$ behavior but fails to predict the main issue (in our opinion), and that is the cusp on $x=0$. To explore the non - analytical behavior  one needs to go to the next order terms in the expansion to include paths with a transition between states. Then as we have shown the equilibrium initial condition yields a uniform distribution of the occupation fraction Eq.~\eqref{eq:fppQNst}. It is this fact that brings the non - analytical behavior in the final result for $P(x,t)$ Eq.~\eqref{eq:Pxstsxc1}, graphically represented by a ``tent'' see Figure~\ref{fig:pdfxUG}, Figure~\ref{fig:pdfxSR} and Figure~\ref{fig:pdfxDR}. It follows that, the exponential conspiracy, that distribution of diffusion
constants is exponential,  is not a necessary condition for a cusp like behavior of $P(x,t)$.  We further remark the non - analytical behavior is found also in the context of normal diffusion in \cite{Luo2018,Luo2019,Post2020} and within the  anomalous one at ~\cite{BOU1990,Barkai2001,Month2003,BB2011PRL,BB2012PRE,Regev2016,Radice2019,Luo2019}.

\subsubsection{Time Average MSD}

We note that in single molecule experiments the time average mean squared displacement (TAMSD) is used in many
cases to estimate the distribution of diffusion constants ~\cite{hapca2009,granick2012,spako2017}. Since time averages are
recorded over a finite measurement time, the time average fluctuates.  Hence
we have naturally a distribution of the estimator for the diffusion parameters. And the mentioned two delta peak distribution of $D$, \textit{i.e.} on $D_{+}$ and on $D_{-}$, is expected to be smeared out. This topic was extensively studied in a wide variety of models ~\cite{BHMB2011,MHCB2014}.

\begin{figure}[H]
\centering
\includegraphics[width=6.5 cm]{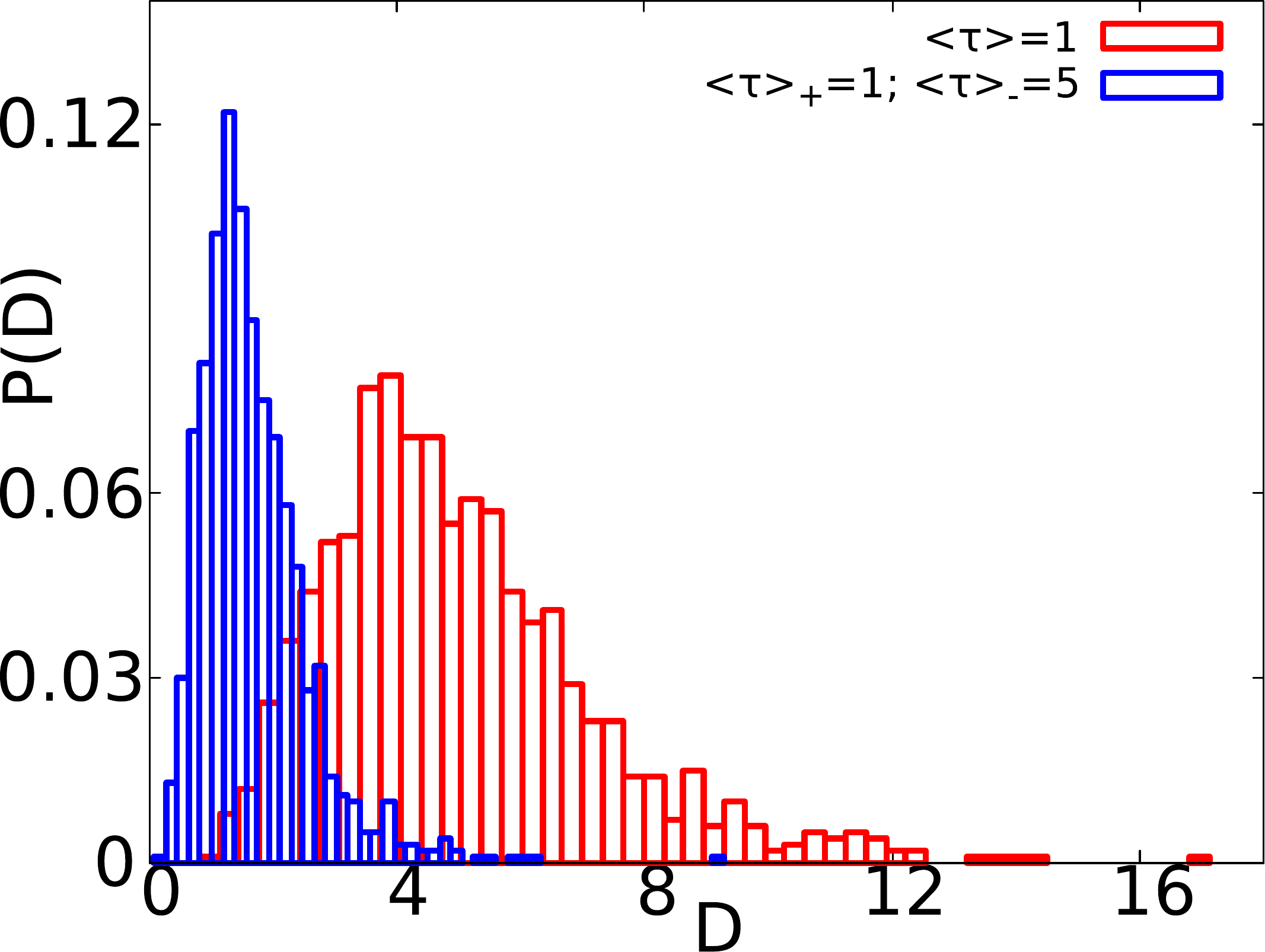}
\caption{Distribution of diffusion coefficients $P(D)$ obtained via TAMSD analysis of simulated trajectories of a two state system with  $ D_{+}=10$, $D_{-}=0$ and exponentially distributed waiting times. From the linear  plots of  the TAMSD versus the lag time  estimates of $D$ were extracted. We show two cases, the first  for a system with same mean waiting times  $\langle \tau \rangle_{+} = \langle \tau \rangle_{-}= \langle \tau \rangle =1$ (red boxes). And the PDF of $D$  for a system with different mean waiting times with  $\langle \tau \rangle_{+} =1$ and  $\langle \tau \rangle_{-}=5$ is also shown (blue boxes). For the system with same mean waiting times the average diffusivity found in the simulations is $\langle D \rangle=4.98$ and for the case of different mean waiting times we have $\langle D \rangle = 1.69$. In both cases we used $t=1000$ and $1000$  trajectories. \label{fig:pdfD}}
\end{figure} 

We now investigate the fluctuations of the time averaged  diffusivities in a two state model and their implications in the distribution of  diffusion coefficients obtained from real experimental data. For a further analysis of the time average diffusivity within a two state system see ~\cite{Greb2019, WangCH_2020}. 

We note that in different single particle tracking experiments with non-Gaussian propagators, the recorded distribution of the diffusion coefficient $D$ (obtained by  means of  TAMSD analysis) is relatively broad and peaked close to the origin ~\cite{hapca2009,granick2012,spako2017}. Those experimental distributions of $D$ are typically fitted by exponential ~\cite{spako2017} or gamma ~\cite{hapca2009} distributions. Within the two state model  the diffusivity   takes only two possible values $D_{-}$ or $D_{+}$, but the respective TAMSD analysis gives values of $D$ around $D_{-}$ and $D_{+}$ ~\cite{Greb2019}.  The average $D$ is given by $ \langle D \rangle = (D_{+}\langle \tau \rangle_{+} + D_{-}\langle \tau \rangle_{-})/(\langle \tau \rangle_{+} + \langle \tau \rangle_{-})$. 
So how different is the distribution of the diffusivities, extracted via TAMSD techniques,  in a two state model  compared with the one present in single molecule experiments? As we show next this will be determined by the values of $D_{\pm}$ and $\langle \tau \rangle _{\pm}$. 
In Figure ~\ref{fig:pdfD} we show the distribution of the diffusion coefficients obtained by means of TAMSD analysis for $D_{+}=10$, $D_{-}=0$ and exponentially distributed waiting times. We show  two different cases, the first one with  the same mean waiting times $\langle \tau \rangle_{+} = \langle \tau \rangle_{-} =1$ (see red boxes). And the second one with different mean waiting times, such that $\langle \tau \rangle_{+} =1$ and $\langle \tau \rangle_{-}=5$ (see blue boxes). As we can see in Figure ~\ref{fig:pdfD},  when the difference between the diffusion coefficients is large, as in our case $D_{+}=10 > D_{-}=0 $,  $P(D)$ is relatively broad. Nonetheless, for the case with $\langle \tau \rangle_{+} =1$ and $\langle \tau \rangle_{-}=5$  the peak of $P(D)$ is closer to the origin compared to the case with $\langle \tau \rangle =1$.

This difference between mean waiting times in each state is the second factor that determines the shape of $P(D)$. For instance  when this difference is such that  $ \langle \tau \rangle_{+} < \langle\tau \rangle_{-} $, it is straightforward that the more the process spends in the state ``$-$'', the more the observed values of $D$ will be closer to $D_{-}$. In this latter case the distribution of $D $ is peaked close to the origin since $D_{-} < D_{+}$. Thus we can say that when the differences between the diffusivities (and the mean waiting times) in the different states are pronounced, \textit{i.e.} $D_{-}<< D_{+}$ and $\langle \tau \rangle_{+} << \langle \tau \rangle_{-} $, $P(D)$ in the two sate model resembles the distributions found in single molecule experiments ~\cite{hapca2009,granick2012,spako2017}.

%%%%%%%%%%%%%%%%%%%%%%%%%%%%%%%%%%%%%%%%%%
\section{Conclusions\label{sec:5}}

From symmetry of the density of spreading particles   $P(x,t)=P(-x,t)$,  we expect an analytical expansion of the propagator as  $P(x,t) \sim K_{1} - K_{2} x^2 + \ldots$, with $K_{1},K_{2}$ constants. Instead in the two state model treated along this work,  we get an expansion that is linear in $\vert x\vert$, see Eq.~\eqref{eq:Pxstsxc1}. This is a non analytical expansion graphically  represented by a tent like structure, see Fig \ref{fig:pdfxUG}, Figure~\ref{fig:pdfxSR} and Figure~\ref{fig:pdfxDR}. As mentioned above,  Laplace in 1774 considered a similar non-analytical PDF, $P(x)=\exp(-\vert x\vert)/2$ for $-\infty <x< \infty$ \cite{Laplace1774,wilson1923}. However, the expression we find is clearly non-exponential, see Eq.\eqref{eq:pxstc1}. Further for large $x$ we get a Gaussian behavior for $P(x,t)$. It should be noted that a non-analytical behavior is found only if $D_{-}=0$, see Appendix~\ref{sec:A0} for further details. In practice we may approach the non - analytical   features of $P(x,t)$, as $D_{-}$  is getting small. 

Recently a very general theory was developed for the non-Gaussian spreading of packets of particles. Using a CTRW framework it was shown, that for any analytical PDF of waiting times, for large $x$ limit  $P(x,t) \sim \exp( - C \vert x \vert \ln \vert x \vert)$, with $C$ a constant ~\cite{bb2019}. In the former model we thus find exponential tails for large $x$, while here the anomaly, \textit{i.e.} the cusp or tent like feature of $P(x,t)$, comes from the small $x$ limit.

Recently, Postnikov \textit{et al.} ~\cite{Post2020},  investigated a model of diffusion in a quenched disordered setting, where the diffusive field is spatially varying.  They showed that  equilibrium initial conditions  plays a major role stating: "within the class of models with quenched disorder, the It\^{o} model under equilibrium conditions is the only promising candidate for the description of Brownian Non Gaussian  diffusion (BnG)." Note that here the definition of BnG means a model or system  where the MSD is increasing linearly \textit{for all times} and the propagator is non - Gaussian. Our model uses a time dependent diffusivity, and we showed  that equilibrium initial conditions are indeed  a key requirement. Here we note that BnG does not imply a cusp, and vice versa. Namely we may find a system  where the MSD is increasing linearly in time, for all the span of time, with or without a cusp for $P(x,t)$ at $x=0$.  
The main focus of our work is the presence of a cusp for $P(x,t)$.  Regarding the behavior of the MSD, it can be shown  that when equilibrium initial conditions are applied  $\langle T_+ \rangle = (\langle \tau \rangle_{+} t)/[\langle \tau \rangle_{+} + \langle \tau \rangle_{-}]$, for all times $t$ (see Appendix ~\ref{sec:A4}). Then by Eq.~\eqref{eq:x22d} and Eq.~\eqref{eq:avTpf} the MSD is provided by 
\begin{equation}\label{eq:MSDS}
\langle x^2 (t) \rangle =\Bigg( \frac{D_{+}\langle \tau \rangle_{+} + D_{-}\langle \tau \rangle_{-}}{\langle \tau \rangle_{+} + \langle \tau \rangle_{-}} \Bigg)t,
\end{equation}
for any time $t$. Thus if the process starts from equilibrium the MSD grows linearly for all times and we have BnG. Nevertheless  we would like to emphasize that our model is exhibiting BnG, but specifically $P(x,t)$ has a cusp only if $D_{-}=0$, and practically when $D_{-} << D_{+}$.

To summarize, we emphasize that we have shown, by means of the statistics of the temporal occupation, that there is  a universality   for the PDF of the temporal occupation fraction  in a two state model. For PDFs of waiting times with finite first moments, $g_{t}(p_{+})$ can be approximated by a uniform distribution following Eq.~\eqref{eq:fppQNst}. This leads to tent like decaying propagators (Eq.~\eqref{eq:pxstc1}) similar to those found in many experimental systems.  We corroborate our results by solving analytically a two state system with exponentially distributed waiting times. We have shown that either for short or long times the distribution of displacements $P(x,t)$ has a general form, either a ``tent'' or a Gaussian bell curve. These two  endpoints of the positional PDF are independent of the actual form of the distribution of waiting times. The crucial point within our framework, is the generality of the behavior of the PDF of the occupation fraction $p_+$, being a uniform distribution for short times  and   a delta peak for long times. The former was found for a system with equilibrium initial conditions. We note that for certain types of non - equilibrium initial conditions we can still get a uniform PDF for the fraction occupation time, however this is not generic (see details in Appendix~\ref{sec:AB}). Therefore  the non - Gaussian features are readily present in our model within the short time regime, and regardless of the specifics of the waiting times. 

Mathematically, we presented an expansion in terms of the number of transitions  from state $+$ to $-$  and backwards.  Naturally, for very short times, the leading contribution to the packet comes from the paths with zero transitions, and then the packet is simply a sum of two Gaussian curves with diffusion coefficients $D_{+}$ and $D_{-}$. However, we showed that by going to next order terms in the expansion, namely considering the paths with a single jump, we get the cusp like shape, found in the limit $D_{-} \rightarrow 0$. Thus the whole effect is achieved by using a perturbation approach obtaining the leading order correction to the trivial behavior.  Put differently, widely popular  super statistical approach is found to miss one of  the main issues of the field, namely the cusp in $P(x,t)$. Super - statistical approach ~\cite{BECK2003,chechk2017} uses a distribution of diffusivities, which in our model is a sum of two delta functions, at  $D_{-}=0$ and $D_{+}$. This does not give the cusp, as it is merely the zeroth order  of the perturbation theory developed here. 

%#############################################################
%#############################################################

\section{Acknowledgments}
E.B and M.H.S. thanks the support of the Israel Science Foundation Grant No. 1898/17. S.B. thanks the support of  the Pazy foundation grant No. 61139927 and the Israel Science Foundation Grant No. 2796/20. 

\nocite{*}

\appendix

\section{\label{sec:A0} A two state model with $D_{+}>D_{-}>0$}
When $D_{+}>D_{-}>0$ the process for the displacements becomes
\begin{eqnarray}\label{eq:x2d}
x(t)= \sqrt{2D_{+} T_{+}} \xi_{1} + \sqrt{2D_{-} (t- T_{+})} \xi_{2},
\end{eqnarray}
with $\xi_{1}$ and $\xi_{2}$ each one i.i.d. Gaussian variables. In this case the form of the conditioned PDF is given by
\begin{eqnarray}
P(x,t \vert T_{+}) = \frac{e^{- \frac{x^{2}}{4[D_{+}T_{+} + D_{-} (t-T_{+})] }}}{\sqrt{4\pi [ D_{+} T_{+} + D_{-} (t-T_{+})] }}.
\end{eqnarray}
Then the marginal distribution for the displacements follows
\begin{eqnarray}\label{eq:PXTdr}
P(x,t)= \displaystyle \int \limits_{0}^{1} \frac{e^{- \frac{x^{2}}{4t[D_{+}p_{+} + D_{-} (1-p_{+})] }}}{\sqrt{4\pi t[ D_{+} p_{+} + D_{-} (1-p_{+})] }} g_{t}(p_{+}) dp_{+}.
\end{eqnarray}
%\subsection{}
\subsection{\label{sec:A01} $P(x,t)$ for arbitrary waiting times}

\begin{figure}[H]
\centering
\includegraphics[width=6.5 cm]{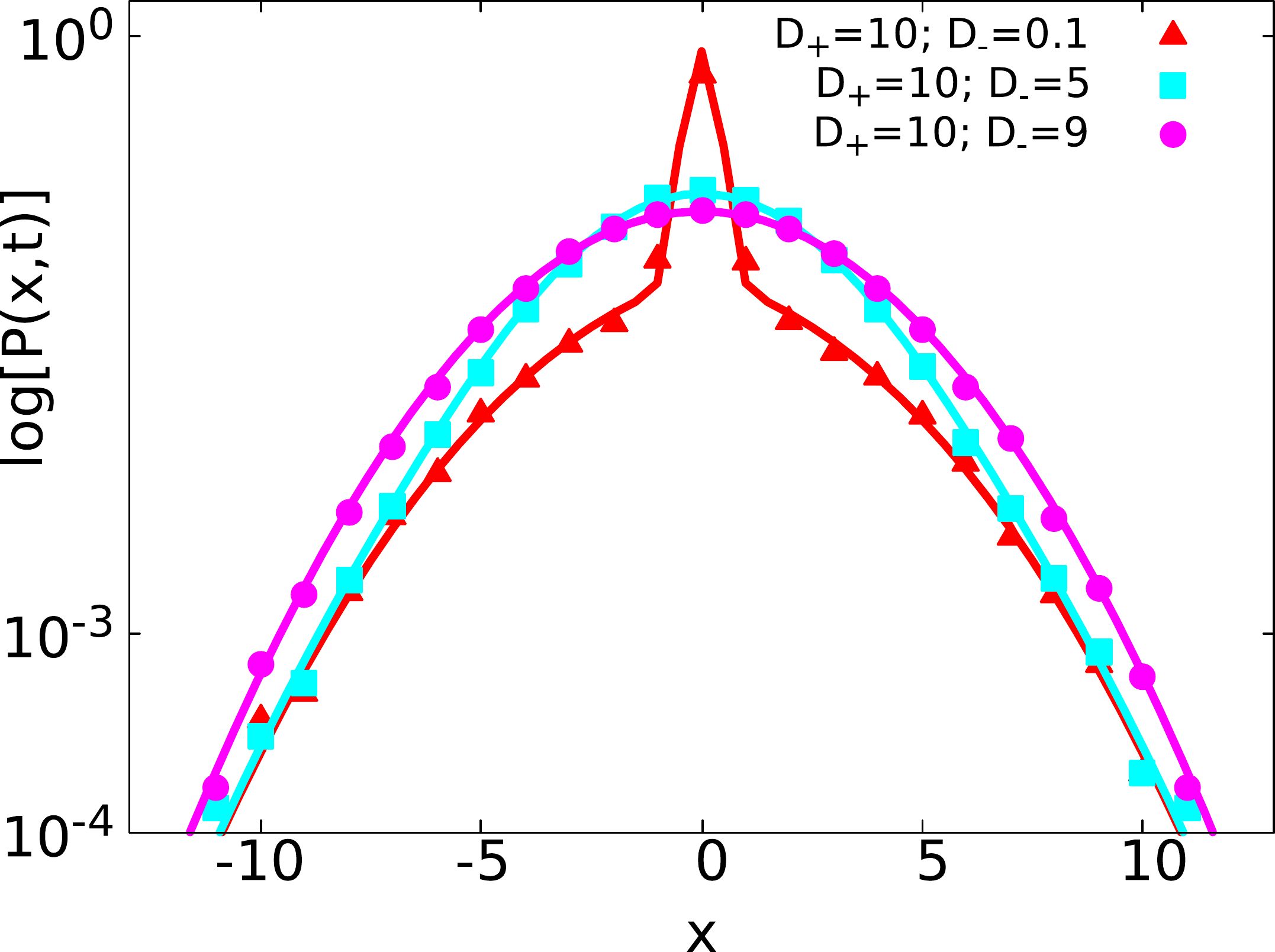}
\caption{Distribution of displacements $P(x,t)$ obtained by simulations of a two state system with $D_{+}> D_{-}>0$ and gamma distributed waiting times $\tau\sim Gamma(3,1)$ at $D_{+}$ and $\tau\sim Gamma(6,1)$ at $D_{-}$ following Eq.~\eqref{eq:taugamma}. We compare with Eq.~\eqref{eq:PxGwstc2} (solid lines) with $t=0.5$, $\langle \tau \rangle_{+}=3$, $\langle \tau \rangle_{-}=6$, $D_{+}=10$. For  $D_{-}=0.1$ (red triangles), $D_{-}=5$ (cyan squares) and $D_{-}=9$ (magenta circles).  Exponential like decaying is present at small values for $x$, when $D_{+}=10>>D_{-}=0.1$ (red solid line). In the cases when $D_{-} \longrightarrow D_{-}$ (cyan and magenta solid lines) $P(x,t)$ follows a full Gaussian distribution. \label{fig:pdfDC}}
\end{figure} 

As we did in section ~\ref{sec:31}, using the general forms obtained above, i.e.  Eq.~\eqref{eq:fTpQN}, Eq.~\eqref{eq:fTpN1p}, Eq.~\eqref{eq:fTpN1m}, Eq.~\eqref{eq:fTpN12} and Eq.~\eqref{eq:fpltC1}, we can analyze $P(x,t)$ in the short and long times limits.

\paragraph{\textbf{Short time regime}\\}
Substituting  Eq.~\eqref{eq:fppQNst}  in  Eq.~\eqref{eq:PXTdr} we get 
\begin{eqnarray}\label{eq:PxGwstc2}
P(x,t)&=&\frac{\langle\tau \rangle_{+} }{ \langle \tau \rangle_{+}+\langle \tau \rangle_{-} }\Bigg( 1- \frac{t}{\langle \tau \rangle_{+}}\Bigg)\frac{e^{-\frac{x^{2}}{4D_{+}t}}}{\sqrt{4\pi D_{+}t}}+ \frac{\langle\tau \rangle_{-} }{ \langle \tau \rangle_{+}+\langle \tau \rangle_{-} }\Bigg( 1- \frac{t}{\langle \tau \rangle_{-}}\Bigg)\frac{e^{-\frac{x^{2}}{4D_{-}t}}}{\sqrt{4\pi D_{-}t}} \nonumber \\
&+& \frac{2 }{\sqrt{\pi} (  \langle \tau\rangle_{+} +\langle \tau \rangle_{-}  ) [ D_{-} - D_{+} ]} \Bigg\lbrace \sqrt{ D_{-}t }   e^{-\frac{x^{2}}{4D_{-}t}}-  \sqrt{D_{+}t}   e^{-\frac{x^{2}}{4D_{+}t}} \nonumber \\
&-& \frac{\sqrt{\pi} \vert x \vert}{2} \Bigg[ Erf\Bigg(  \frac{\vert x \vert}{\sqrt{4D_{+}t} } \Bigg) - Erf\Bigg(  \frac{\vert x \vert}{\sqrt{4D_{-}t} } \Bigg)\Bigg] \Bigg\rbrace.
\end{eqnarray}
In Figure~\ref{fig:pdfDC} we compare Eq.~\eqref{eq:PxGwstc2} (in solid lines) and $P(x,t)$ obtained by simulations of a two state model for a fixed value of $D_{+}$ and different values of $D_{-}$, such that $D_{+}>D_{-}$. In all cases we used  gamma distributed waiting times $\tau\sim Gamma(3,1)$ for the state with $D_{+}$ and $\tau\sim Gamma(6,1)$ for the state with $D_{-}$ (the gamma distribution is defined by  Eq.~\eqref{eq:taugamma}). 
As we can see when $D_{+} >> D_{-}$, \textit{e.g.} $D_{+}=10$ and $D_{-}0.1$ (red triangles), the PDF of the displacements at small values of $x$ has a non-Gaussian peak, thereafter for large values of $x$  it follows a Gaussian distribution. When the values of $D_{-}$ approach to $D_{+}$ (cyan squares and magenta circles), $P(x,t)$ is fully described by Gaussian statistics even in the short time limit.

\paragraph{\textbf{Long time regime}\\}
In the long time limit the PDF of temporal occupation fraction is provided by Eq.~\eqref{eq:fpltC1}, then according to Eq.~\eqref{eq:PXTdr} the PDF of the displacements is determined by 
\begin{eqnarray}\label{eq:PXdrlt}
P(x,t)\sim \sqrt{\frac{ \langle \tau\rangle_{+} +\langle \tau \rangle_{-} }{4\pi t [D_{+} \langle \tau\rangle_{+} + D_{-} \langle \tau \rangle_{-}]}}e^{- \frac{x^{2}(\langle \tau \rangle_{+} + \langle \tau \rangle_{-})}{4t[D_{+} \langle \tau\rangle_{+} + D_{-} \langle \tau \rangle_{-}] }}.
\end{eqnarray}  
Thus the Gaussian limit is also restored.

\subsection{\label{sec:A02}  $P(x,t)$ for exponentially distributed waiting times with $ \langle \tau \rangle_{+} \neq \langle \tau \rangle_{-} $ }

In the short time regime we can use the uniform approximation  Eq.~\eqref{eq:fppudr}  in Eq.~\eqref{eq:PXTdr}, the distribution for the displacements yields 
\begin{eqnarray}\label{eq:PXustdr}
P(x,t)&\sim &\frac{\langle \tau\rangle_{+}}{ \langle \tau \rangle_{+} + \langle \tau \rangle_{-} } \frac{e^{- \frac{t}{\langle \tau \rangle_{+}}- \frac{x^{2}}{4 D_{+}t}}}{\sqrt{4\pi D_{+} t}} + \frac{\langle \tau \rangle_{-}}{ \langle \tau \rangle_{+} + \langle \tau \rangle_{-} } \frac{e^{- \frac{t}{\langle \tau \rangle_{-}}- \frac{x^{2}}{4 D_{-}t}}}{\sqrt{4\pi D_{-} t}}  \nonumber \\
&+& \frac{1}{(\langle \tau \rangle_{+} + \langle \tau \rangle_{-}) (D_{-}-D_{+}) \sqrt{\pi}} \Bigg\lbrace \sqrt{4D_{-}t}e^{-\frac{x^{2}}{4D_{-}t}} - \sqrt{4D_{+}t}e^{-\frac{x^{2}}{4D_{+}t}} \nonumber \\
&+& \pi \Bigg[ Erf\Big( \frac{\vert x \vert}{\sqrt{4D_{-} t}}\Big) -  Erf\Big( \frac{\vert x \vert}{\sqrt{4D_{+} t}}\Big)   \Bigg] \Bigg\rbrace .
\end{eqnarray}   
For the long time regime we use $g_{t}(p_{+})$ provided by Eq.~\eqref{eq:fpltC1}, then according to Eq.~\eqref{eq:PXTdr} the PDF of the displacements follows Gaussian statistics  described by Eq.~\eqref{eq:PXdrlt}.

\section{\label{sec:AB} A complementary deduction of $f^{\pm}_{t}(T_{+}\vert 1)$}
In this section we obtain Eq.~\eqref{eq:fTpN12} from the definition of conditional probability. The conditional probability of $T_{+}$, given that $N$ jumps have been made,  follows
\begin{eqnarray}\label{eq:condfTp1}
f^{\pm}_{t}(T_{+}\vert N)=\frac{f^{\pm}_{t}(T_{+},N) }{Q^{\pm}_{t}(N)},
\end{eqnarray}
with the distribution of jumps defined by ~\cite{GL2001} 
\begin{eqnarray}\label{eq:QNGL}
Q^{\pm}_{t}(N)=\langle \mathbbm{1}_{(t_{N},t_{N+1})}(t) \rangle,
\end{eqnarray}
with $\mathbbm{1}_{(a,b)}(t)$ the indicator function, such that is equal to one if $t\in (a,b)$ and zero if $t\notin (a,b)$. The average $\langle \cdot \rangle$ is over all the the values of $\tau_{i}$'s, with $i\in\lbrace1,2,\ldots,N+1 \rbrace$. And $f^{\pm}_{t}(T_{+},N)$ is the joint probability of $T_{+}$ and $N$, which satisfies ~\cite{GL2001}
\begin{eqnarray}\label{eq:fTpNJGL}
f^{\pm}_{t}(T_{+},N)=\langle \delta(y-T_{+})\mathbbm{1}_{(t_{N},t_{N+1})}(t)  \rangle,
\end{eqnarray}
with $t_{N}=\tau_{1}+\ldots +\tau_{N}$ and the average $\langle\cdot \rangle$ defined as above. 

Let us find Eq.~\eqref{eq:QNGL}, Eq.~\eqref{eq:fTpNJGL}   and therefore Eq.~\eqref{eq:condfTp1}  for the case $N=1$. Important to notice, is that for the case of equilibrium initial conditions as the one treated in section ~\ref{sec:3}, when  $N=1$ the corresponding average on $\tau_{1}$ is given by the forward recurrence distribution Eq.~\eqref{eq:pdfforward}. Following Eq.~\eqref{eq:QNGL}, and taking the  Laplace transform  defined as $\hat{Q}^{\pm}_{s}(N) = \int ^{\infty}_{0}  e^{-st} Q^{\pm}_{t}(N) dt$, after simple manipulations we obtain
\begin{eqnarray}\label{eq:Qs1}
\hat{Q}^{\pm}_{s}(1) &=& \displaystyle \int \limits _{0}^{\infty} e^{-s\tau_{1}}f^{\pm}_{eq}(\tau_{1})d\tau_{1}\displaystyle \int \limits _{0}^{\infty} \Big(\frac{1-e^{-s\tau_{2}}}{s}\Big)\psi_{\mp}(\tau_{2})d\tau_{2},\nonumber\\
&=& \Bigg( \frac{1-\hat{\psi}_{\pm}(s)}{\langle \tau \rangle_{\pm} s} \Bigg) \Bigg( \frac{1-\hat{\psi}_{\mp}(s) }{s} \Bigg),
\end{eqnarray}
which is already the result shown in Eq.~\eqref{eq:QsNS}.
%%%%%%%%%%%%%%%%%%%%%%%%%%%%%%%%%%%%%%%%%%

For the joint distribution $f^{\pm}_{t}(T_{+},1)$, following Eq.~\eqref{eq:fTpNJGL} and taking the double Laplace transform defined as $\hat{f}^{\pm}_{s}(u,N) = \int ^{\infty}_{0} e^{-uT_{+}} \int ^{\infty} _{0} e^{-st} f^{\pm}_{t}(T_{+},N) dt dT_{+}$,  after performing the corresponding integrals in the case we started from "+" we get 
\begin{eqnarray}\label{eq:fsu1}
\hat{f}^{+}_{s}(u,1) &=& \displaystyle \int \limits _{0}^{\infty} e^{-(s+u)\tau_{1}}f^{+}_{eq}(\tau_{1})d\tau_{1}\displaystyle \int \limits _{0}^{\infty} \Big(\frac{1-e^{-s\tau_{2}}}{s}\Big)\psi_{-}(\tau_{2})d\tau_{2},\nonumber\\
&=& \Bigg( \frac{1-\hat{\psi}_{+}(s+u)}{\langle \tau \rangle_{+} (s+u)} \Bigg) \Bigg( \frac{1-\hat{\psi}_{-}(s) }{s} \Bigg).
\end{eqnarray}
Following the same procedure for the case when the process started from "-", we obtain
\begin{eqnarray}\label{eq:fsu1m}
\hat{f}^{-}_{s}(u,1) =  \Bigg( \frac{1-\hat{\psi}_{-}(s)}{\langle \tau \rangle_{-} (s)} \Bigg) \Bigg( \frac{1-\hat{\psi}_{+}(s+u) }{s+u} \Bigg).
\end{eqnarray}

Next we show the connection of the joint PDf Eq.~\eqref{eq:fTpNJGL} with the uniform distribution of the occupation times Eq.~\eqref{eq:fTpQNst}. Let $\hat{f}_{s}^{eq}(u,1)$ be the double Laplace transform of $f_{t}^{eq}(T_{+},1)$, \textit{i.e.} the joint PDF of $T_{+}$ and one single jump, starting from equilibrium. Clearly the former follows  
\begin{eqnarray}\label{eq:fequ1}
\hat{f}_{s}^{eq}(u,1)=\frac{\langle \tau \rangle_{+}}{\langle \tau \rangle_{+}+\langle \tau \rangle_{-} } \hat{f}^{+}_{s}(u,1)+\frac{\langle \tau \rangle_{-}}{\langle \tau \rangle_{+}+\langle \tau \rangle_{-} } \hat{f}^{-}_{s}(u,1).
\end{eqnarray}
Using Eq.~\eqref{eq:fsu1} and Eq.\eqref{eq:fsu1m} in Eq.~\eqref{eq:fequ1}, we get  
\begin{eqnarray}\label{eq:fsum1eqa}
\hat{f}_{s}^{eq}(u,1)= \frac{2}{\langle \tau \rangle_{+} + \langle \tau \rangle_{-}}\Big( \frac{1-\hat{\psi}_{+}(s+u)}{s+u}\Big) \Big( \frac{1-\hat{\psi} (s)}{s}\Big).
\end{eqnarray}
One of the key features of our paper is found when we consider both $u$ and $s$ to be large. This corresponds to the short time limit, when $T_{+}$ and $t$ are of the same order.  Then we may use $\hat{\psi}_{+}(s+u),\hat{\psi}_{-}(s) \longrightarrow 0$ in Eq.~\eqref{eq:fsum1eqa}, yielding to
\begin{eqnarray}\label{eq:fsum1eqapp}
\hat{f}_{s}^{eq}(u,1)\sim \frac{2}{[\langle \tau \rangle_{+} + \langle \tau \rangle_{-}](s+u)s}.
\end{eqnarray}
Eq.~\eqref{eq:fsum1eqapp} is easy to invert, and  we find in the short time limit 
\begin{eqnarray}\label{eq:fJTp1eqf}
f_{t}^{eq}(T_{+},1)\sim \frac{2}{\langle \tau \rangle_{+} + \langle \tau \rangle_{-}}; \,\,\ for \,\,\ T_{+}<t.
\end{eqnarray}
This is the short time uniformity we have found that in turn, as explained in section ~\ref{sec:3}, gives the cusp like shape in $P(x,t)$. Eq.~\eqref{eq:fJTp1eqf} is the last term in Eq.~\eqref{eq:fTpQNst}, corresponding to $N=1$ (the first two terms in Eq.~\eqref{eq:fTpQNst} are contributions from $N=0$).

Now we are interested  to invert Eq.~\eqref{eq:Qs1}, Eq.~\eqref{eq:fsu1}, Eq.~\eqref{eq:fsu1m} and then apply the definition of  conditional probability Eq.~\eqref{eq:condfTp1}. But first, since we are dealing with  the short time limit $t\longrightarrow 0$, in the Laplace space this corresponds to the limit of $s\longrightarrow \infty$ and $u \longrightarrow \infty$. Thus for this particular approximation, due to the definition of the Laplace transform $\hat{\psi}_{\pm}(s)=\int^{\infty}_{0}e^{-st}\psi_{\pm}(t)dt$, we have that $\lim_{s \to \infty}\hat{\psi}_{\pm}(s)\longrightarrow 0$ for a general $\psi_{\pm}(\tau)$. In this case Eq.~\eqref{eq:Qs1}, Eq.~\eqref{eq:fsu1} and Eq.~\eqref{eq:fsu1m} are approximated by 
\begin{eqnarray}\label{eq:Qs1A}
\hat{Q}^{\pm}_{s}(1) &\sim & \frac{1}{\langle \tau\rangle_{\pm} s^{2}},\\
\hat{f}^{\pm}_{s}(u,1) &\sim & \frac{1}{\langle \tau \rangle_{\pm} (s+u)s}.\label{eq:fsu1A} 
\end{eqnarray}
Inverting Eq.~\eqref{eq:Qs1A} with respect to $s$ and Eq.~\eqref{eq:fsu1A} with respect to $u$ and $s$ with $0<T_{+}<t$, we obtain 
\begin{eqnarray}\label{eq:Qt1A}
Q^{\pm}_{t}(1) &\sim & \frac{t}{\langle \tau\rangle_{\pm}},\\
f^{\pm}_{t}(T_{+},1) &\sim & \frac{1}{\langle \tau \rangle_{\pm}}.\label{eq:ftTp1A} 
\end{eqnarray}
Now substituting Eq.~\eqref{eq:Qt1A} and Eq.\eqref{eq:ftTp1A} in the conditional probability Eq.\eqref{eq:condfTp1} for $N=1$ we obtain 
\begin{eqnarray}
f^{\pm}_{t}(T_{+}\vert 1) \sim \frac{1}{t}. 
\end{eqnarray}
Which is the same result shown in Eq.~\eqref{eq:fTpN12} of section~\ref{sec:3}. As expected, since the joint distribution $f^{\pm}_{t}(T_{+},N)$ Eq.~\eqref{eq:ftTp1A} does not depend on the time or any other variable, when is used for computing the PDF of the occupation time it gives the uniform distribution Eq.~\eqref{eq:fTpQNst}. The same procedure for values of $N\geq 2$ gives a joint distribution $f^{\pm}_{t}(T_{+},N)$ such that in the double Laplace space is defined as
\begin{eqnarray}
\hat{f}^{+}_{s}(u,N)&=&\Bigg( \frac{1-\hat{\psi}_{+}(s+u)}{\langle \tau \rangle_{+} (s+u)} \Bigg) \hat{\psi}_{-}^{k}(s)\hat{\psi}_{+}^{k}(s+u) \Bigg( \frac{1-\hat{\psi}_{-}(s) }{s} \Bigg); \nonumber \\
& if & \,\,\ N=2k+1. \label{eq:fsuNGo}\\
\hat{f}^{+}_{s}(u,N)&=&\Bigg( \frac{1-\hat{\psi}_{+}(s+u)}{\langle \tau \rangle_{+} (s+u)} \Bigg) \hat{\psi}_{+}^{k-1}(s+u)\hat{\psi}_{-}^{k}(s) \Bigg( \frac{1-\hat{\psi}_{+}(s+u) }{s+u} \Bigg); \nonumber \\ 
& if & \,\,\ N=2k.\label{eq:fsuNGe} \\
\hat{f}^{-}_{s}(u,N)&=&\Bigg( \frac{1-\hat{\psi}_{-}(s)}{\langle \tau \rangle_{-} (s)} \Bigg) \hat{\psi}_{-}^{k}(s)\hat{\psi}_{+}^{k}(s+u) \Bigg( \frac{1-\hat{\psi}_{+}(s+u) }{s+u} \Bigg); \nonumber \\
& if & \,\,\ N=2k+1. \label{eq:fsuNGom}\\
\hat{f}^{-}_{s}(u,N)&=&\Bigg( \frac{1-\hat{\psi}_{-}(s)}{\langle \tau \rangle_{-} (s)} \Bigg) \hat{\psi}_{-}^{k-1}(s)\hat{\psi}_{+}^{k}(s+u) \Bigg( \frac{1-\hat{\psi}_{-}(s) }{s} \Bigg); \nonumber \\ 
& if & \,\,\ N=2k.\label{eq:fsuNGem} 
\end{eqnarray}
In order to analyze the joint we have to deal with the analytical expression of $\psi_{\pm}(\tau)$   defined by Eq.\eqref{eq:STpsi}. For instance for $N=2$,  after substituting Eq.~\eqref{eq:psils}  in Eq.~\eqref{eq:fsuNGe} and Eq.~\eqref{eq:fsuNGem}, we have that the dominant term is $\hat{f}^{+}_{s}(u,2)\sim \ [\Gamma(A_{-}+1) C^{-}_{A_{-}}]/[\langle \tau \rangle_{+}(s+u)^{2} s^{A_{-}+1} ] $,  and $\hat{f}^{-}_{s}(u,2)\sim \ [\Gamma(A_{+}+1) C^{+}_{A_{+}}]/[\langle \tau \rangle_{-}s^{2} (s+u)^{A_{+}+1} ] $ respectively. Inverting the double Laplace transform, in each case it gives a positive power of $T_{+}$ and therefore with respect to $t$, \textit{e.g.} $f^{\pm}_{t}(T_{+},2)\sim   T_{+}^{A_{\mp}+1}$. $A_{\mp}\geq 0$ is a positive integer number, this correction term for $t\longrightarrow 0$ ($T{+}\longrightarrow0$) is negligible, and also the remaining terms in $f^{\pm}_{t}(T_{+},N)$ with $N>2$. We conclude that for the case of equilibrium initial conditions the uniformity in the short time limit, for the PDF of the occupation/fraction time is always preserved, as long  $\psi_{\pm}(\tau)$ is analytical.  

\subsection{\label{sec:AB2}  Non - equilibrium initial conditions }
Still, by employing the joint distribution $f^{\pm}_{t}(T_{+},N)$ it can be shown, as follows, that for non - equilibrium initial conditions.  When  $\psi_{\pm}(\tau)$ in the Laplace space is approximated by  $\hat{\psi}_{\pm}(s) \sim 1 / s$ for short times. It can lead to a uniform distribution in the occupation time and therefore to a tent shape in $P(x,t)$. 

In the case of non - equilibrium initial conditions, either starting just from $D_{+}$ or from $D_{-}$. The averages over $\tau_{1}$, within $Q^{\pm}_{t}(N)$ Eq.~\eqref{eq:QNGL} and the joint distribution $f^{\pm}_{t}(T_{+},N)$ Eq.~\eqref{eq:fTpNJGL}, are no longer given by $f^{\pm}_{eq}(\tau_{1})$ Eq.~\eqref{eq:pdfforward}. Now in the non - equilibrium case these corresponding averages are performed using the waiting time PDF $\psi_{\pm}(\tau_{1})$.  Following the same procedure as above, for the case $N=1$  the double Laplace transform of the joint distribution $f^{\pm}_{t}(T_{+},1)$ yields to 
\begin{eqnarray}\label{eq:jSUneq}
\hat{f}^{\pm}_{s}(u,1)=\hat{\psi}_{\pm}(s+u)\Bigg( \frac{1-\hat{\psi}_{\mp}(s)}{s}\Bigg). 
\end{eqnarray}
For a system with non - equilibrium initial conditions, in order to recover the uniform distribution in the PDF of $T_{+}$, is enough to ask that for large $s$ (short times), the PDF of the waiting times in the Laplace space follows 
\begin{eqnarray}\label{eq:Psilsn}
\hat{\psi}_{\pm}(s)\sim \frac{1}{s}.
\end{eqnarray}
Substituting Eq.~\eqref{eq:Psilsn} in Eq.~\eqref{eq:jSUneq}, we have that $\hat{f}^{\pm}_{s}(u,1)\sim 1/[(s+u)s]$. Which implies in the real space, for $0<T_{+}<t$, that the joint distribution follows $f^{\pm}_{t}(T_{+},1)\sim 1$, and therefore because of  Eq.\eqref{eq:pdfTp}, we also have a uniform distribution for $T_{+}$. As an example of a model in which $\hat{\psi}_{\pm}(s)$ goes as Eq.~\eqref{eq:Psilsn} and for non - equilibrium initial conditions $\psi_{\pm}(\tau) \neq f^{\pm}_{eq}(\tau_{1}) $ as expected, we have the case in which the PDF of waiting times is the sum of two exponential functions, \textit{e.g.} $\psi_{\pm}(\tau)= (1/2)\lbrace[\exp(-\tau / C_{1\pm})/C_{1\pm} ] + [\exp(-\tau / C_{2\pm})/C_{2\pm} ]\rbrace$, with $C_{1\pm},C_{2\pm}>0$.  In this case for $s\longrightarrow \infty$, $\hat{\psi}_{\pm}(s)\sim [C_{1\pm}+ C_{2\pm} ]/[2 C_{1\pm}C_{2\pm} s]$ and $f^{\pm}_{eq}(\tau_{1})= [\exp(-\tau_{1}/C_{1\pm})+ \exp(-\tau_{1}/C_{2\pm})]/[C_{1\pm} + C_{2\pm}  ]$. For this latter case,  since $\hat{\psi}_{\pm}(s)$ satisfies Eq.~\eqref{eq:Psilsn}, following the same analysis as above we find that the joint distribution $f^{\pm}_{t} (T_{+},1)$ is uniform.   

In Appendix~\ref{sec:A1} we show that for a system with non - equilibrium conditions and exponentially distributed waiting times with equal and different mean values, the distribution of the occupation/fraction time is also uniform. This is mainly because for exponentially distributed waiting times the forward recurrence time distribution in equilibrium $f^{\pm}_{eq}(\tau_{1})$ Eq.~\eqref{eq:pdfforward} is equal to $\psi_{\pm}(\tau_{1})$, as in the non - equilibrium case. Furthermore for exponentially distributed sojourn times Eq.~\eqref{eq:Psilsn} is also satisfied, its PDF in the Laplace space for $s\longrightarrow \infty$ follows $\hat{\psi}_{\pm}(s)\sim 1/[\langle \tau \rangle_{\pm} s]$. By using  Eq.\eqref{eq:pdfTp}, this gives the uniform distribution shown in Eq~\eqref{eq:fppneqsrun} and Eq.~\eqref{eq:fppneqdrun}. 

For $\hat{\psi}_{\pm}(s)$ given in  Eq.~\eqref{eq:Psilsn} the correction terms when $N\geq 2$, following the same analysis as in the case of equilibrium initial conditions, yield to  elements  of the form  $f^{\pm}_{t}(T_{+},N)\sim T^{N-1}_{+}$, which are negligible for $t \longrightarrow 0 \Longleftrightarrow T_{+} \longrightarrow 0$. So they do not contribute in the PDF of the fraction occupation time.

\section{\label{sec:AC} $P(x,t)$ from simulations with uniform and gamma distributed waiting times within the complete range of $x$}

\begin{figure}[H]
\centering
\includegraphics[width=6.5 cm]{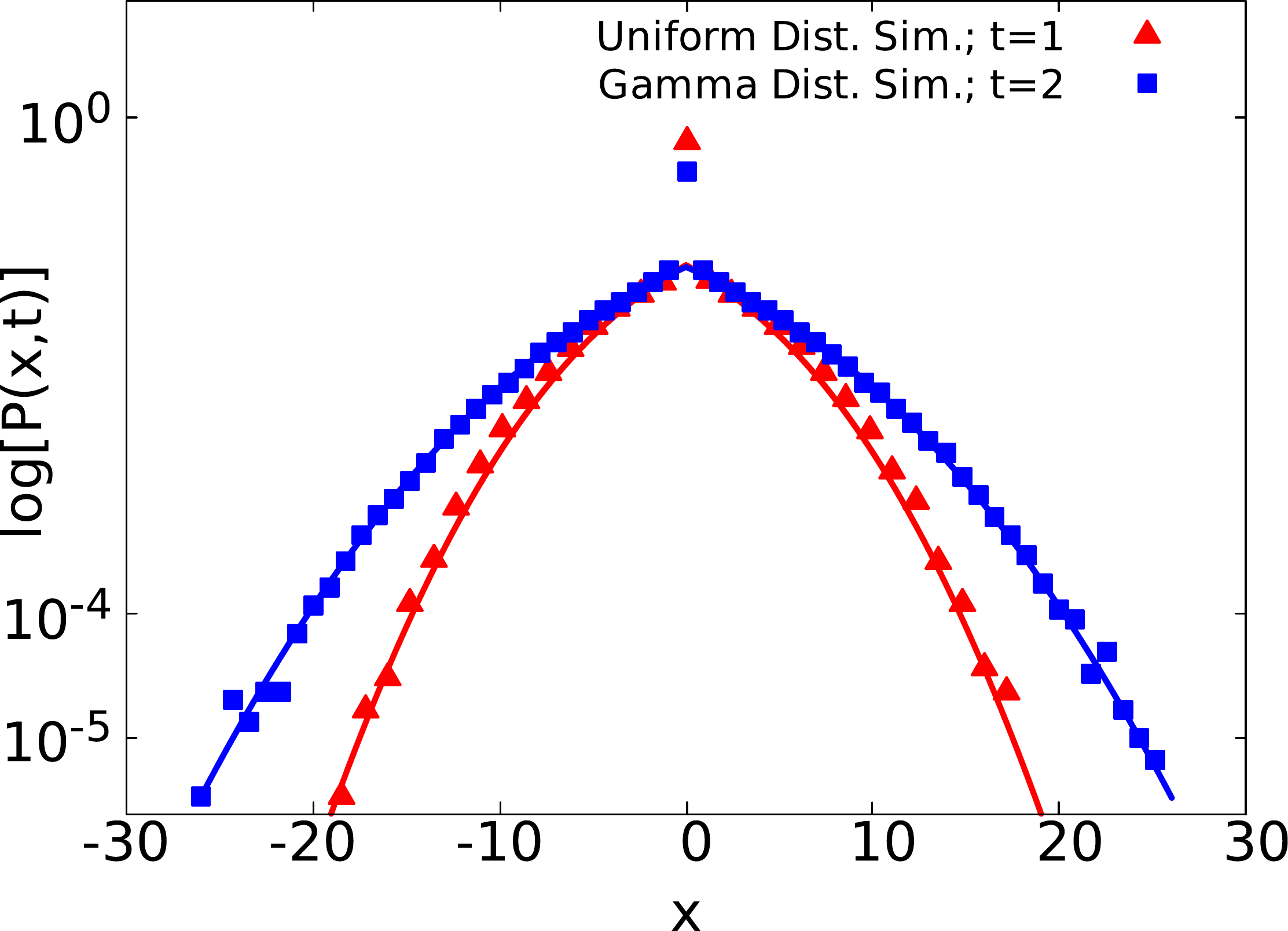}
\caption{Distribution of displacements $P(x,t)$ in semi-log scale, obtained from simulations, of a two state system with  uniform and  gamma distributed waiting times within the short time limit and displaying the whole span of $x$. $P(x,t)$ for  uniformly distributed waiting times is shown in  red triangles. And the case of gamma distributed waiting times is shown in blue squares. We employed the same set of parameters as those used in Fig.~\ref{fig:pdfxUG} at the left panel.  Both cases fit with Eq.~\eqref{eq:pxstc1} (red and blue solid lines).
\label{fig:pdfxUGC}}
\end{figure}

Following Fig.~\ref{fig:pdfxUG} at the left panel in which, for short time and displacements,  the cusp of $P(x,t)$  is displayed. Now from simulations (with the same parameters as above) of a two state model,   with uniform (red triangles) and gamma (blue squares) distributed waiting times. In Fig.~\ref{fig:pdfxUGC} we show $P(x,t)$ in semi-log scale but for the whole span of $x$. In each case, we compare the normalized histogram of the simulation data  with the short time analytical formula Eq.~\eqref{eq:pxstc1}, finding a perfect agreement. As we can see the cusp is located at the origin, and for large displacements Gaussianity is recovered. 

\section{\label{sec:A1} PDF of occupation times for exponentially distributed waiting times and non equilibrium initial conditions}
\begin{figure}[H]
\centering
\includegraphics[width=6.5 cm]{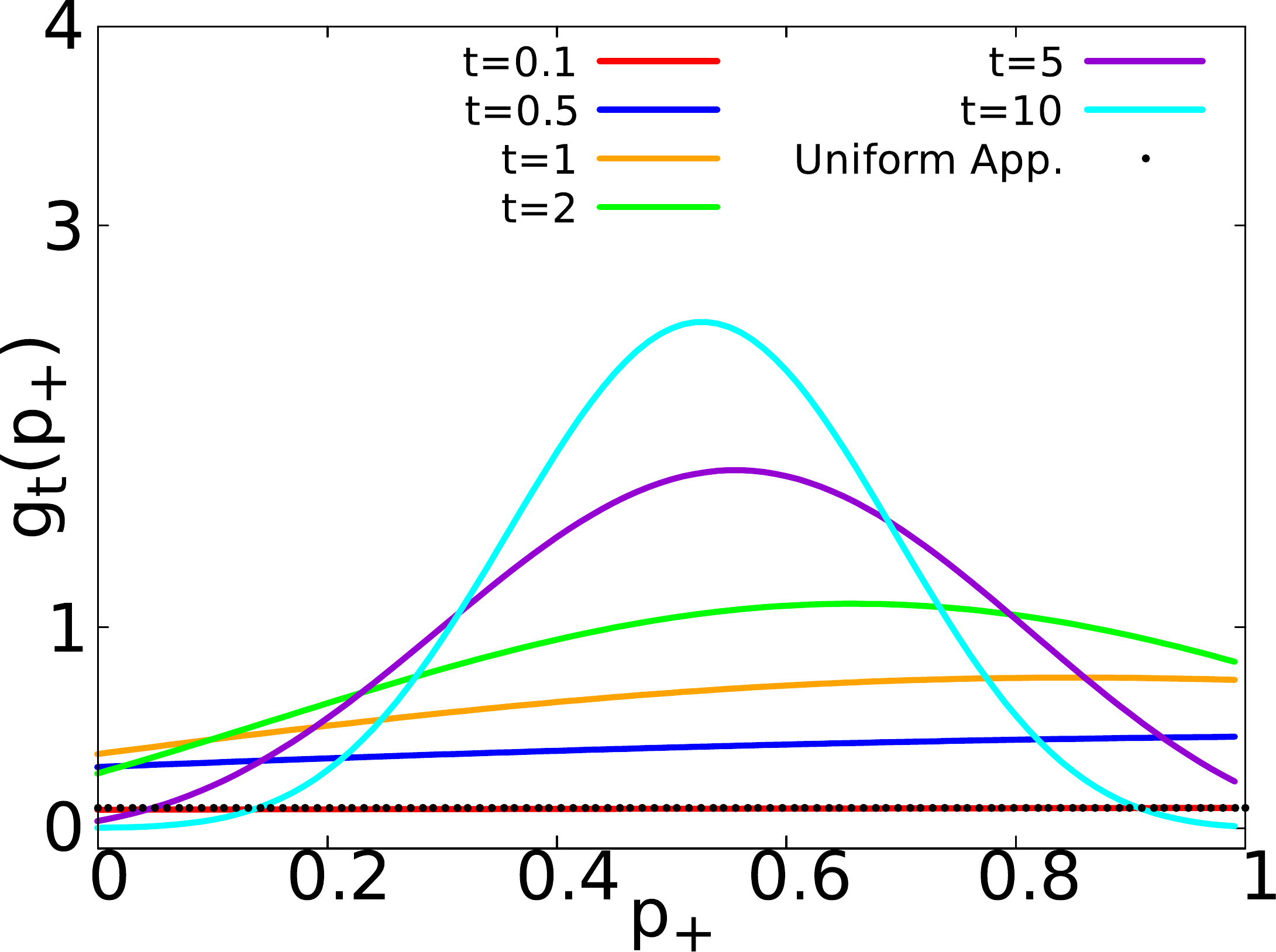}
\includegraphics[width=6.5 cm]{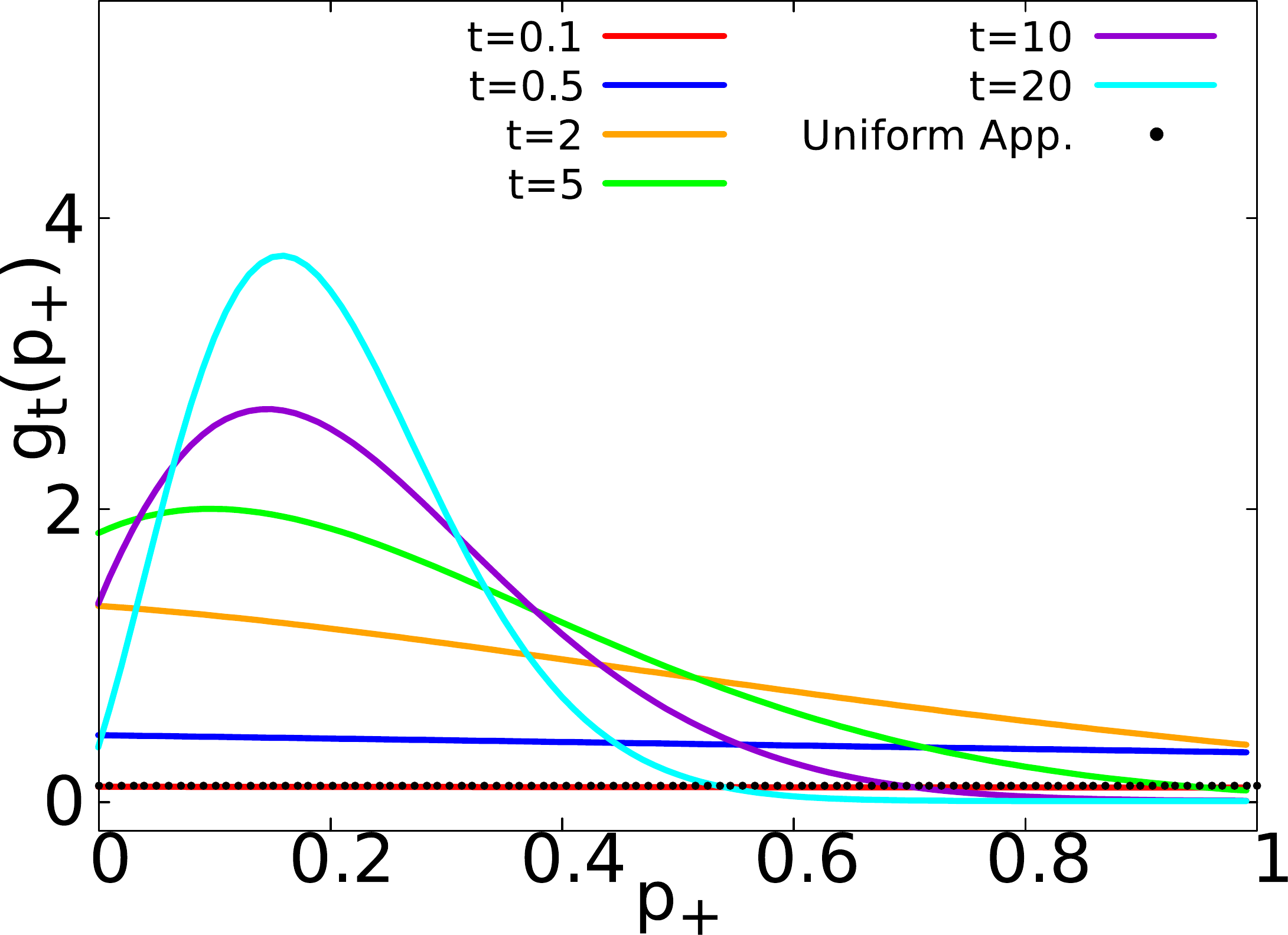}
\caption{Left: $g_{t}(p_{+})$ Eq.~\eqref{eq:fppneqsr}  for $\langle \tau \rangle=1$ and $t\in\lbrace 0.1,0.5,1,2,5,10\rbrace$ and non-equilibrium initial conditions (starting from state ``+''). The uniform approximation of $g_{t}(p_{+})$ Eq.~\eqref{eq:fppneqsrun} for $t=0.1$ is shown in black circles. Right: $g_{t}(p_{+})$ Eq.~\eqref{eq:fppneqdr}  for $\langle \tau \rangle_{+}=1$, $\langle \tau \rangle_{-}=5$  and $t\in\lbrace 0.1,0.5,2,5,10,20\rbrace$ and non-equilibrium initial conditions (starting from state ``+''). The uniform approximation of $g_{t}(p_{+})$ Eq.~\eqref{eq:fppneqdrun} for $t=0.1$ is shown in black circles.\label{fig:pdfpsrneq}}
\end{figure}

We consider  the case of a system with exponentially distributed waiting times, with $\langle \tau \rangle_{+}= \langle \tau \rangle_{-}$ in Eq.~\eqref{eq:pdfm}. Here we address the situation with non - equilibrium initial conditions. Particularly the initial conditions are such that the probability of starting at the state with $D_{+}$ is $1$ and the probability of starting from the state with $D_{-}$ is $0$. The PDF of $T_{+}$ then satisfies 
%%%%%%%%%%%%%%%%%%%%%%%%%%%
\begin{eqnarray}\label{eq:fTpneq}
f_{t}(T_{+})=f^{+}_{t}(T_{+}) = \displaystyle \sum \limits ^{\infty} _{N=0} f^{+}_{t}(T_{+},N).
\end{eqnarray}
% this is the full derivation
With  $f^{+}_{t}(T_{+},N)$ given by Eq.~\eqref{eq:fTpNJGL} explicitly for this case we have ~\cite{GL2001}
\begin{eqnarray}\label{eq:fTPNk}
f_{t}^{+}(T_{+},2k+1)&=& \displaystyle \int \ldots \int  \delta\Big(T_{+} - \sum ^{2k+1}_{i=1(odd)}\tau_{i}\Big) \mathbbm{1}_{(t_{2k+1},t_{2k+2})}(t) \psi(\tau_{1})\psi(\tau_{2}) \nonumber \\
&\ldots & \psi(\tau_{2k+2})d\tau_{1}d\tau_{2}\ldots d\tau_{2k+2}\,\,\,\ if \,\,\,\ N=2k+1,\nonumber \\
f_{t}^{+}(T_{+},2k)&=& \displaystyle \int \ldots \int \delta\Big(T_{+} - \sum ^{2k-1}_{i=1(odd)} \tau_{i} - \tau^{\ast}\Big) \mathbbm{1}_{(t_{2k},t_{2k+1})}(t)\psi(\tau_{1})\psi(\tau_{2}) \nonumber\\
&\ldots & \psi(\tau_{2k+1})d\tau_{1}d\tau_{2}\ldots d\tau_{2k+1}\,\,\,\ if \,\,\,\ N=2k,
\end{eqnarray}
with $\mathbbm{1}_{(a,b)}(t)$ the indicator function equal to $1$ if $t\in(a,b)$ and $0$ if $t\notin(a,b)$. We work with the double Laplace transform $\mathcal{L}\Big\lbrace f^{+}_{t}(T_{+},N) \Big\rbrace= f^{+}_{s}(u,N)$ with $t \Longleftrightarrow s$ and $u \Longleftrightarrow T_{+}$, which is given by $\hat{f}^{+}_{s}(u,N) = \int ^{\infty}_{0} e^{-uT_{+}} \int ^{\infty} _{0} e^{-st} f^{+}_{t}(T_{+},N) dt dT_{+}$. So taking the double Laplace transform of Eq.~\eqref{eq:fTPNk},  after substitution of $\psi(\tau)$,  we have
\begin{eqnarray} \label{eq:futkN}
\hat{f}^{+}_{s}(u,2k+1) &=& \hat{\psi}^{k+1} (s+u)\hat{\psi}^{k}(s)\Big( \frac{1-\hat{\psi}(s)}{s}\Big) \,\,\,\ if \,\,\,\ N=2k+1 \nonumber\\
\hat{f}^{+}_{s}(u,2k) &=& \hat{\psi} ^{k} (s+u)\hat{\psi}^{k}(s)\Big( \frac{1-\hat{\psi}(s+u)}{s+u}\Big) \,\,\,\ if \,\,\,\ N=2k.
\end{eqnarray}
Thus using Eq.~\eqref{eq:futkN} for summing over all the values of $N$  in Eq.~\eqref{eq:fTpneq} we get
\begin{eqnarray}\label{eq:fsusr}
\hat{f}_{s}(u) &=& \Bigg( \hat{\psi}(s+u) \frac{1-\hat{\psi}(s)}{s} + \frac{1-\hat{\psi}(s+u)}{s+u} \Bigg) \frac{1}{1-\hat{\psi}(s+u) \hat{\psi}(s)}.
\end{eqnarray}
%%%%%%%%%%%%%%%%%%%%%%%%%
For exponentially distributed waiting times  $\hat{\psi}(s)=1/(1+\langle \tau \rangle s)$, by substituting $\hat{\psi}(s)$ in Eq.~\eqref{eq:fsusr}  we get that  the double Laplace transform of Eq.~\eqref{eq:fTpneq}  is given by
\begin{eqnarray}\label{eq:fTuneq}
\hat{f}_{s}(u)= \frac{2+ \langle \tau \rangle s}{2s + \langle \tau \rangle s^{2} +(1+\langle \tau \rangle s) u}.
\end{eqnarray}
By the same procedures used in Appendix ~\ref{sec:A3}, the inversion of the double Laplace transform of Eq.~\eqref{eq:fTuneq} yields %%%%%%%%%%%%%%%%%%%%%%
\begin{eqnarray}\label{eq:fTpneqsr}
f_{t}(T_{+})&=&\delta(t-T_{+})e^{-\frac{t}{\langle \tau \rangle}}+\frac{e^{- \frac{t}{\langle \tau \rangle}}}{\langle \tau \rangle}{}_{0}\tilde{F}_{1}\Bigg(;1;\frac{T_{+}(t-T_{+})}{\langle \tau \rangle ^{2}} \Bigg)\nonumber \\
&+& \frac{e^{- \frac{t}{\langle \tau \rangle}}}{\langle \tau \rangle ^{2} } T_{+}{}_{0}\tilde{F}_{1}\Bigg(;2;\frac{T_{+}(t-T_{+})}{\langle \tau \rangle ^{2}} \Bigg).
\end{eqnarray} 
%%%%%%%%%%%%%%%%%%%%%%%%%%%%%
Employing the identity $I_{\nu}(y)=(y/2)^{\nu}{}_{0}\tilde{F}_{1}(;\nu+1;y^{2}/4)$ ~\cite{BesselM} and changing variables we obtain the PDF of the occupation fraction, which  follows
%%%%%%%%%%%%%%%%%%%%%%%%
\begin{eqnarray}\label{eq:fppneqsr}
g_{t}(p_{+})&=&\delta(1-p_{+})e^{-\frac{t}{\langle \tau \rangle}} + \frac{t}{\langle \tau \rangle} \Bigg\lbrace I_{0}\Bigg( \frac{2t}{\langle \tau \rangle} \sqrt{p_{+}(1-p_{+})}\Bigg) \nonumber \\
&+& p_{+} \frac{I_{1}\Big( \frac{2t}{\langle \tau \rangle} \sqrt{p_{+}(1-p_{+})}\Big)}{\sqrt{p_{+}(1-p_{+})}}\Bigg\rbrace e^{-\frac{t}{\langle \tau \rangle}}.
\end{eqnarray}
%%%%%%%%%%%%%%%%%%%%%%%%%%
By taking the series expansion of Eq.~\eqref{eq:fppneqsr} in the limit  $t\longrightarrow 0$, the PDF of $p_{+}$ can be approximated by
%%%%%%%%%%%%%%%%%%%%%%%%%%
\begin{eqnarray}\label{eq:fppneqsrun}
g_{t}(p_{+}) \sim \delta(1-p_{+})e^{-\frac{t}{\langle \tau \rangle}} + \frac{t}{\langle \tau \rangle}.
\end{eqnarray}
%%%%%%%%%%%%%%%%%%%%%%%%%%%%
Therefore for $1>p_{+}>0$ the  PDF of $p_{+}$ follows a uniform distribution (see left panel of Figure~\ref{fig:pdfpsrneq}), as in the case of equilibrium initial conditions (Eq.~\eqref{eq:fppu}).

$P(x,t)$ is obtained by exploiting the uniform approximation of $g_{t}(p_{+})$ in Eq.~\eqref{eq:fppneqsrun}, i.e 
%%%%%%%%%%%%%%%%%%%%%%%%%%%
\begin{eqnarray}\label{eq:pxsruneq}
P(x,t)\sim  \frac{e^{-\frac{t}{ \langle\tau \rangle } - \frac{x^{2}}{4D_{+}t}}}{\sqrt{4\pi D_{+}t}} + \frac{t}{ \langle\tau\rangle}\Bigg\lbrace \frac{2 e^{-\frac{x^{2}}{4D_{+}t}}}{ \sqrt{4\pi D_{+} t}}- \frac{\vert x \vert}{2 D_{+} t}\Bigg[ 1 - Erf\Bigg( \frac{\vert x \vert}{\sqrt{4D_{+}t}}\Bigg)  \Bigg] \Bigg\rbrace.
\end{eqnarray} 
%%%%%%%%%%%%%%%%%%%%%%%%%%
Eq.~\eqref{eq:pxsruneq} follows the same structure as Eq.~\eqref{eq:pxsru}, i.e., the case with equilibrium initial conditions.

When  $\langle \tau \rangle_{+} \neq \langle \tau \rangle_{-}$, we have to include $\psi_{\pm}(\tau)$  and $\hat{\psi}_{\pm}(s)$ in   Eq.~\eqref{eq:fTPNk} and Eq.~\eqref{eq:futkN}.  Summing the resulting expressions in Eq.~\eqref{eq:fTpneq} we obtain 
\begin{eqnarray}\label{eq:fsudr}
\hat{f}_{s}(u) &=& \Bigg( \hat{\psi}_{+}(s+u) \frac{1-\hat{\psi}_{-}(s)}{s} + \frac{1-\hat{\psi}_{+}(s+u)}{s+u} \Bigg) \frac{1}{1-\hat{\psi}_{+}(s+u) \hat{\psi}_{-}(s)}.
\end{eqnarray}
By employing $\hat{\psi}(s)=1/(1+\langle \tau \rangle_{\pm} s)$ in Eq.~\eqref{eq:fsudr} we obtain that the double Laplace transform of the PDF of $T_{+}$ is provided  by \cite{BBel2005}
%%%%%%%%%%%%%%%%%%%%%%%%%%
\begin{eqnarray}\label{eq:fTpusnqdr}
\hat{f}_{s}(u)= \frac{\langle \tau\rangle_{+} + \langle \tau \rangle_{-} +  \langle \tau \rangle_{+} \langle \tau \rangle_{-} s }{ \langle \tau \rangle_{-} s + \langle \tau \rangle_{+} (1 + \langle \tau \rangle_{-} s)(s+u)}.
\end{eqnarray} 
%%%%%%%%%%%%%%%%%%%%%%%%%%
The inverse Laplace transform of Eq.~\eqref{eq:fTpusnqdr} is obtained by the same procedures explained above and in Appendix~\ref{sec:A3}, eventually
%%%%%%%%%%%%%%%%%%%%%%%%%%%%
\begin{eqnarray}\label{eq:fppneqdr}
g_{t}(p_{+})&=& \delta(1-p_{+})e^{-\frac{t}{\langle \tau \rangle_{+}}}+ \frac{t}{\langle \tau \rangle_{+}} \Bigg\lbrace  I_{0}\Bigg(2t \sqrt{\frac{p_{+}(1-p_{+})}{\langle \tau \rangle_{+} \langle \tau \rangle_{-}}} \Bigg)\nonumber \\
&+& \sqrt{\frac{\langle \tau \rangle_{+}}{\langle \tau \rangle_{-}}} p_{+}  \frac{I_{1} \Big( 2t \sqrt{\frac{p_{+}(1-p_{+})}{\langle \tau \rangle_{+} \langle \tau \rangle_{-}}} \Big)}{\sqrt{p_{+}(1-p_{+})}}\Bigg\rbrace e^{- \frac{tp_{+}}{\langle \tau \rangle_{+}} - \frac{t(1-p_{+})}{\langle \tau \rangle_{-}}}.
\end{eqnarray} 
%%%%%%%%%%%%%%%%%%%%%%%%%%%%
We recover Eq.~\eqref{eq:fppneqsr} when $\langle \tau \rangle_{+} = \langle \tau \rangle_{-}= \langle \tau \rangle$. In the short time limit Eq.~\eqref{eq:fppneqdr} follows as well a uniform distribution for $1>p_{+} >0$, see right panel of Figure~\ref{fig:pdfpsrneq}. In this case the PDF of $p_{+}$ is  given by
\begin{eqnarray}\label{eq:fppneqdrun}
g_{t}(p_{+})\sim \delta(1-p_{+})e^{-\frac{t}{\langle \tau \rangle_{+}}}+ \frac{t}{\langle \tau \rangle_{+}}. 
\end{eqnarray}
%%%%%%%%%%%%%%%%%%%%%%%%%%%%%
Thus for exponentially distributed waiting times, for equilibrium and non-equilibrium conditions, in the short time regime the PDF of the occupation fraction is always uniform. This feature only valid for exponentially distributed waiting times, and it is not necessarily fulfilled for  other distributions of waiting times.  

$P(x,t)$ for the specific case of non-equilibrium initial conditions and exponentially distributed waiting times is 
%%%%%%%%%%%%%%%%%%%%%%%%%%%
\begin{eqnarray}\label{eq:pxdruneq}
P(x,t)\sim \frac{e^{-  \frac{t}{\langle \tau \rangle_{+} } - \frac{x^{2}}{4D_{+}t} } }{\sqrt{4\pi D_{+}t}} + \frac{t}{\langle \tau \rangle_{+}} \Bigg\lbrace \frac{e^{- \frac{x^{2}}{4D_{+}t}}}{\sqrt{\pi D_{+}t}}+ \frac{\vert x \vert}{2D_{+}t} \Bigg[1- Erf\Bigg(\frac{\vert x \vert}{\sqrt{4D_{+}t}} \Bigg) \Bigg]\Bigg \rbrace. 
\end{eqnarray}
%%%%%%%%%%%%%%%%%%%%%%%%%%%%%%

\section{\label{sec:A2} Deduction of $g_t(p_+)$ for waiting times with similar mean waiting times}

We can use the results found in ~\cite{MW1996,ML2017} for  inverting  the double Laplace transform of  $S_{t}$ as provided by  Eq.~\eqref{eq:fSvs1}. 
For a Fourier-Laplace transform   
$\hat{\kappa}(\omega,s)=\int_{-\infty}^{\infty} \int_{0}^{\infty}e^{i\omega\tilde{x}}e^{-st}\kappa(\tilde{x},t) dsdx$
of the form 
\begin{eqnarray}\label{eq:MWp}
\hat{\kappa}(\omega,s)=\frac{2\tilde{\lambda} + s}{s^{2}+2\tilde{\lambda} s + c^{2}\omega^{2}},
\end{eqnarray}
the inversion yields the result~\cite{MW1996,ML2017}
%%%%%%%%%%%%%%%%%%%%%%%%%%%%%%%%%%
\begin{eqnarray}\label{eq:Pxtwm}
\kappa(\tilde{x},t)=\frac{1}{2}e^{- \frac{t}{ \tilde{\lambda}}}\Big \lbrace \delta(\tilde{x}-ct)+\delta(\tilde{x}+ct)\Big\rbrace+\frac{1}{2\tilde{\lambda} c}\Theta(ct-\vert \tilde{x} \vert)\big[ I_{0}(z(t))+\frac{t }{ \tilde{\lambda} z(t)}I_{1}(z(t)) \big],
\end{eqnarray}
%%%%%%%%%%%%%%%%%%%%%%%%%%%%%%%%%%%%%
with $z(t)=\frac{1}{\tilde{\lambda} c}\sqrt{c^{2}t^{2}-\tilde{x}^{2}}$. 

Now we can compare the double Laplace transform of $S_{t}$ given in Eq.\eqref{eq:fSvs1} with the result shown in   Eq.~\eqref{eq:MWp}. Concretely  we can relate the  Laplace variable $v$ with the Fourier variable $\omega $ in $\hat{\kappa}(\omega,s)$. Since the Laplace transform and the Fourier transform are exponential operators,  by setting $v=i c \omega$ we can make the former  equivalent to a Fourier transform and we  use the expression given by Eq.\eqref{eq:Pxtwm} for inverting $\phi_{s}(v)$. In our case $c=1$, so  $S_{t}\Leftrightarrow \omega$ are Fourier conjugates and therefore the inversion of $\phi_{s}(v)$ is results in
%%%%%%%%%%%%%%%%%%%%%%%%%%%%%%%%%%%%%%%
\begin{eqnarray}\label{eq:pdfSsr}
\phi_{t}(S_{t})&=&\frac{1}{2}e^{-\frac{t}{\langle \tau \rangle}}\Bigg\lbrace \delta(S_{t}-t)+ \delta(S_{t}+t) \Bigg\rbrace \nonumber \\
&+&\frac{\Theta(t-\vert S_{t}\vert)}{2\langle \tau \rangle}\Bigg[ I_{0}\Big(\frac{\sqrt{t^{2}-S^{2}_{t}}}{\langle \tau \rangle}\Big)+ \frac{tI_{1}\Big( \frac{\sqrt{t^{2}-S^{2}_{t}}}{\langle \tau \rangle}\Big)}{\sqrt{t^{2}-S^{2}_{t}}}\Bigg].
\end{eqnarray} 
%%%%%%%%%%%%%%%%%%%%%%%%%%%%%%%%%%%%%%%
By changing variables, $S_{t}=2T_{+}-t=2p_{+}t-t$, we obtain Eq.~\eqref{eq:MWpdfp} in a straightforward manner.
The result  in Eq.~\eqref{eq:MWpdfp} can also be obtained by  the inversion of the double Laplace transform   ($t\Leftrightarrow s $ and $T_{+} \Leftrightarrow u$) of the PDF of $T_{+}$ Eq.~\eqref{eq:pdfTp} (see Appendix ~\ref{sec:A3}). In this case the double Laplace transform of the PDF of $T_{+}$ is given by 
%%%%%%%%%%%%%%%%%%%%%%%%%%%%%%%%%%%%
\begin{eqnarray}\label{eq:fsuTgl}
\hat{f}_{s}(u)= \frac{4+2 \langle  \tau \rangle s +\langle  \tau \rangle u}{2 \langle  \tau \rangle s^{2} + 4s + (2+2\langle  \tau \rangle s)u}.
\end{eqnarray}
%%%%%%%%%%%%%%%%%%%%%%%%

Finally, we mention that the moments of $T_{+}$ and therefore $p_{+}$, can be obtained by expanding Eq.~\eqref{eq:fsuTgl} in powers of $u$ as 
\begin{eqnarray}\label{eq:fsuTglmom}
\hat{f}_{s}(u)= \frac{1}{s}- \frac{1}{2s^{2}}u+ \frac{1+ \langle \tau \rangle s}{2s^{3} (2+\langle \tau \rangle s)} u^{2} + O(u^{3}).
\end{eqnarray}
The first two moments of $T_{+}$ are then
%%%%%%%%%%%%%%%%%%%%%%%%%%%%%%%%%
\begin{eqnarray}
\langle T_{+} \rangle &\sim & \frac{t}{2},\\
\langle T^{2}_{+} \rangle &\sim & \Big( \frac{\langle \tau \rangle }{4} + \frac{t}{4}\Big)t + \frac{\langle \tau \rangle ^{2}}{8}\Big( e^{-\frac{2t}{\langle \tau \rangle}} -1\Big).
\end{eqnarray}
%%%%%%%%%%%%%%%%%%%%%%%%%%%%%%%%%
For $\langle p_{+} \rangle= \langle T_{+} \rangle /t$, we obtain 
%%%%%%%%%%%%%%%%%%%%%%%%%%%%%%%%%%%
\begin{eqnarray}
\langle p_{+} \rangle &\sim & \frac{1}{2},\\
\langle p^{2}_{+} \rangle &\sim &\frac{1}{4} + \frac{\langle \tau \rangle }{4t} +  \frac{\langle \tau \rangle ^{2}}{8t^{2}}\Big( e^{-\frac{2t}{\langle \tau \rangle}} -1\Big),\\
Var(p_{+})&\sim & \frac{\langle \tau \rangle }{4t} +  \frac{\langle \tau \rangle ^{2}}{8t^{2}}\Big( e^{-\frac{2t}{\langle \tau \rangle}} -1\Big). 
\end{eqnarray}
%%%%%%%%%%%%%%%%%%%%%%%%%%%%%%%%%%%%%%%
\section{\label{sec:A3} Deduction of $g_t(p_+)$ for waiting times with $\langle \tau \rangle_{+}\neq \langle \tau \rangle_{-} $}
Here we show the procedure for obtaining Eq.~\eqref{eq:pdfpdr} in section ~\ref{sec:42}. Starting from the double Laplace transform of the PDF of $T_{+}$ given by Eq.~\eqref{eq:fsudrm}, first by inverting with respect to $u\Longleftrightarrow T_{+}$ we get

\begin{eqnarray}\label{eq:fTpsdr}
\hat{f}_{s}(T_{+})&=& \frac{\langle \tau \rangle_{-} ^{2} \delta(T_{+})}{(\langle \tau\rangle_{+} + \langle \tau \rangle_{-} )(1+ \langle \tau\rangle_{-} s)} \nonumber \\
&+& \frac{(\langle \tau\rangle_{+} + \langle\tau\rangle_{-} + \langle \tau \rangle_{+} \langle \tau\rangle_{-} s)^{2}}{ \langle \tau\rangle_{+} ( \langle \tau\rangle_{+} + \langle \tau \rangle_{-}  ) (1+ \langle \tau \rangle_{-}  s)^{2}} e^{-T_{+}s \Big(\frac{\langle \tau \rangle_{+} + \langle \tau \rangle_{-} + \langle \tau\rangle_{+} \langle \tau \rangle_{-} s }{\langle \tau\rangle_{+} + \langle \tau \rangle_{+} \langle \tau \rangle_{-} s}\Big)},\nonumber \\
\end{eqnarray}
the exponent in Eq.~\eqref{eq:fTpsdr} can be written as $-T_{+}s \Big(1+ \frac{\langle \tau \rangle_{-}}{ \langle \tau \rangle_{+} + \langle \tau\rangle_{+} \langle \tau \rangle_{-}  s}\Big)$. The inversion of Eq.~\eqref{eq:fTpsdr} with respect to $s \Leftrightarrow t$ can be  expressed as
\begin{eqnarray}\label{eq:fTptdr}
\hat{f}_{s}(T_{+})&=& \frac{\langle \tau \rangle_{-} e^{- \frac{t}{\langle \tau \rangle_{-}}}}{\langle \tau \rangle_{+} + \langle \tau \rangle_{-}}\delta(T_{+})+ \mathcal{L}^{-1}\lbrace \hat{q}(s) \hat{h}(s)\rbrace.
\end{eqnarray}
So the inversion of the second term in Eq.~\eqref{eq:fTptdr} is given by the convolution theorem, following $\mathcal{L}^{-1}\lbrace \hat{q}(s) \hat{h}(s) \rbrace= \int ^{t} _{0} \mathcal{L}^{-1}\lbrace \hat{q}(s) \rbrace \vert_{t-t^{\prime}} \mathcal{L}^{-1} \lbrace \hat{h}(s) \rbrace\vert_{t^{\prime}} dt^{\prime} $. With $\hat{q}(s)=\frac{(\langle \tau \rangle_{+} + \langle \tau \rangle_{-} + \langle \tau \rangle_{+} \langle \tau \rangle_{-}   s)^{2}}{\langle \tau \rangle_{+} (\langle \tau \rangle_{+}  + \langle \tau \rangle_{-} )(1+ \langle \tau \rangle_{-}  s)^{2}}$ and $\hat{h}(s)=e^{-T_{+}s} e^{- \frac{T_{+}\langle \tau \rangle_{-} s}{\langle \tau \rangle_{+} + \langle \tau\rangle_{+} \langle \tau \rangle_{-} s}}$. The inverse Laplace transform of $\hat{q}(s)$ is given by
\begin{eqnarray}\label{eq:fsilt}
\mathcal{L}^{-1} \lbrace \hat{q}(s)\rbrace = \frac{2e^{-\frac{t}{\langle \tau \rangle_{-} }}}{ \langle \tau \rangle_{+} + \langle \tau \rangle_{-}}+  \frac{te^{-\frac{t}{\langle \tau \rangle_{-} }}}{ \langle \tau \rangle_{+} (\langle \tau \rangle_{+} + \langle \tau \rangle_{-})}+ \frac{\langle \tau \rangle_{+} \delta(t)}{ \langle \tau \rangle_{+} + \langle \tau \rangle_{-} }.
\end{eqnarray}
The inverse Laplace transform  of $\hat{h}(s)$  can be obtained by rewriting the exponent in the second term of $\hat{h}(s)$ as $- \frac{T_{+}\langle \tau \rangle_{-} s}{\langle \tau \rangle_{+} + \langle \tau\rangle_{+} \langle \tau \rangle_{-} s}=-\frac{T_{+}}{\langle \tau \rangle_{+}} + \frac{T_{+} \langle \tau \rangle_{-} }{ \langle \tau \rangle_{+}  \langle \tau\rangle_{-} + \langle \tau \rangle_{-} ^{2} \langle \tau\rangle_{+}  s}$, then we obtain
\begin{eqnarray}\label{eq:gsilt}
 \mathcal{L}^{-1} \lbrace \hat{h}(s)\rbrace = \mathcal{L}^{-1} \Big \lbrace e^{-\frac{T_{+} \langle \tau \rangle_{-} s}{ \langle \tau \rangle_{+} + \langle \tau \rangle_{+} \langle \tau \rangle_{-} s}} \Big \rbrace \Big \vert_{t-T_{+}}\Theta(t - T_{+}) \nonumber \\
 = e^{- \frac{T_{+}}{ \langle \tau \rangle_{+} }} \mathcal{L}^{-1} \Big \lbrace e^{\frac{T_{+} \langle \tau \rangle_{-} }{ \langle \tau \rangle_{+}  \langle \tau\rangle_{-} + \langle \tau \rangle_{-} ^{2} \langle \tau\rangle_{+}  s}} \Big \rbrace \Big \vert_{t-T_{+}}\Theta(t - T_{+}) \nonumber \\
= e^{- \frac{T_{+}}{ \langle \tau \rangle_{+} }} \Bigg[ e^{-\frac{(t-T_{+}) }{\langle \tau \rangle_{-}}} \sqrt{\frac{T_{+}}{ \langle \tau \rangle_{+} \langle \tau \rangle_{-}(t - T_{+})}} I_{1}\Bigg(2\sqrt{\frac{T_{+}(t-T_{+})}{\langle \tau \rangle_{+} \langle \tau \rangle_{-}}} \Bigg) \nonumber \\
 + \frac{e^{-\frac{(t-T_{+})}{\langle \tau \rangle_{-}}}}{\langle \tau \rangle_{+} \langle \tau \rangle_{-} ^{2}} \delta \Bigg(\frac{t -T_{+}}{\langle \tau \rangle_{+} \langle \tau \rangle_{-} ^{2} } \Bigg) \Bigg]\Theta(t - T_{+}).
\end{eqnarray}
%%%%%%%%%%%%%%%%%%%%
Substituting Eq.~\eqref{eq:fsilt} and Eq.~\eqref{eq:gsilt} in Eq.~\eqref{eq:fTptdr}, and after integration we obtain
%%%%%%%%%%%%%%%%%%%%%%%%%%
\begin{eqnarray}\label{eq:fTpdrfin}
f_{t}(T_{+})&=& \frac{\langle \tau \rangle_{-} e^{-\frac{t}{\langle \tau \rangle_{-}}}}{ \langle \tau \rangle_{+} + \langle \tau \rangle_{-}}\delta(T_{+}) \nonumber \\
&+& \frac{\langle \tau \rangle_{+} e^{-\frac{t}{\langle \tau \rangle_{+}}}}{ \langle \tau \rangle_{+} + \langle \tau \rangle_{-}}\delta(t-T_{+})  +\Bigg\lbrace \frac{2}{ \langle \tau \rangle_{+} + \langle \tau \rangle_{-}}  {}_{0}\tilde{F}_{1}\Bigg(;1;\frac{T_{+}(t-T_{+})}{\langle \tau \rangle_{+} \langle \tau \rangle_{-} } \Bigg) \nonumber \\ 
&+& \Bigg[ \frac{t-T_{+}}{\langle \tau \rangle_{+} (\langle \tau \rangle_{+} + \langle \tau \rangle_{-})}  + \frac{T_{+}}{\langle \tau \rangle_{-} (\langle \tau \rangle_{+} + \langle \tau \rangle_{-})} \Bigg] {}_{0}\tilde{F}_{1}\Bigg(;2;\frac{T_{+}(t-T_{+})}{\langle \tau \rangle_{+} \langle \tau \rangle_{-} } \Bigg) \Bigg\rbrace e^{-\frac{T_{+}}{\langle \tau \rangle_{+}} - \frac{(t-T_{+})}{\langle \tau \rangle_{-} }}.\nonumber\\
\end{eqnarray}
%%%%%%%%%%%%%%%%%%%%%%%%%%
Employing  the identity $I_{\nu}(y)=(y/2)^{\nu}{}_{0}\tilde{F}_{1}(;\nu+1;y^{2}/4)$ ~\cite{BesselM} and changing variables we obtain the form of $g_t(p_+)$ as provided by Eq.~\eqref{eq:pdfpdr}. This procedure can be employed  for a system with the same mean waiting times,  inverting the double Laplace transform Eq.~\eqref{eq:fsuTgl} and obtaining $g_{t}(p_{+})$ shown in Eq.~\eqref{eq:MWpdfp}. And also for the non - equilibrium cases treated in Appendix ~\ref{sec:A1}, see Eq.~\eqref{eq:fppneqsr} and Eq.~\eqref{eq:fppneqdr}.

Finally we show the corresponding  first two  moments of $T_{+}$ and $p_{+}$.  As we proceeded in section ~\ref{sec:A2} we obtain the moments of $T_{+}$ by expanding in powers of $u$ Eq.~\eqref{eq:fsudr},
which yields
%%%%%%%%%%%%%%%%%%%%%%%%%%
\begin{eqnarray}
& \langle T_{+}\rangle \sim \frac{\langle \tau\rangle_{+} t}{ \langle \tau\rangle_{+} + \langle \tau \rangle_{-}},\\
& \langle T^{2}_{+}\rangle \sim \frac{\langle \tau\rangle_{+} ^{2} t^{2}}{ (\langle \tau\rangle_{+} + \langle \tau \rangle_{-})^{2}} +\frac{2\langle \tau\rangle_{+} ^{2} \langle \tau \rangle_{-} ^{2} t}{ (\langle \tau\rangle_{+} + \langle \tau \rangle_{-})^{3}} + \frac{2 \langle \tau \rangle_{+} ^{3} \langle \tau \rangle_{-} ^{3}}{ ( \langle \tau \rangle_{+} + \langle \tau \rangle_{-})^{3}}\Big(e^{- \frac{(\langle \tau\rangle_{+} + \langle \tau \rangle_{-}) t}{ \langle\tau \rangle_{+} \langle \tau \rangle_{-}}}-1 \Big) . 
\end{eqnarray} 
%%%%%%%%%%%%%%%%%%%%%%%%%%%%%%
Therefore the moments of $p_{+}$ are 
\begin{eqnarray}\label{eq:ppav}
& \langle p_{+}\rangle \sim \frac{\langle \tau\rangle_{+}}{ \langle \tau\rangle_{+} + \langle \tau \rangle_{-}},\\
& \langle p^{2}_{+}\rangle \sim \frac{\langle \tau\rangle_{+} ^{2} }{ (\langle \tau\rangle_{+} + \langle \tau \rangle_{-})^{2}} +\frac{2\langle \tau\rangle_{+} ^{2} \langle \tau \rangle_{-} ^{2} }{ t(\langle \tau\rangle_{+} + \langle \tau \rangle_{-})^{3}} + \frac{2 \langle \tau \rangle_{+} ^{3} \langle \tau \rangle_{-} ^{3}}{ t^{2}( \langle \tau \rangle_{+} + \langle \tau \rangle_{-})^{3}}\Big(e^{- \frac{(\langle \tau\rangle_{+} + \langle \tau \rangle_{-}) t}{ \langle\tau \rangle_{+} \langle \tau \rangle_{-}}}-1 \Big), \\
& Var(p_{+})\sim \frac{2\langle \tau\rangle_{+} ^{2} \langle \tau \rangle_{-} ^{2} }{ t(\langle \tau\rangle_{+} + \langle \tau \rangle_{-})^{3}} + \frac{2 \langle \tau \rangle_{+} ^{3} \langle \tau \rangle_{-} ^{3}}{ t^{2}( \langle \tau \rangle_{+} + \langle \tau \rangle_{-})^{3}}\Big(e^{- \frac{(\langle \tau\rangle_{+} + \langle \tau \rangle_{-}) t}{ \langle\tau \rangle_{+} \langle \tau \rangle_{-}}}-1 \Big).
\end{eqnarray} 

\section{\label{sec:A4} Deduction of the MSD in a two state model with $\langle \tau \rangle _{+} \neq \langle \tau \rangle _{-}$ }

From Eq.~\eqref{eq:x2d} we can compute the second moment of $x(t)$ as
\begin{eqnarray}\label{eq:x22d}
\langle x^{2}(t)\rangle &=& \Big \langle \Big(\sqrt{2D_{+} T_{+}} \xi_{1} + \sqrt{2D_{-} (t- T_{+})} \xi_{2} \Big)^{2} \Big\rangle \nonumber,\\
&=& 2D_{+}\langle T_{+}\rangle + 2D_{-}(t-\langle T_{+} \rangle),
\end{eqnarray}
In the second line of Eq.~\eqref{eq:x22d} we have employed the linearity of $\langle \cdot \rangle$, and the properties of independent standard normal random variables, \textit{i.e.}  $\langle \xi_{i}^{2}\rangle=1$ and $\langle \xi_{i}\xi_{j}\rangle=0$ (with $i,j\in\lbrace 1,2 \rbrace$ and $i \neq j$). So now we just have to find $\langle T_{+}\rangle$. In order to do that, we start from the definition of average occupation time  
\begin{eqnarray}\label{eq:avTPdL}
\langle T_{+} \rangle &=&\displaystyle \int \limits _{0}^{\infty} T_{+}f_{t}(T_{+})dT_{+}, \nonumber\\
&=& \displaystyle \int \limits _{0}^{\infty}f_{t}(T_{+}) \Big(-\frac{d}{du}e^{-uT_{+}}\Big)\Big\vert _{u=0} dT_{+}, \nonumber \\
&=& -\underset{u \to 0}{\lim} \Big( \frac{d}{du}\hat{f}_{t}(u)\Big).
\end{eqnarray}
With $\hat{f}_{t}(u)=\int_{0}^{\infty} f_{t}(T_{+})e^{-uT_{+}}$, and $T_{+} \Leftrightarrow u$ Laplace conjugates. Now for equilibrium initial conditions, the PDF of the occupation  time is given by 
\begin{eqnarray}\label{eq:PDFTpJ}
f_{t}(T_{+})= \frac{\langle \tau \rangle_{+}}{ \langle \tau \rangle_{+}  + \langle \tau \rangle_{-} } \displaystyle \sum \limits _{N=0}^{\infty}f_{t}^{+}(T_{+},N)+ \frac{\langle \tau \rangle_{-}}{ \langle \tau \rangle_{+}  + \langle \tau \rangle_{-} } \displaystyle \sum \limits _{N=0}^{\infty}f_{t}^{-}(T_{+},N), 
\end{eqnarray}
with $f_{t}^{\pm}(T_{+},N)$ the joint PDF of the occupation times at $D_{+}$ and the number of jumps between states during $t$, once started from $D_{\pm}$. When starting from $D_{+}$ and having $N=2k+1$ or $N=2k$ jumps, the occupation time in each case satisfies Eq.~\eqref{eq:Tpo}. In the case when the initial state is at $D_{-}$ we have 
\begin{eqnarray}\label{eq:Tpodm}
T_{+}&=&\tau_{2}+\tau_{4}+\ldots+\tau^{\ast} \nonumber\,\,\,\ if \,\,\,\ N=2k+1,\\
T_{+}&=&\tau_{2}+\tau_{4}+\ldots+\tau_{N}\,\,\,\,\,\,\,\,\,\,\,\,\,\,\,\,\,\,\,\,\,\,\,\ if \,\,\,\ N=2k,
\end{eqnarray}
with $\tau^{\ast}=t-t_{N}$, the backward recurrence time. The definition of the joint PDF $f_{t}^{\pm}(T_{+},N)$ is already given in Eq.~\eqref{eq:fTpNJGL}. And its double Laplace transform  $\hat{f}_{s}^{\pm}(u,N)= \int_0^\infty\int_0^\infty f_{t}(T_{+},N)\exp(-uT_+ -st)\,dT_+\,dt$, is shown in Eq.~\eqref{eq:fsuNGo}- Eq.~\eqref{eq:fsuNGem}. When $N=0$, we have
\begin{eqnarray}\label{eq:pdftpz}
\hat{f}^{+}_{s}(u,0)&=&\frac{1-\Big(\frac{1-\hat{\psi}_{+}(s+u) }{\langle \tau \rangle_{+} (s+u)} \Big)}{s+u}, \nonumber \\
\hat{f}^{-}_{s}(u,0)&=&\frac{1-\Big(\frac{1-\hat{\psi}_{-}(s) }{\langle \tau \rangle_{-} s} \Big)}{s}.
\end{eqnarray}
Now for obtaining $\hat{f}_{s}(u)$ we compute the double Laplace transform of Eq.~\eqref{eq:PDFTpJ} and then we  sum Eq.~\eqref{eq:fsuNGo}- Eq.~\eqref{eq:fsuNGem} and Eq.~\eqref{eq:pdftpz} for all values of $N$. Thereafter we compute the derivative of $\hat{f}_{s}(u)$ with respect to $u$ and its corresponding limit when $u\longrightarrow 0$. Following  algebraic simplifications we yield to 
\begin{eqnarray}\label{eq:limfsu}
\underset{u \to 0}{\lim} \Big( \frac{d}{du}\hat{f}_{s}(u)\Big) = -\frac{\langle \tau \rangle _{+}}{\langle \tau \rangle_{+} + \langle \tau \rangle_{-}}\frac{1}{s^{2}}. 
\end{eqnarray}
For obtaining  the average occupation time Eq.~\eqref{eq:avTPdL}, we  invert Eq.\eqref{eq:limfsu} with respect to $s$, having
\begin{eqnarray}\label{eq:avTpf}
\langle T_{+} \rangle= \Bigg(\frac{\langle \tau \rangle_{+} }{ \langle \tau \rangle_{+} + \langle \tau \rangle_{-} } \Bigg) t. 
\end{eqnarray}
Finally substituting Eq.~\eqref{eq:avTpf} in Eq.~\eqref{eq:x22d} we get Eq.~\eqref{eq:MSDS}, which indicates  that the MSD is linear with respect to $t$, for any value of time $t$.

%\bibliography{main1}% Produces the bibliography via BibTeX.


%%%%%%%%%%%%%%%%%%%%%%%%%%%%%%%%%%%%%%%%%%%%%%%%%%%%%%%%%%%%%%%%%%%%%%%%%%%

%%%%%%%%%%%%%%%%%%%%%%%%%%%%%%%%%%%%%%%%%%%%%%%%%%%%%%%%%%%%%%%%%%%%%%%%%%%%%

\end{document}